\documentclass[11pt,a4paper]{article} 
\pdfoutput=1
\usepackage{epsfig}
\usepackage{graphicx}
\usepackage{jheppub}
\usepackage{amsmath}
\usepackage{amsfonts}
\usepackage{amssymb}
\usepackage{graphicx}

\usepackage{color}

\def\beq{\begin{equation}}
\def\eeq{\end{equation}}
\def\beqa{\begin{eqnarray}}
\def\eeqa{\end{eqnarray}}
\newcommand{\nn}{\nonumber}

\def\eq#1{Eq.~(\ref{#1})}
\newcommand\Eqns[2]{Eqs.\,(\ref{#1}) and~(\ref{#2})}
\newcommand\Eqnss[2]{Eqs.\,(\ref{#1})--(\ref{#2})}

\newcommand{\secn}[1]{section~\ref{#1}}

\newcommand\Appx[1]{appendix~\ref{#1}}

\newcommand{\ord}{{\cal O}}
\newcommand{\RE}{{\rm Re}}
\newcommand{\IM}{{\rm Im}}

\def\ifm{\ifmmode}

\definecolor{darkgreen}{rgb}{0.0, 0.4, 0.13}

\def \as {\relax\ifmmode\alpha_s\else{$\alpha_s${ }}\fi}

\def \al #1 {\frac {\as({#1})}{\pi} }
\def \ds #1 {\ooalign{$\hfil/\hfil$\crcr$#1$}}

\def \a{\alpha}

\def\eps{\varepsilon}
\def\eq#1{Eq.~(\ref{#1})}

\allowdisplaybreaks[1]

\title{Analyzing high-energy factorization beyond next-to-leading logarithmic accuracy}

\author[a]{Vittorio Del Duca,}
\author[b]{Giulio Falcioni,}
\author[b]{Lorenzo Magnea,}
\author[b]{Leonardo Vernazza}

\affiliation[a]{INFN, Laboratori Nazionali Frascati, \\
00044 Frascati (Roma), Italy}
\affiliation[b]{Dipartimento di Fisica Teorica, Universit{\`a} di Torino, 
and \\ INFN, Sezione di Torino, Via P. Giuria 1, I-10125 Torino, Italy}

\emailAdd{delduca@lnf.infn.it}
\emailAdd{giulio.falcioni@unito.it}
\emailAdd{lorenzo.magnea@unito.it}
\emailAdd{vernazza@to.infn.it}

\abstract{We provide a complete and detailed study of the high-energy limit of four-parton 
scattering amplitudes in QCD, giving explicit results at two loops and higher orders, and 
going beyond next-to-leading logarithmic (NLL) accuracy. Building upon recent results, we 
use the techniques of infrared factorization to investigate the failure of the simplest form of 
Regge factorization, starting at next-to-next-to-leading logarithmic accuracy (NNLL) in 
$\ln(s/|t|)$. We provide detailed accounts and explicit expressions for the terms responsible 
for this breaking in the case of two-loop and three-loop quark and gluon amplitudes in QCD; 
in particular, we recover and explain a known non-logarithmic double-pole contribution at 
two-loops, and we compute all non-factorizing single-logarithmic singular contributions at 
three loops. Conversely, we use high-energy factorization to show that the hard functions 
of infrared factorization vanish in $d = 4$ to all orders in the coupling, up to NLL accuracy 
in $\ln(s/|t|)$. This provides clear evidence for the infrared origin of high-energy logarithms. 
Finally, we extend earlier studies to $t$-channel exchanges of color representations beyond 
the octet, which enables us to give predictions based on the dipole formula for single-pole 
NLL contributions at three and four loops.}


\begin{document}
\begin{flushright}
\vspace*{-25pt}
\end{flushright}
\maketitle
\allowdisplaybreaks 


\section{Introduction}
\label{intro}

In the high-energy limit, in which the squared centre-of-mass energy $s$ is much 
larger than the typical momentum transfer $(-t)$, so that $|s/t|\to \infty$ with $t$ held 
fixed, a four-point tree-level gauge theory scattering amplitude acquires a factorized
structure, given by a $t$-channel propagator, associated with the highest-spin particle
in the theory (in the case of QCD, a gluon), connecting two emission vertices, termed 
{\it impact factors}, which characterize the particles undergoing the scattering. The 
impact factors depend on the specific scattering process, while the $t$-channel 
propagator is process independent.

When loop corrections are included, the $t$-channel propagator gets dressed 
according to the schematic form~\cite{Balitsky:1979ap},
\beq
  \frac1{t} \, \to \, \frac1{t} \, \left( \frac{s}{-t}\right)^{\alpha(t)} \, ,
\label{eq:regge}
\eeq
where $\alpha(t)$ is a function of the coupling constant and of the momentum 
transfer $t$, which  can be expanded perturbatively at weak coupling. The 
expansion of \eq{eq:regge} in powers of the coupling then generates the leading 
logarithmic corrections to the amplitude in $\ln(s/|t|)$. Because of the analytic structure
of \eq{eq:regge}, which is typical of Regge theory, $\alpha(t)$ is called {\it Regge
trajectory}.

For the real part of the amplitude, the $t$-channel picture, often termed {\it 
high-energy factorization}, is in fact accurate at leading and at next-to-leading 
logarithmic (NLL) accuracy in $\ln(s/|t|)$~\cite{Fadin:2006bj}. Because the 
amplitude has a $t$-channel ladder-like structure, we can assume it to be 
even under $s \leftrightarrow u$ exchange. As a consequence, it must be 
composed of kinematic and color parts which are either both even or both odd 
under $s \leftrightarrow u$ exchange, a feature commonly referred to as `signature'
of the amplitude in the literature on Regge theory. As an example, let us consider 
the amplitude for gluon-gluon scattering. In this case, for the process $g(k_1) + 
g(k_2) \to g(k_3) + g(k_4)$, one may write~\cite{Fadin:1993wh}
\beqa
  && {\cal M}^{g g \rightarrow g g}_{a_1 a_2 a_3 a_4} \left( \frac{s}{\mu^2}, 
  \frac{t}{\mu^2}, \alpha_s (\mu^2) \right)
  \, = \, 4 \pi \alpha_s (\mu^2)  \, \, \frac{s}{t} \, \bigg[ \left({\bf T}^b 
  \right)_{a_1 a_3} C_{\lambda_1\lambda_3} (k_1, k_3) \bigg] \nonumber \\ 
  && \hspace{2cm} \times \, \left[ \left( 
  \frac{s}{- t} \right)^{\alpha(t)} \!\! + \left( \frac{- s}{- t} 
  \right)^{\alpha(t)} \right] \bigg[ \left( {\bf T}_b \right)_{a_2 a_4} 
  C_{\lambda_2\lambda_4}  (k_2, k_4) \bigg] \, ,
\label{Mgg2}
\eeqa
where $a_j$ and $k_j$ are the color index and momentum of gluon $j$, and 
${\bf T}^b$ is a color generator in the adjoint representation, so that $\left( {\bf T}^a 
\right)_{b c} = - {\rm i} f_{a b c}$. The impact factors, $C_{\lambda_i\lambda_j} (k_i, k_j)$, 
depend on the  helicities of the gluons, but are independent of the squared centre-of-mass 
energy $s$. In the weak coupling limit, both the impact factors and the Regge 
trajectory can be expanded in powers of the renormalized coupling $\alpha_s (\mu^2)$: 
they are then affected by infrared and collinear divergences, which are (implicitly) 
regularized by dimensional regularization in \eq{Mgg2}.

Beyond leading order, and for the real part of the amplitude beyond the NLL accuracy 
in $\ln(s/|t|)$, one should consider also the exchange of two or more reggeized gluons. 
Accordingly, one should include the contribution to the amplitude in which the kinematic 
and color parts are both even under $s \leftrightarrow u$ exchange, and in particular 
the case in which a color singlet is exchanged.

The process independence of $t$-channel gluon exchange implies that one 
can write formulae similar to \eq{Mgg2} for quark-quark and quark-gluon scattering,
differing only for the presence of the quark impact factor instead of the gluon impact 
factor. Considering them together with gluon-gluon scattering, as given by \eq{Mgg2}, 
one obtains a system of three equations, which can be used to determine the impact 
factors for quark and gluon scattering. In fact, one gets an over-constrained system 
of three equations and two unknowns. One can use two of the equations to determine 
the one-loop impact factors, and the third to perform a consistency check on 
high-energy factorization. Because the Regge trajectory and the impact factors 
can be expanded as a series in the coupling, this procedure can be repeated in 
principle at each loop order, although it is unwarranted for terms beyond the NLL 
accuracy. Specifically, the expansion of the Regge trajectory and of the impact factors 
at one loop shows, as expected, that each equation has a term proportional to $\ln(s/|t|)$, 
which is the same for all three amplitudes. That term gives the one-loop Regge trajectory, 
and the fact that is the same for all three equations shows its universality, {\it i.e.} its 
independence of the particular scattering process under consideration. Conversely, 
the term independent of $\ln(s/|t|)$ is different for each equation. One can then use 
two of the equations to determine the one-loop gluon and quark impact factors, and 
use the third to check the consistency of high-energy factorization.

Repeating the procedure above at two loops, one can use the terms proportional to 
$\ln(s/|t|)$ to determine the two-loop Regge trajectory and verify its universality, and 
the terms independent of $\ln(s/|t|)$ to compute the two-loop impact factors and check 
that high-energy factorization holds. Such a check, however, fails~\cite{DelDuca:2001gu}, 
due to the presence of a term proportional to $\alpha_s^2 \pi^2/\epsilon^2$, which 
therefore invalidates high-energy factorization, making the determination of the 
two-loop impact factors ambiguous. This is not totally unexpected, since terms 
independent of $\ln(s/|t|)$ at two loops are beyond NLL accuracy. It is however 
interesting to notice that the violation originates only in the term proportional to 
$\pi^2/\epsilon^2$, and not in the single pole nor in the finite part of the would-be 
impact factors. Furthermore, it must be emphasized that, in the context of Regge 
theory, \eq{Mgg2} is only an approximation, based on the assumption that only Regge 
poles appear in the angular momentum plane. Regge theory arguments predating 
QCD (see, for example, Ref.~\cite{Collins:1977jy}) suggest that this approximation 
is likely to break down, for logarithmic terms, at the three-loop level, at NNLL accuracy 
and for non-planar contributions to the amplitude. At this accuracy, one may in fact 
envisage contributions to the amplitude due to Regge cuts in the angular momentum 
plane, which are absent in expressions such as \eq{Mgg2}. These corrections to 
Regge-pole-based high-energy factorization were never previously pinned down 
in any specific computation of a scattering amplitude in the high-energy limit. The 
violation of universality observed at two loops in Ref.~\cite{DelDuca:2001gu}, as 
we show in the present paper, is a harbinger of precisely such phenomena at the 
three-loop level.

In recent years, a general approach to the high-energy limit of scattering amplitudes 
based on the universal properties of their infrared singularities has been developed 
in~\cite{DelDuca:2011ae,Bret:2011xm}, following the earlier results 
of~\cite{Korchemsky:1993hr,Korchemskaya:1994qp,Korchemskaya:1996je}. 
This approach suggests, in particular, that the violation of high-energy factorization 
reported in~\cite{DelDuca:2001gu} at order $\alpha_s^2$ and at next-to-next-to-leading 
logarithmic accuracy in $\ln(s/|t|)$ is due to the amplitude becoming non-diagonal 
in the $t$-channel-exchange basis. Such a violation iterates then at three loops 
in the $\alpha_s^3$ term proportional to $\ln(s/|t|)$, invalidating the universality of 
the three-loop Regge trajectory. Thus, the eventual definition of a universal three-loop 
Regge trajectory requires additional conditions.

In Refs.~\cite{DelDuca:2013ara,DelDuca:2013dsa} we have further developed 
the approach above, identifying the origin of the high-energy factorization violation 
discovered in~\cite{DelDuca:2001gu} at two loops. In order to be able to define 
unambiguously a universal Regge trajectory and the related impact factors beyond 
the NLL accuracy, we have proposed a way to isolate factorization-breaking terms at 
three loops and beyond. This goal can be achieved introducing a non-factorizing contribution 
to the amplitude, whose infrared and collinear divergent parts can then be unambiguously 
predicted using the tools described in~\cite{DelDuca:2011ae,Bret:2011xm}. We believe 
that a framework for consistently  identifying factorizing and non-factorizing contributions 
to high-energy amplitudes can be useful both in practical finite-order calculations, 
to assess the reliability of high-energy resummations, and for theoretical developments. 
Indeed, a precise expression for the discrepancy between pole-based Regge factorization 
and the actual perturbative results for the amplitude may be useful at least as a boundary 
condition for future attempts to extend high-energy factorization to include the contributions 
of Regge cuts. Furthermore, our results are a first step in the direction of systematically 
combining informations on amplitudes which arise from infrared factorization, which is exact 
to all orders in perturbation theory for all singular contributions to the amplitudes, with those 
arising from Regge factorization, which applies also to finite contributions to the amplitudes, 
but has limited validity in terms of logarithmic accuracy. 

In this paper, we extend the analysis of Refs.~\cite{DelDuca:2013ara,DelDuca:2013dsa} 
beyond leading poles, beyond three loops and beyond the leading color amplitude. 
Furthermore, we provide a more flexible framework for combining infrared factorization 
with the high-energy limit, which is better suited to disentangle the various color components
of the amplitudes. We are then able to provide detailed predictions for singular terms 
contributing to the high-energy limit of quark and gluon amplitudes in QCD up to three loops, 
and furthermore we are able to derive towers of constraints on real and imaginary parts of 
finite contributions to the amplitudes, valid to all orders in perturbation theory, up to NLL 
accuracy. These constraints show that, to the stated accuracy, the hard (infrared-finite) parts 
of the amplitudes can be chosen to vanish, so that {\it all} high energy logarithms (up to 
NLL included) are generated by the infrared operators arising from infrared factorization.
This result lends support to the conjecture that all high-energy logarithms may be
understood  as originating from a special class of infrared enhancements, as suggested
in~\cite{Korchemsky:1993hr,Korchemskaya:1994qp,Korchemskaya:1996je,Balitsky:1998ya,
Balitsky:2001gj,Kucs:2003ei}, and also in agreement with the recently proposed approach 
of Ref.~\cite{Caron-Huot:2013fea}. 

The paper is organized as follows. In \secn{irhe}, we discuss the general features of 
infrared factorization, and we review the results of Ref.~\cite{DelDuca:2011ae}. In 
\secn{improved}, we provide a general parametrization of four-point scattering amplitudes 
in the high-energy limit, and we introduce an improved organization of infrared 
operators, which better adapts to the color structure and symmetry properties of 
scattering amplitude in the high-energy limit. In \secn{coirf} we give a detailed 
comparison of infrared and high-energy factorizations up to three-loop order.
This allows us to recover the results of Refs.~\cite{DelDuca:2001gu,DelDuca:2013ara,
DelDuca:2013dsa}, and to provide definite and complete predictions for factorization-breaking 
terms at three loops. Furthermore, we examine the coefficients of the hard functions in 
the high-energy limit to all orders in the coupling, up to NLL accuracy in $\ln(s/|t|)$.
In \secn{even-even}, we analyze the $t$-channel exchange of color representations 
other than the octet, which do not admit a high-energy factorization as in \eq{Mgg2}, 
and we provide a comparison, based on the dipole formula for the soft anomalous dimension 
matrix, to similar studies performed in terms of Wilson lines in Ref.~\cite{Caron-Huot:2013fea}. 
In \secn{discu}, we briefly discuss our results and the prospects for future developments. 
Finally, several technical details which, we believe, will be useful for future high-order 
calculations of high-energy quark and gluon amplitudes, are given in the appendices. 
In \Appx{AppColor}, we provide the color bases we use for four-point scattering amplitudes, 
in \Appx{AppAnDim} we give the expressions for all relevant anomalous dimensions up to 
three loops, while in \Appx{AppHard} we write down the high-energy limit of the hard functions 
of four-point QCD scattering amplitudes, up to two-loop accuracy, using the exact four-point 
two-loop amplitudes provided in the literature~\cite{Bern:2002tk,Bern:2003ck,DeFreitas:2004tk,
Glover:2003cm,Glover:2004si}. In addition, as an example, in \Appx{singlet} we provide 
expressions for the singularities of singlet exchange amplitudes up to two-loop accuracy.


\section{Infrared factorization and the high-energy limit}
\label{irhe}

Matrix elements for quark and gluon scattering have a non-trivial color structure, which 
is best understood by assembling them into vectors in the space of color configurations
available for the process at hand. In general one writes, for $n$-parton scattering,
\beq
  {\cal M}_n \left(\frac{p_i}{\mu}, \as(\mu^2) \right) \, = \,
  \sum_{j} {\cal M}^{[j]}_n \left( \frac{p_i}{\mu}, \as(\mu^2) \right) \,
  c_{[j]}^{(n)} \, ,
\label{GenColExp}
\eeq
where the $c_{[j]}^{(n)}$'s are color tensors of rank $n$, with indices (not shown here) in 
the color representations of the external partons undergoing the scattering, while the 
index $[j] = 1, \ldots, r$ enumerates the color representations which can be exchanged 
in intermediate states in a selected channel. For a detailed discussion of how such 
tensors can be enumerated and constructed, when the external particles are in arbitrary
color representations, we refer the reader to~\cite{Beneke:2009rj,DelDuca:2011ae,
Beenakker:2013mva,Sjodahl:2012nk}: briefly, having selected for example an $s$-channel 
basis, one must construct the tensor product of the initial state representations, and take 
its intersection with the tensor product of final state representations. We note that, as in 
the rest of the paper, in \eq{GenColExp} we left implicit the dependence on the infrared 
regulator $\epsilon = 2 - d/2 < 0$.

Infrared and collinear singularities in the vector ${\cal M}_n$ are known to 
factorize\footnote{See~\cite{Catani:1998bh,Sterman:2002qn,Dixon:2008gr} and 
references therein for discussions of matrix element factorization. An analysis in 
the context of SCET was recently proposed in Ref.~\cite{Feige:2014wja}.}, so that 
the matrix element can be written as
\beq 
  {\cal M}_n \left(\frac{p_i}{\mu}, \as (\mu^2) \right) \, = \, 
  {\cal Z}_n \left(\frac{p_i}{\mu}, \as (\mu^2) \right)
  {\cal H}_n \left(\frac{p_i}{\mu}, \as (\mu^2) \right) \, .
\label{IRfact}
\eeq
Here ${\cal H}$ is a color vector, which is finite as $\epsilon \to 0$, and represents a
matching condition, to be determined order by order in perturbation theory after the 
subtraction of divergent contributions. The infrared operator ${\cal Z}_n$, on the other 
hand, is an $r \times r$ matrix in color space, generating all infrared and collinear 
singularities of the amplitude; it satisfies a (matrix) renormalization group equation, 
whose general solution can be written in the form
\beq
  {\cal Z}_n \left(\frac{p_i}{\mu}, \as (\mu^2) \right) \, = \,  
  {\cal P} \exp \left[ \frac{1}{2} \int_0^{\mu^2} \frac{d \lambda^2}{\lambda^2} \, \,
  \Gamma_n \left(\frac{p_i}{\lambda}, \alpha_s(\lambda^2) \right) \right] \, ,
\label{RGsol}
\eeq
where ${\cal P}$ denotes path ordering in color space. Note that all poles in $\epsilon$ 
are generated~\cite{Magnea:1990zb} through the integration of the $d$-dimensional 
running coupling down to vanishing scale, $\lambda \to 0$. For massless particles, the 
results of Refs.~\cite{Aybat:2006wq,Aybat:2006mz} showed that, up to two loops, the 
$n$-parton soft anomalous dimension matrix has a remarkably simple form, proportional 
to the one-loop result, regardless of the number of partons involved. This stimulated 
further investigations, and led to the proposal of the `dipole formula'~\cite{Becher:2009cu,
Gardi:2009qi,Becher:2009qa,Gardi:2009zv} as an all-order ansatz for $\Gamma_n$. 
This takes the form
\beq
  \Gamma_n^{\rm dip}  \left(\frac{p_i}{\lambda}, \alpha_s(\lambda^2) \right) \, = \,
  \frac{1}{4} \, \widehat{\gamma}_K \left(\alpha_s (\lambda^2) \right) \,
  \sum_{(i,j)} \ln \left(\frac{- s_{i j}}{\lambda^2} 
  \right) {\bf T}_i \cdot {\bf T}_j \, - \, \sum_{i = 1}^n
  \gamma_i \left(\alpha_s (\lambda^2) \right) \, .
\label{sumodipoles}
\eeq
The basic feature of \eq{sumodipoles} is that the color structure, expressed in 
terms of the color-insertion operator ${\bf T}^i$ for parton $i$, remains the same as
at leading order, and therefore it is expressed as a sum over color dipoles, with 
all higher-order multipoles vanishing exactly. Color and kinematics are tightly 
correlated, since momentum dependence occurs only through the `dipole' 
invariants $s_{i j} = (p_i + p_j)^2$, where for the sake of simplicity we have 
taken all momenta as outgoing. An important consequence of the simple color 
structure of \eq{sumodipoles} is that the path ordering symbol in \eq{RGsol} can 
be dropped, since scale dependence through the coupling is confined to colorless 
anomalous dimensions. These are defined as follows. Denoting by $\gamma_K^{[i]}$
the cusp anomalous dimension~\cite{Korchemsky:1985xj,Korchemsky:1987wg}
in representation $[i]$, and by ${\cal C}_{[i]}$ the corresponding quadratic Casimir 
eigenvalue, we assume $\gamma_K^{[i]}$ to be proportional to ${\cal C}_{[i]}$ 
through a universal function $\hat{\gamma}_K$, so that $\gamma_K^{[i]} = {\cal 
C}_{[i]} \hat{\gamma}_K$. This is known to be true at least up to three loops.
The functions $\gamma_i$, on the other hand, are collinear anomalous dimensions 
which can be extracted from form factor data~\cite{Becher:2009qa,Dixon:2008gr,
Magnea:1990zb}.

The dipole formula, \eq{sumodipoles}, arises as the simplest solution to a set 
of exact equations satisfied by the soft anomalous dimension, which can be 
understood as anomaly equations. Indeed, correlators of semi-infinite straight 
Wilson lines have a classical symmetry under independent rescalings of all
four-velocities $\beta_i$, which is broken by quantum corrections for light-like
lines, as a consequence of collinear divergences. The kinematic dependence
of the soft anomalous dimension in this case is constrained by the cancellation
of this anomaly in physical matrix elements. Eq.~(\ref{sumodipoles}) is exact up to 
two loops for massless partons, as first shown in Refs.~\cite{Aybat:2006wq,Aybat:2006mz}. 
The advantage of having exact equations for $\Gamma_n$ is that one can study
possible corrections to \eq{sumodipoles} in a systematic way~\cite{Becher:2009qa,
Dixon:2009ur,Ahrens:2012qz}. One finds that possible corrections could come only 
from two sources. They can  take the form of conformal cross-ratios of kinematic 
invariants, starting at three loops and with at least four hard partons, which 
however are very tightly constrained by symmetry requirements and by 
known properties of scattering amplitudes, including their high-energy behavior. 
Alternatively, they can arise as a consequence of violations of Casimir scaling 
for the cusp anomalous dimension, which can happen in principle starting at four 
loops. The complete calculation of the three-loop soft anomalous dimension 
matrix $\Gamma_n$ is a very challenging project, and recent progress to this 
end has recently been discussed in~\cite{Gardi:2013saa,Falcioni:2014pka}. 
Also recently, evidence for the existence of corrections to \eq{sumodipoles} at 
the four-loop level, and at NLL accuracy in the high-energy limit, was uncovered 
in Ref.~\cite{Caron-Huot:2013fea}. Finally, the results of the recent calculation of 
the three-loop non-light-like cusp anomalous dimension in Ref.~\cite{Grozin:2014hna} 
led to conjecture a possible violation of the Casimir scaling of $\gamma_K (\alpha_s)$,
at four loops and for contributions proportional to $n_f$.

For the rest of this paper, we will simply assume that the dipole formula is 
correct. Essentially all of the results given below are in any case not affected
by possible corrections. When this is not the case, for example for single pole
terms at three loops, we will explicitly note how the results could change. In the 
remainder of this section, we discuss the implications of the dipole formula for
the high-energy limit, specializing to four-point amplitudes, which are the simplest 
and most studied case. Here we summarize the results of Refs.~\cite{DelDuca:2011ae,
Bret:2011xm,DelDuca:2013ara,DelDuca:2013dsa}, while in \secn{improved} we 
propose an improved organization of infrared factors, which yields a more transparent 
comparison with high-energy factorization.

The main result of Refs.~\cite{DelDuca:2011ae,Bret:2011xm,DelDuca:2013ara,
DelDuca:2013dsa} is that, in the high energy limit, the infrared factor ${\cal Z}_n$
takes a simple factorized form, which is valid to leading power in $|t|/s$ and to all 
logarithmic accuracies. In the case of four-point amplitudes, one can write
\beq
  {\cal Z} \left(\frac{s}{\mu^2}, \frac{t}{\mu^2}, \as \right) \, = \, 
  \exp \left[ - {\rm i} \, \frac{\pi}{2} \, K \left( \as \right)
  {\cal C}_{\rm tot} \right] \,
  {\cal Z}_{1, {\bf R}} \left(\frac{t}{\mu^2}, \as \right) \,   
  \widetilde{\cal Z} \left(\frac{s}{t}, \as \right) + \ord \left(\frac{t}{s} \right) \, ,
\label{Zfact1}
\eeq
where, for simplicity, we omit henceforth the label $n = 4$, and where ${\cal C}_{\rm tot} 
\equiv \sum_{i = 1}^{4} {\cal C}_{[i]}$ is the sum of the Casimir eigenvalues of the external 
particles. The main ingredient of \eq{Zfact1}  is the matrix  $\widetilde{\cal Z}$, which
encodes the dependence on high-energy logarithms, and carries non-trivial color
information, which will be crucial for our discussion. It is given by
\beq
  \widetilde{{\cal Z}} \left(\frac{s}{t}, \as \right)
  \, = \, \exp \left\{ K( \as )
  \left[ \log \left( \frac{s}{-t } \right) {\bf T}_t^2 + i \pi {\bf T}_s^2\right] \right\} \, .
\label{widetildeZ}
\eeq
${\cal Z}$ is responsible for generating all high-energy logarithms of the amplitude which 
are accompanied by infrared poles. In \eq{widetildeZ} we also introduced `Mandelstam' 
combinations of color-insertion operators ${\bf T}_s = {\bf T}_1 + {\bf T}_2$ and 
${\bf T}_t = {\bf T}_1 + {\bf T}_3$. The coefficients of the high-energy logarithms 
are determined by the function
\beq
  K \left( \as \right) = 
  - \frac{1}{4} \int_0^{\mu^2} \frac{d \lambda^2}{\lambda^2} \,
  \hat{\gamma}_K \left( \alpha_s (\lambda^2) \right) \,,
\label{cusp}
\eeq
which is a scale integral over the cusp anomalous dimension. This integral is well 
known in perturbative QCD: it enters the resummation of infrared poles in the quark 
form factor~\cite{Magnea:1990zb} and in planar multi-parton scattering 
amplitudes~\cite{Bern:2005iz}; it was recursively computed to all orders, in terms of 
the perturbative coefficients of $\beta(\as)$ and $\gamma_K (\as)$, in~\cite{Magnea:2000ss}; 
in the context of the high-energy limit, a slightly different form of \eq{cusp} was shown 
to give the all-order infrared part of the Regge trajectory in~\cite{Korchemskaya:1996je}. 
In \eq{cusp} the singular $\epsilon$ dependence is generated through the integration of 
the $d$-dimensional version of the running coupling, so that the result is a pure counterterm. 
To three-loop order one finds\footnote{We choose to expand all functions in powers of 
$\alpha_s/\pi$. The explicit expressions for the perturbative coefficients of the various 
anomalous dimensions we use, up to three loops, are given in Appendix~\ref{AppAnDim}.
Note that normalizations must be changed appropriately when comparing with the
literature, for example~\cite{DelDuca:2001gu}, where perturbative expansions are often 
in powers of $\alpha_s/(4 \pi)$.}
\beqa
  K (\as) & = & \frac{\alpha_s}{\pi} \, 
  \frac{\widehat{\gamma}_K^{(1)}}{4 \epsilon} \, + \left(\frac{\alpha_s}{\pi}\right)^2 \,
  \left( \frac{\widehat{\gamma}_K^{(2)}}{8 \epsilon} -
  \frac{b_0 \, \widehat{\gamma}_K^{(1)}}{32 \epsilon^2} \right) \nonumber \\
  && \, + \left(\frac{\alpha_s}{\pi}\right)^3 \left( \frac{\widehat{\gamma}_K^{(3)}}{12 \epsilon} -
  \frac{b_0 \, \widehat{\gamma}_K^{(2)} + b_1 \, \widehat{\gamma}_K^{(1)}}{48 \epsilon^2} 
  + \frac{b_0^2 \, \widehat{\gamma}_K^{(1)}}{192 \epsilon^3} \right) \, + \ord (\alpha_s^4) \, , 
\label{KNNLO}
\eeqa
The final ingredient of \eq{Zfact1} is the function ${\cal Z}_{1, {\bf R}}$, which is a 
singlet in color space and real in the physical region. Its explicitly expression is
\beq
  {\cal Z}_{1, {\bf R}} \left(\frac{t}{\mu^2}, \as \right) \, = \,
  \exp \Bigg\{  \frac{1}{2} \Bigg[ K \left( \as \right) \log \left( \frac{-t}{\mu^2} 
  \right) + D \left( \as \right) \Bigg] {\cal C}_{\rm tot}  
  + \, \sum_{i = 1}^4 B_i \left( \as \right) \Bigg\} \,, 
\label{Z1}
\eeq
where the functions $D(\as)$ and $B(\as)$, just like $K(\alpha_s)$, are given by scale integrals 
of the cusp and collinear anomalous dimensions, and they similarly yield a perturbative 
series of pure counterterms, representing infrared and collinear divergences. Explicitly,
\beqa
  D \left( \as \right) & = &
  - \frac{1}{4} \int_0^{\mu^2} \frac{d \lambda^2}{\lambda^2} \,
  \widehat{\gamma}_K \left( \alpha_s (\lambda^2) \right) 
  \log \left(\frac{\mu^2}{\lambda^2}\right), \nonumber\\
  B_i \left( \as \right) & = &
  - \frac{1}{2} \int_0^{\mu^2} \frac{d \lambda^2}{\lambda^2} \,
  \gamma_i \left(\alpha_s (\lambda^2) \right) \, .
\label{intandim}
\eeqa
Because of the extra logarithm, the function $D(\alpha_s)$ is responsible for double poles
combining infrared and collinear singularities. An important property of the operator 
${\cal Z}_{{\bf 1}, {\bf R}}$, relevant for high-energy factorization and manifest in \eq{Z1}, 
is that it can be written to all orders in perturbation theory as the product of four factors,
each one associated with one of the external hard partons. Labeling the partons involved 
in the $2 \to 2$ scattering process by means of indices $\{ {\rm r}, {\rm s} \}$, with $\{ {\rm r}, 
{\rm s} \} = \{q, g \}$ for quarks and gluons respectively, so that  ${\cal Z}_{{\bf 1}, {\bf R}} 
\to {\cal Z}_{{\bf 1}, {\bf R}}^{\, {\rm r s}} $, one may write
\beq
  {\cal Z}_{{\bf 1}, {\bf R}}^{\, {\rm r s}}  \left( \frac{t}{\mu^2}, \as \right) 
  \, = \,  \left({\cal Z}_{{\bf 1},{\bf R}}^{\, {\rm r}} \left( \frac{t}{\mu^2}, \as \right)\right)^2
           \left({\cal Z}_{{\bf 1},{\bf R}}^{\, {\rm s}} \left( \frac{t}{\mu^2}, \as \right)\right)^2\, .
\label{jetfactors}
\eeq
Each factor ${\cal Z}_{{\bf 1},{\bf R}}^{\, {\rm i}}$ can be thought of as a `jet' operator, and 
one may expect these jet operators to combine naturally to yield the divergent parts 
of the impact factors. We will see below that this is indeed the case.


\section{High-energy color structure and the signature of the amplitude}
\label{improved}

One of the key feature of high-energy factorization, as exemplified in \eq{Mgg2}, is
the `signature' of reggeized gluon exchange, derived from the expected symmetry
of ladder diagrams contributing to high-energy logarithms. We now show that this
feature of high-energy amplitudes emerges naturally from infrared factorization, with 
a simple rearrangement of \eq{Zfact1}. To this end, we must first generalize \eq{Mgg2}
to include the scattering of quarks as well as gluons. In order to do so, we need to take 
into account the fact that the color factor for quark-quark scattering does not have 
a definite symmetry property under $s \leftrightarrow u$. In that case, therefore, the 
symmetric and the antisymmetric parts of the kinematic factor must be weighted
differently. Furthermore, we must write the result in a notation compatible with our 
discussion of infrared factorization, noting that high-energy factorization, as described 
in \eq{Mgg2}, applies only to the exchange of the octet representation in the $t$ 
channel. Choosing therefore a $t$-channel exchange basis, we can generalize 
\eq{Mgg2} as
\beqa
  {\cal M}_{\, {\rm r s}} ^{[8]} \left(\frac{s}{\mu^2}, \frac{t}{\mu^2}, \as \right)
  & = & 2 \pi \alpha_s \, H^{(0),[8]}_{\, {\rm r s}}   
  \nonumber \\ && \hspace{-2cm} \times \,\,
  \Bigg\{
  C_{\rm r} \left(\frac{t}{\mu^2}, \as \right)
  \bigg[ A_+ \left(\frac{s}{t}, \as \right) + \, \kappa_{\, {\rm r s}}  \,
  A_- \left(\frac{s}{t}, \as \right) \bigg]
  C_{\rm s} \left(\frac{t}{\mu^2}, \as \right) \nonumber \\
  & & \, + \, \, {\cal R}_{\, {\rm r s}} ^{[8]} \left(\frac{s}{\mu^2}, \frac{t}{\mu^2}, \as \right)
 + \ord \left( \frac{t}{s} \right) \Bigg\} \, ,
\label{ReggeFact}
\eeqa
where, as before, the indices ${\rm r,s}$ label the parton species (quark or gluon), and
the Regge trajectory appears in the combinations
\beq
  A_\pm \left(\frac{s}{t}, \as \right) \, = \, \left( \frac{- s}{- t} \right)^{\alpha(t)}
  \pm \left( \frac{s}{-t} \right)^{\alpha(t)} \, , 
\label{ReggeStructure}
\eeq
with  $\kappa_{gg} = \kappa_{qg} = 0$, while $\kappa_{qq} = (4 - N_c^2)/N_c^2$.
In \eq{ReggeFact} we have also introduced a non-factorizing remainder function
${\cal R}_{\, {\rm r s}} $, which is expected to receive contributions starting at NNLL 
and which will be discussed in detail in what follows. Finally, $H^{(0)[8]}_{\, {\rm r s}}$ 
represents the tree-level amplitude, which depends on the process, and includes 
the factor $s/t$ which appeared explicitly in \eq{Mgg2}.

In order to accurately match \eq{Zfact1} with \eq{ReggeFact}, the presence of the 
`Coulomb phase' factor proportional to ${\cal C}_{\rm tot}$ in \eq{Zfact1} is crucial. 
Indeed, using the relation
\beq
  {\bf T}_s^2 + {\bf T}_t^2 + {\bf T}_u^2 \, = \, {\cal C}_{\rm tot} \, , 
\label{casimirsum}  
\eeq
it is possible to combine the Coulomb phase in \eq{Zfact1} with the matrix 
$\widetilde{{\cal Z}}$ to define a new infrared matrix
\beqa
  \widetilde{{\cal Z}}_{\rm S} \left( \frac{s}{t}, \as \right)
  & \equiv &   \exp \left( - {\rm i} \, \frac{\pi}{2} \, K \left( \as \right)
  {\cal C}_{\rm tot} \right) \, \widetilde{\cal Z} \left(\frac{s}{t}, \as \right) \,  \\
  && \hspace{- 2.0cm} = \, \exp \Bigg\{ K( \as ) \bigg[ \bigg( \log \left( \frac{s}{-t } 
  \right) - {\rm i} \, \frac{\pi}{2} (1+ \kappa_{\, {\rm r s}}  ) \bigg){\bf T}_t^2
  + {\rm i} \, \frac{\pi}{2} \, \left({\bf T}_s^2 - {\bf T}_u^2 + 
  \kappa_{\, {\rm r s}} {\bf T}_t^2  \right) \bigg] \Bigg\} \, .    
  \nonumber
\label{Zfact2} 
\eeqa
Factorizing the matrix ${\cal Z}$  in terms of $\widetilde{{\cal Z}}_{\rm S}$ and 
${\cal Z}_{1, {\bf R}}$, it is easy to realise that the coefficient of ${\bf T}_t^2$ in 
\eq{Zfact2} correctly reproduces not only the energy logarithms, $\ln(s/(-t))$, but also 
the correct symmetry properties under $s \leftrightarrow u$ exchange, that is to 
say the correct signature of the amplitude (hence the label $S$ attributed to the new
infrared matrix $\widetilde{{\cal Z}}_{\rm S}$). This becomes more evident by rewriting 
$\widetilde{{\cal Z}}_{\rm S}$ as 
\beqa
  \widetilde{{\cal Z}}_{\rm S}  \left(\frac{s}{t}, \as \right)
  & = & \exp \Bigg\{ K( \as ) \Bigg[  \frac{1}{2} \bigg[ \log \left( \frac{s + {\rm i} \eta }{-t } 
  \right) + \log \left(\frac{- s - {\rm i} \eta }{- t }\right) 
  \label{Zfact3} \\ 
  & & \hspace{-1.5cm} 
  + \, \kappa_{\, {\rm r s}}  \left( \log \left(\frac{- s - {\rm i} \eta }{- t }\right)  
  - \log \left( \frac{s + {\rm i} \eta }{- t } \right) \right) \bigg]{\bf T}_t^2 \, 
  + \, {\rm i} \, \frac{\pi}{2} \,
  \Big({\bf T}_s^2 - {\bf T}_u^2 + \kappa_{\, {\rm r s}} {\bf T}_t^2  \Big) \Bigg] \Bigg\} 
  \, , \nonumber     
\eeqa
and comparing with the structure of the scattering amplitude in the high-energy limit,
as given in \eq{ReggeFact}. It's easy to see that at the leading logarithmic level (and 
at NLL for the real part of the amplitude) the terms proportional to ${\bf T}_t^2$ in the
exponent of the infrared operator $\widetilde{{\cal Z}}_{\rm S}$ reproduce the structure 
of the high-energy amplitude with signature, given by \eq{ReggeFact}. The breaking of 
high-energy factorization arises from the last term in \eq{Zfact3}, which is a color- and
process-dependent phase independent of Mandelstam invariants. This is in accordance 
with the expectation that the failure of high-energy factorization should come from the
mixing of different color amplitudes: indeed, while the color operator ${\bf T}_t^2$ is 
diagonal in a $t$-channel color basis, the operator ${\bf T}_s^2 - {\bf T}_u^2 + 
\kappa_{\, {\rm r s}}{\bf T}_t^2$ is not, and will induce mixing between different color 
components of the amplitude. 

A more detailed analysis of \eq{Zfact3} readily leads to the expected conclusion that
high-energy factorization begins to break down at NNLL level. In order to expand 
\eq{Zfact2} in order of increasing logarithmic accuracy, we need to make use of the
Zassenhaus formula
\beq
  {\rm e}^{k \, (X + Y)} \,\, = \,\, {\rm e}^{k X} \, \, {\rm e}^{k Y} \,  \,
  {\rm e}^{- \frac{k^2}{2} [X, Y]} \, \, {\rm e}^{\frac{k^3}{3!} \big( 2 \, [Y, [X, Y] ] \, + \,
  [X, [X, Y] ] \big)}  \, \, {\rm e}^{{\cal O} (k^4)} \, ,
\label{Zassenhaus}
\eeq
with the identifications
\beqa
  k & = & K( \as ), \nonumber \\
  X & = &   \bigg( \log \left( \frac{s}{- t } \right) - {\rm i} \, \frac{\pi}{2}(1 + 
  \kappa_{\, {\rm r s}}  ) \bigg){\bf T}_t^2 , \\
  Y & = &  {\rm i} \, \frac{\pi}{2} \,
  \left({\bf T}_s^2 - {\bf T}_u^2 + \kappa_{\, {\rm r s}} {\bf T}_t^2  \right) \, .
\eeqa
Clearly, since the function $K(\as)$ begins at ${\cal O} (\as)$, all leading logarithms 
are generated by the first exponential in \eq{Zassenhaus}, ${\rm e}^{k X}$. Next-to-leading
logarithms arise from the infinite sequence of multiple commutators involving only
one power of $Y$, and an arbitrary number of powers of $X$, and so forth. In order 
to continue the analysis, it is convenient to introduce a shorthand notation for color
operators. We define
\beqa
  {\bf O}_t & = & {\bf T}_t^2 \, , \nonumber  \\
  {\bf O}_{s - u} & = & {\bf T}_s^2 - {\bf T}_u^2 + \kappa_{\, {\rm r s}}  {\bf T}_t^2
  \, = \, 2 {\bf T}_s^2 + (1 + \kappa_{\, {\rm r s}} ) {\bf T}_t^2 - {\cal C}_{\rm tot} 
  \, , \nonumber  \\
  {\bf O}_{t, s} & = & \big[ {\bf T}_t^2, {\bf T}_s^2 \big] 
  \, , \label{coperator} \\
  {\bf O}_{t, t, s} & = & \Big[ {\bf T}_t^2, \big[ {\bf T}_t^2, {\bf T}_s^2 \big] \Big] 
  \, ,  \nonumber \\
  {\bf O}_{s,t,s} &=& \Big[ {\bf T}_s^2, \big[ {\bf T}_t^2, {\bf T}_s^2 \big] \Big] 
  \, , \nonumber  
\eeqa
with the natural generalizations to higher order commutators. In terms of these
color operators, the exponents of the various factors on the right-hand side of 
\eq{Zassenhaus} read
\beqa
  k X & = & K ( \as ) \bigg[ \log \left( \frac{s}{- t } \right) - {\rm i} \,
  \frac{\pi}{2} \left(1 + \kappa_{\, {\rm r s}}  \right) \bigg] {\bf O}_t \, ,  \nonumber \\  
  k Y & = & {\rm i} \, \frac{\pi}{2} \, K ( \as ) \,  {\bf O}_{s - u} \, , 
  \label{Zassexp} \\
  - \frac{k^2}{2} \big[ X, Y \big]  & = &  K^2( \as ) 
  \bigg[ - {\rm i} \, \frac{\pi}{2} \log \left( \frac{s}{- t } \right) - \, \frac{\pi^2}{4} 
  \left(1 + \kappa_{\, {\rm r s}}  \right) \bigg] {\bf O}_{t, s} \, , \nonumber \\
  \frac{k^3}{3!} \bigg( 2 \Big[ Y, \big[ X, Y \big] \Big] + \Big[X, \big[X, Y \big] \Big] \bigg)  
  & = & \frac{K^3( \as )}{3!} \Bigg\{ \bigg[ - 2 \pi^2  \log \left( \frac{s}{- t} \right) +
  {\rm i} \pi^3 \left( 1+ \kappa_{\, {\rm r s}} \right) \bigg] {\bf O}_{s, t, s} \nonumber \\
  & & + \, \left[ {\rm i} \pi  \log^2 \left( \frac{s}{- t} \right) + {\rm i} \frac{\pi^3}{4} 
  \left(1 + \kappa_{\, {\rm r s}}  \right)^2 \right] {\bf O}_{t, t, s} \Bigg\} \, . \nonumber
\eeqa
Starting with \eq{Zassexp}, one can verify that all color operators appearing at 
NLL, namely ${\bf O}_{s - u}$, ${\bf O}_{t, s}$, ${\bf O}_{t, t, s}$, and more generally 
${\bf O}_{t, \ldots, t, s}$ give vectors with a vanishing octet component, when acting 
on pure octet matrix elements, and thus in particular on the tree-level quark and gluon
amplitudes at leading power in $|t|/s$. Indeed, one finds that
\beqa
  \Big[ {\bf O}_{s - u} \Big]^{[8], [8]} & = & 
  \Big[ 2 {\bf T}_s^2 + \left(1 + \kappa_{\, {\rm r s}} \right) {\bf T}_t^2 - 
  {\cal C}_{\rm tot} \Big]^{[8], [8]} \nonumber \\
  & = & 2 \big[ {\bf T}_s^2 \big]^{[8],[8]} + \left( 1 + \kappa_{\, {\rm r s}} \right) C_A 
  - {\cal C}_{\rm tot} \, = \, 0 
  \, , \label{ottott} \\
  \Big[ {\bf O}_{t, \ldots, t, s} \Big]^{[8],[8]} & = & \Big[ {\bf T}_t^2, \big[ {\bf T}_t^2, 
  \ldots, \big[ {\bf T}_t^2, {\bf T}_s^2 \big] \ldots \big] \Big]^{[8],[8]}  \, = \, 0 \, . \nonumber
\eeqa
It is important to notice that, while the second identity in \eq{ottott} is a simple consequence 
of the fact that ${\bf T}_t^2$ can be replaced by its eigenvalues in a $t$-channel basis,
the first identity is non-trivial and to some extent surprising: it is a necessary condition
for the reggeization of next-to-leading logarithms, and, as such, it is a consequence of 
known properties of the high-energy limit; it embodies symmetry properties of the 
amplitudes, related to crossing symmetry, and indeed it could be used as a definition
of the symmetry factor $\kappa_{\, {\rm r s}}$; finally, we have explicitly checked that it is 
verified for quark and gluon amplitudes with the choice of color basis described in 
\Appx{AppColor}. On the other hand, operators like ${\bf O}_{s,t,s}$ in \eq{Zassexp}, 
which induce non-trivial mixing between different color amplitudes, appear only starting 
at NNLL, so that only at this level one expects a breakdown of high-energy factorization.


\section{Comparing infrared and high-energy factorizations for octet exchange}
\label{coirf}

We now get to the central goal of this paper, which is the comparison of the two 
different factorizations that we have described, given respectively by \eq{IRfact}, together
with the information on the high-energy limit collected in \secn{irhe}, and \eq{ReggeFact}. 
The two factorizations differ substantially in scope and accuracy: infrared factorization
for amplitudes organizes only infrared divergent contributions, but it is exact to all orders 
in perturbation theory; furthermore, the high-energy limit of the infrared operator
${\cal Z}$ discussed in \secn{improved} is accurate to leading power in $t/s$. On the 
other hand, high-energy factorization targets finite contributions to the amplitude, but 
it is only expected to work to a limited logarithmic accuracy. Comparing the two approaches,
we are going to extract constraints on the amplitude, which will eventually enable us to
make predictions based on one of the two factorizations, when the second one is not 
applicable.

Our first task is to systematically expand our factorized expressions in powers of the 
coupling and, where present, of the high-energy logarithm, $\ln(s/|t|)$. For example, to 
leading power in $t/s$, each color component of the amplitude can be organized as 
a double expansion, writing
\beq
  {\cal M}^{[j]} \left(\frac{s}{\mu^2}, \frac{t}{\mu^2}, \as \right) \, = \, 4 \pi \as \, 
  \sum_{n = 0}^\infty \sum_{i = 0}^n
  \left( \frac{\as}{\pi} \right)^n \ln^i \left( \frac{s}{- t} \right)
  M^{(n), i, [j]} \left( \frac{t}{\mu^2} \right) \, ,
\label{AmpExpansion}
\eeq
with corrections suppressed by powers of $t/s$. The color components of the finite hard
vector ${\cal H}$ can be similarly expanded as
\beq
  {\cal H}^{[j]} \left( \frac{s}{\mu^2}, \frac{t}{\mu^2}, \as \right) \, = \, 
  4 \pi \as \,\sum_{n = 0}^\infty \sum_{i = 0}^n
  \left( \frac{\as}{\pi} \right)^n \ln^i \left( \frac{s}{-t} \right)
  H^{(n), i, [j]} \left( \frac{t}{\mu^2} \right) \, .
\label{HExpansion}
\eeq
In this notation, since the tree-level matrix element has no logarithms, as well as obviously
no divergences, one has $H^{(0), [8]} = H^{(0), 0, [8]}$. The octet remainder ${\cal R}_{\, 
{\rm r s}}^{[8]}$ defined in \eq{ReggeFact} is also expanded as
\beq
  {\cal R}_{\, {\rm r s}}^{[8]} \left(\frac{s}{\mu^2}, \frac{t}{\mu^2}, \as \right)
  \, = \, \sum_{n = 2}^{\infty} \sum_{k = 0}^{n - 2}
  \left( \frac{\as}{\pi} \right)^n \ln^k \left( \frac{s}{- t} \right)
  R_{\, {\rm r s}}^{\, (n), k, [8]} \left(\frac{t}{\mu^2} \right) \, .
\label{Rexp}
\eeq
Notice that we have included the information that the remainder function must start
at NNLL and at the two-loop level. In principle, there is a finite, logarithmically subleading
ambiguity in the definition of the remainder function ${\cal R}_{\, {\rm r s}}^{[8]}$: we will 
see, however, that the knowledge of the structure of the amplitude which comes from 
infrared factorization suggests a natural choice of `high-energy factorization scheme', 
and therefore a natural choice for the non-factorizing remainder.

Quantities that do not depend on the center-of-mass energy $s$ are just expanded in 
perturbation theory. For example for the Regge trajectory and for the impact factors 
we write
\beq
  \alpha(t) \, = \, \sum_{n = 1}^{\infty} \left( \frac{\as}{\pi} \right)^n \alpha^{(n)}(t) \, ,
  \qquad C_{\, \rm r} \left(\frac{t}{\mu^2}, \as \right) \, = \, \sum_{n = 0}^{\infty} 
  \left( \frac{\as}{\pi} \right)^n C_{\, \rm r}^{(n)} \left(\frac{t}{\mu^2} \right)  \, ,
\label{expalc}
\eeq
and we choose the normalization so that $C_{\rm r}^{(0)} = 1$. In practice, in all 
subsequent calculations we will set the renormalization scale $\mu^2 = - t$, so that 
the perturbative coefficients of these functions will be just numbers. Notice also that 
in the literature on high-energy amplitudes~\cite{DelDuca:2001gu,Caron-Huot:2013fea,
DelDuca:1998kx} certain universal $\epsilon$-dependent factors are reabsorbed in the 
coupling, and the perturbative expansion is effectively in powers of a shifted coupling 
$\tilde{\alpha}_s = c_\Gamma \alpha_s$, where
\beq
  c_\Gamma \, = \, {\rm e}^{\eps \, \gamma_{\rm E}} \, 
  \frac{\Gamma(1 - \eps)^2 \Gamma(1 + \eps)}{\Gamma(1 - 2 \eps)} \, .
\label{cgamma}
\eeq
We will not follow this practice in our calculations below, since we want our results to be
expressed in terms of the standard $\overline{MS}$ coupling, to be readily comparable
with finite-order calculations. Since the two definitions begin to differ at ${\cal O}(\eps^2)$, 
some of our results for subleading poles at high-orders would change if the other scheme 
was adopted.

We now proceed with our comparison, order by order, beginning at one loop, where
everything is well known, in order to set up our convention and illustrate  our procedure
in a simple setting.


\subsection{One-loop matrix elements}
\label{1loopme}

We begin by expanding Eqs.~(\ref{AmpExpansion}) and (\ref{ReggeFact}), to first 
order in $\as$. For simplicity, we will omit the parton indices ${\rm r, s}$ whenever 
they are not specifically needed. Infrared factorization yields the expressions
\beqa
  M^{(1),0} & = & \left\{ Z_{1, {\bf R}}^{(1)} + {\rm i} \, \frac{\pi}{2} K^{(1)} 
  \Big[ {\bf O}_{s - u} - \left( 1 + \kappa_{\, {\rm r s}} \right) {\bf O}_t \Big] 
  \right\} H^{(0)} + H^{(1),0} \, , \nn \\
  M^{(1),1} & = &  K^{(1)} {\bf O}_t \, H^{(0)} + H^{(1),1} \, , 
\label{AmpCoeff1}
\eeqa
which are still vectors in color space. For the octet component of these vectors, 
high-energy factorization provides the expressions
\beqa
  M^{(1),0,[8]} & = & \left[ C_a^{(1)} + C_b^{(1)} - {\rm i} \, \frac{\pi}{2} \left( 1 + 
  \kappa_{\, {\rm r s}} \right) \alpha^{(1)} \right] H^{(0),[8]} \, , \nonumber \\
  M^{(1),1, [8]} & = & \alpha^{(1)} H^{(0), [8]} \, .
\label{ReggeCoeff1}
\eeqa
One of the constraints of Regge factorization is the fact that the Regge trajectory
and the impact factors are required to be real: in other words, the imaginary part 
of the amplitude is completely determined by the signature properties under the
exchange $s \leftrightarrow u$, as given in \eq{ReggeFact} and in \eq{ReggeStructure}.
There is therefore interesting information to be extracted about the imaginary parts
of the amplitude when comparing results such as \eq{AmpCoeff1} and \eq{ReggeCoeff1}.
Comparing first one-loop terms proportional to $\ln(s/(-t))$, and noting that the second of
Eqs.~(\ref{ReggeCoeff1}) is real, we immediately see that 
\beq
  \IM \Big[ H^{(1),1,[8]}_{\, {\rm r s}} \Big] \, = \, 0 \,. 
\label{eq:h118}
\eeq
In addition, it is known~\cite{DelDuca:1998kx} that
\beq
  \RE \Big[ H^{(1),1,[8]}_{\, {\rm r s}} \Big] \, = \, \ord( \eps ) \, .
\label{reh118}
\eeq
This simple one-loop result, as we will see, bootstraps to higher orders and has
important consequences on finite parts of higher-order amplitudes. To begin with, 
we can now write the one-loop Regge trajectory as
\beq
  \alpha^{(1)} \, = \, \frac{K^{(1)} {\bf T}_t^2 H^{(0)}}{H^{(0),[8]}} + \ord(\eps) \, .
\label{alpha1}
\eeq
In the high-energy limit, for all parton species, the tree-level amplitude at leading 
power in $|t|/s$ is a pure color octet in the $t$-channel, and therefore it is an 
eigenvector of the ${\bf T}_t^2$ operator with eigenvalue $C_A$. As expected, 
the Regge trajectory then becomes
\beq
  \alpha^{(1)} \, = \, C_A K^{(1)} + \ord(\eps) \, = \, \frac{C_A}{2 \eps} + \ord(\eps) \, ,
\label{alpha1qqqggg}
\eeq
which confirms the universality of the one-loop Regge trajectory~\cite{Tyburski:1975mr,
Fadin:1975cb,Lipatov:1976zz,Kuraev:1976ge,Mason:1976fr,Cheng:1977gt,Kuraev:1977fs} 
to $\ord(\eps)$. Notice that in the context of one-loop calculations these $\ord(\eps)$
terms can be safely neglected. Here however we allow for the possibility that $\ord(\eps)$
contributions might interfere with higher-order poles coming from the expansion of
the infrared operators beyond one loop. In the present case, $\ord(\eps)$ contributions
come exclusively from the factor $c_\Gamma$ in \eq{cgamma}.

Turning to non-logarithmic contributions to the matrix elements in Eqs.~(\ref{AmpCoeff1}) 
and (\ref{ReggeCoeff1}), and comparing their imaginary parts, we obtain
\beqa
  \IM \Big[ H^{(1),0,[8]}_{\, {\rm r s}} \Big] & = & - \frac{\pi}{2} \left(1 + \kappa_{\, {\rm r s}} 
  \right) \, \alpha^{(1)} H^{(0),[8]}_{\, {\rm r s}} \nonumber \\ 
  & & - \, \frac{\pi}{2} \, K^{(1)} 
  \left\{ \Big[ {\bf O}_{s - u} - \left( 1 + \kappa_{\, {\rm r s}} \right) {\bf O}_t \Big] H^{(0)} 
  \right\}^{[8]}  . 
   \label{imoneloop}
\eeqa
Using the form of the Regge trajectory, \eq{alpha1}, and the identity in \eq{ottott},
it is easy to see that
\beq
  \IM \Big[ H^{(1),0,[8]}_{\, {\rm rs}} \Big] \, = \, - \frac{\pi}{2} 
  \left(1 + \kappa_{\, {\rm r s}} \right) \RE \Big[ H^{(1),1,[8]}_{\, {\rm r s}} \Big]  
  \, = \, \ord(\eps) \, .
\label{H10}
\eeq
Notice that the vanishing of the octet-octet component of the operator ${\bf O}_{s - u}$,
noted in \eq{ottott}, is crucial for the compatibility of \eq{imoneloop} with infrared factorization:
if that matrix element were to be non-vanishing, the right-hand side of \eq{imoneloop}
would have a leftover uncancelled IR divergence, which would be incompatible with
the definition of ${\cal H}$ as the finite remainder of the matrix element. A combined
consequence of Reggeization and infrared factorization is thus that imaginary parts of 
one-loop amplitudes are completely fixed in terms of the real parts. The key element 
which guarantees that this can happen is precisely the fact that the operator ${\bf O}_{s - 
u}$, appearing in \eq{imoneloop}, gives a vector with a vanishing octet component, when 
applied to an octet amplitude, as shown in \eq{ottott}.

Finally, comparing the real parts of the non-logarithmic contributions to 
Eqs.~(\ref{AmpCoeff1}) and (\ref{ReggeCoeff1}), and considering separately the 
quark-quark and the gluon-gluon scattering amplitudes, we can determine the 
respective impact factors. One finds that
\beq
  C_{\rm r}^{(1)} \, = \, \frac{1}{2} Z_{1, {\bf R}, {\rm r}}^{(1)}  + \frac{1}{2} 
  \widehat{H}^{(1),0,[8]}_{\, {\rm r r}} \, ,
\label{imp1}
\eeq
where we defined  $\widehat{H}^{(m), n, [J]}_{\, {\rm r s}} \, = \, H^{(m),n,[J]}_{\, 
{\rm r s}}/H^{(0),[8]}_{\, {\rm r s}}$. The formal coefficients $Z_{1, {\bf R}, {\rm r}}^{(1)}$ 
can be expressed in terms of known anomalous dimensions, given in \eq{hatgammaK} 
and \eq{gammacol}, while hard parts can be read off eqns.~(\ref{Hqqqqtree}), 
(\ref{Hqqqq10}), (\ref{Hgggg0}), and (\ref{Hgggg1nll}). This gives the one loop 
impact factors
\beqa
  C_q^{(1)} & = & - \frac{1}{2} C_F \left( \frac{1}{\eps^2} + \frac{3}{2 \eps} \right)
  + N_c \left(\frac{13}{72}  + \frac{7}{8} \zeta(2) \right)
  + \frac{1}{N_c} \left( 1 - \frac{1}{8} \zeta(2) \right) - \frac{5}{36} n_f + 
  {\cal O}(\eps) \, , \nonumber \\
  C_g^{(1)} & = & - \frac{1}{2 \eps^2} N_c - \frac{b_0}{4 \eps}
  - N_c \left(\frac{67}{72} - \zeta(2) \right) + \frac{5}{36} n_f + {\cal O}(\eps) \, .
\label{oneloopif}
\eeqa
Having determined both  impact factors, one can finally verify the consistency of Regge 
factorization, by constructing the high-energy quark-gluon scattering amplitude. One can 
use the fact that, by virtue of \eq{jetfactors}, the color-singlet infrared operators 
$Z_{1, {\bf R}, {\rm r s}}$ satisfy
\beq
  Z_{1, {\bf R}, qg}^{(1)} \, = \,  \frac{1}{2} \bigg[ Z_{1, {\bf R}, qq}^{(1)} 
 + Z_{1, {\bf R}, gg}^{(1)} \bigg] \, .
\label{qgrelRE1}
\eeq
With this result, borrowed from infrared factorization, requiring Regge factorization leaves a 
constraint on the hard parts of the amplitudes, which must satisfy
\beq
  \RE \left( \widehat{H}^{(1), 0, [8]}_{qg} \right) \, = \, \frac{1}{2} \left[ \RE \left( 
  \widehat{H}^{(1), 0, [8]}_{gg} \right) + \RE \left( \widehat{H}^{(1), 0, [8]}_{qq} \right) 
  \right] \, .
\label{qgrelRE2}
\eeq
Our explicit results for hard parts, extracted from Ref.~\cite{Kunszt:1993sd},
are given in Appendix C, and they are easily verified to satisfy \eq{qgrelRE2}.


\subsection{Two-loop matrix elements}
\label{2loopme}

Repeating the procedure at two loops, one finds more interesting results and, as we 
describe below, at the level of non-logarithmic terms one begins to see the breakdown 
of the high-energy factorization, as given in \eq{ReggeFact}.

As above, we begin by expanding \eq{AmpExpansion}, this time to second order in 
$\as$. We find
\beqa
  M^{(2),0} & = & \bigg[ {\rm i} \, \frac{\pi}{2} \left(K^{(2)} + K^{(1)} Z_{1, {\bf R}}^{(1)} \right) 
  \Big( {\bf O}_{s - u} - \left(1 + \kappa_{\, {\rm r s}} \right) {\bf O}_t \Big) + 
  \, Z_{1, {\bf R}}^{(2)}  \nonumber \\
  & & \hspace{- 1cm} - \, \frac{\pi^2}{8} \left(K^{(1)}\right)^2 \Big(
  {\bf O}_{s - u}^2 + 2 {\bf O}_{t, s} \left(1 + \kappa_{\, {\rm r s}} \right) 
  - 2 \, {\bf O}_t {\bf O}_{s - u} \left( 1 + \kappa_{\, {\rm r s}} \right)
  + {\bf O}_t^2 \left(1 + \kappa_{\, {\rm r s}} \right)^2 \Big) \bigg] H^{(0)} 
  \nonumber \\
  & + & \bigg[ {\rm i} \, \frac{\pi}{2} K^{(1)}  \Big( {\bf O}_{s - u} - 
  \left(1 + \kappa_{\, {\rm r s}} \right) {\bf O}_t \Big)
  + Z_{1, {\bf R}}^{(1)} \bigg] \, H^{(1),0} + H^{(2),0} \, , \nonumber  \\
  M^{(2),1} & = & \bigg[ {\rm i} \, \frac{\pi}{2} \left( K^{(1)} \right)^2 
  \Big(- {\bf O}_{t, s} + {\bf O}_t {\bf O}_{s - u} - \left(1 + \kappa_{\, {\rm r s}} \right)
  {\bf O}_t^2 \Big) + K^{(1)} Z_{1, {\bf R}}^{(1)}  \, {\bf O}_t
  + K^{(2)}  {\bf O}_t^2 \bigg] H^{(0)} \nonumber \\
  & + & K^{(1)} {\bf O}_t \, H^{(1),0}
  + \bigg[ {\rm i} \frac{\pi}{2} K^{(1)} \Big({\bf O}_{s - u} - \left(1 + \kappa_{\, {\rm r s}} \right)
  {\bf O}_t \Big) + Z_{1, {\bf R}}^{(1)} \bigg] H^{(1),1} + H^{(2),1} \, , \nonumber \\
  M^{(2),2} & = & \frac{(K^{(1)})^2}{2} \, {\bf O}_t^2 H^{(0)} + K^{(1)}  \, {\bf O}_t \, 
  H^{(1),1} + H^{(2),2} \, ,
\label{AmpCoeff2}
\eeqa
where each expression is a vector in color space. For the octet component, we can also
expand \eq{ReggeFact} to second order in $\as$, yielding
\beqa
  M^{(2),0,[8]}_{\, {\rm r s}}  & = & \Bigg\{ C_{\rm r}^{(1)} C_{\rm s}^{(1)} 
  + C_{\rm r}^{(2)} + C_{\rm s}^{(2)} - {\rm i} \, \frac{\pi}{2} \left(1 + \kappa_{\, {\rm r s}} \right)
  \bigg[ \left(C_{\rm r}^{(1)} + C_{\rm s}^{(1)}\right) \alpha^{(1)}
  + \alpha^{(2)} \bigg]  \nonumber  \\
  & & - \, \frac{\pi^2}{4} \left(1 + \kappa_{\, {\rm r s}} \right) \left(\alpha^{(1)}\right)^2
  + \frac{1}{2} \, R^{(2),0,[8]}_{\, {\rm r s}}  \Bigg\} \, H^{(0),[8]}_{\, {\rm r s}} \, , 
  \label{ReggeCoeff2} \\
  M^{(2),1,[8]}_{\, {\rm r s}}  & = & \bigg[ \left(C_{\rm r}^{(1)} + C_{\rm s}^{(1)} 
  \right) \alpha^{(1)} - {\rm i} \, \frac{\pi}{2} \left(1 + \kappa_{\, {\rm r s}} \right) 
  \left(\alpha^{(1)}\right)^2 + \alpha^{(2)} \bigg] H^{(0),[8]}_{\, {\rm r s}}  \, , \nonumber  \\
  M^{(2),2,[8]}_{\, {\rm r s}}  & = & \frac{1}{2} \left( \alpha^{(1)} \right)^2 
  H^{(0),[8]}_{\, {\rm r s}} \, , \nonumber
\eeqa
where in the first equation, which contains the non-logarithmic NNLL contribution at two 
loops, we allow for a non-factorizing remainder, as in \eq{Rexp}.

Comparing the coefficients of the leading logarithms, that is the expressions at the 
bottom of \Eqns{AmpCoeff2}{ReggeCoeff2}, one readily verifies that the coefficient of 
the highest power of the energy logarithm is determined by the one-loop result, and in 
particular that the imaginary part of the hard matrix element vanishes, while the real 
part is of higher order in $\eps$,
\beqa
  \IM \Big[ H^{(2),2,[8]}_{\, {\rm r s}} \Big] & = & 0 \, , \nonumber \\
  \RE \Big[ H^{(2),2,[8]}_{\, {\rm r s}} \Big] & = &
  \frac{{\RE \left[ H^{(1),1,[8]}_{\, {\rm r s}} \right]}^2}{2 \, H^{(0),[8]}_{\, {\rm r s}} } 
  \, = \, \ord(\eps^2) \, ,
\label{H22}
\eeqa
as expected from high-energy factorization. 

At the level of single logarithms at two loops, that is the middle expressions in 
\Eqns{AmpCoeff2}{ReggeCoeff2}, we expect to recover the two-loop Regge trajectory,
and we expect high-energy factorization to continue holding. Indeed, one finds that
the imaginary part yields,
\beqa
  \IM \Big[ H^{(2),1,[8]}_{\, {\rm r s}} \Big]  & = & - \frac{\pi}{2} \left( K^{(1)} \right)^2 
  \left[ \big( - {\bf O}_{t,s} + {\bf O}_t {\bf O}_{s - u} - \left(1 + \kappa_{\, {\rm r s}} \right)
  {\bf O}_t^2 \big)  H^{(0)}_{\, {\rm r s}} \right]^{[8]} \nonumber  \\
  & & - \, \frac{\pi}{2} \, K^{(1)} \left[ \big( {\bf O}_{s - u} - \left(1 + \kappa_{\, {\rm r s}} 
  \right){\bf O}_t \big) \RE \big[ H^{(1),1}_{\, {\rm r s}} \big] \right]^{[8]} 
  - Z_{1, {\bf R}}^{(1)} \, \IM \left[ H^{(1),1,[8]}_{\, {\rm r s}} \right]  \nonumber  \\
  & & - \, \frac{\pi}{2} \left(1 + \kappa_{\, {\rm r s}} \right)  
  \left( \alpha^{(1)} \right)^2 H^{(0),[8]}_{\, {\rm r s}}  \, .
\label{imh21}
\eeqa
Substituting the one-loop Regge trajectory, \eq{alpha1}, it is easy to see that \eq{imh21} 
reduces to
\beq 
  \IM \Big[ H^{(2),1,[8]}_{\, {\rm r s}} \Big] \, = \, - \frac{\pi}{2} \left(1 + \kappa_{\, {\rm r s}} 
  \right) \left( \widehat{H}^{(1),1,[8]}_{\, {\rm r s}} \right)^2 H^{(0),[8]}_{\, {\rm r s}} 
  \, = \, \ord(\eps^2) \, ,
\label{imh21b}
\eeq
which is easy to understand, using again \eq{ottott}, and remembering that 
$\IM \left[ H^{(1),1}_{\, {\rm r s}} \right] = 0$. Once again, \eq{ottott} is crucial for
consistency with infrared factorization.

The two loop Regge trajectory~\cite{DelDuca:2001gu,Fadin:1995xg,Fadin:1996tb,
Fadin:1995km,Blumlein:1998ib} is determined from the real part of the single logarithms.
By replacing $\IM \left[H^{(1),1} \right] = 0$ in the expansion of the amplitude and 
introducing the explicit results for the one loop coefficients in the Regge formula we have
\beq
  \alpha^{(2)} \, = \, C_A K^{(2)} + \RE \left[ \widehat{H}^{(2), 1, [8]}_{\, {\rm r s}}  \right]  
  + \ord(\eps) \,.
\label{alpha2}
\eeq
As expected, the divergent part of the two loop Regge traiectory is entirely given by the 
integral of the two loop cusp anomalous dimension and is independent of the specific 
scattering process considered. This is again in perfect agreement with high-energy 
factorization. The requirement that the finite part of the two-loop Regge trajectory found 
in the $qq$ and $gg$ process be independent of the scattering process implies an 
identity for the real part of the amplitude. The requirement is that
\beq
  \RE \left[ \widehat{H}^{(2),1,[8]}_{gg} \right]
  \, = \, \RE \left[ \widehat{H}^{(2),1,[8]}_{qg} \right]
  \, = \, \RE \left[ \widehat{H}^{(2),1,[8]}_{qq} \right] \, ,
\label{obvid2}
\eeq
which is indeed satisfied. We directly check this condition by normalizing 
eqs.~(\ref{Hqqqq2nll}), (\ref{Hgggg2nll}) and (\ref{Hqgqg2nll}) with the 
corresponding tree level amplitudes, given respectively in (\ref{Hqqqqtree}), 
(\ref{Hgggg0}) and (\ref{Hqgqg0}). This gives the universal result
\beq
  \RE \left[ \widehat{H}^{(2),1,[8]}_{\rm r s} \right] \, = \, \left(\frac{101}{108} - 
  \frac{\zeta(3)}{8}\right) N_c^2 - \frac{7}{54} N_c n_f \, ,
\label{reggetrafin1}
\eeq
which, when inserted in \eq{alpha2}, reproduces the well-known result of 
Refs.~\cite{DelDuca:2001gu,Fadin:1995xg,Fadin:1996tb,Fadin:1995km,Blumlein:1998ib}.

Turning finally to non-logarithmic terms, given in the top expressions of 
\Eqns{AmpCoeff2}{ReggeCoeff2}, we see that their imaginary part yields
\beqa
  \IM \left[ H^{(2),0,[8]}_{\, {\rm r s}} \right] & = & - \frac{\pi}{2} \left( K^{(2)} + K^{(1)} 
  Z_{1, {\bf R}}^{(1)} \right) \left[ \big( {\bf O}_{s - u} - \left(1 + \kappa_{\, {\rm r s}} \right)
  {\bf O}_t \big) H^{(0)} \right]^{[8]} \nonumber \\
  & & - \, \frac{\pi}{2} \, K^{(1)} \left[ \big( {\bf O}_{s - u} - \left(1 + \kappa_{\, {\rm r s}} \right)
  {\bf O}_t \big) \RE \big[ H^{(1),0} \big] \right]^{[8]} + Z_{1, {\bf R}}^{(1)} \,
  \IM \Big[ H^{(1),0,[8]} \Big]  \nonumber \\ 
  & & - \, \frac{\pi}{2} \left(1 + \kappa_{\, {\rm r s}} \right)
  \left[ \left(C_{\rm r}^{(1)} + C_{\rm s}^{(1)} \right) \alpha^{(1)}
  + \alpha^{(2)} \right] H^{(0),[8]}_{\, {\rm r s}} \, . 
\label{eq:coeffim20}
\eeqa
A remarkable consequence of \eq{eq:coeffim20} is that high-energy factorization still 
works at NNLL for the imaginary part of the octet amplitude. This is a consequence of the
fact that the effects of the color mixing operators on the imaginary part of the octet 
amplitude are delayed by one order as compared to the real part of the amplitude. 
Specifically, we see that the only color mixing operator appearing in \eq{eq:coeffim20}  
is again ${\bf O}_{s - u}$, which, as noted above, gives a vanishing octet component 
when acting on a color octet state. The remaining terms in the first two lines of 
\eq{eq:coeffim20} combine to cancel exactly the contribution given in the third line, 
leaving the finite remainder 
\beq
  \IM \Big[ H^{(2),0,[8]}_{\, {\rm r s}} \Big] \, = \, - \frac{\pi}{2} \left(1 + \kappa_{\, {\rm r s}} 
  \right) \RE \Big[ H^{(2),1,[8]}_{\, {\rm r s}} \Big] \, ,
\label{eq:coeffim20fin}
\eeq
which is in agreement with the corresponding coefficients of the two-loop 
amplitudes~\cite{DelDuca:2001gu}, given in eqs. (\ref{Hqqqq2nnll}), (\ref{Hgggg2nnll}) 
and (\ref{Hqgqg2nnll}). 

When we consider the real part of NNLL contributions at two loops, given by the 
top expressions of \Eqns{AmpCoeff2}{ReggeCoeff2}, we finally begin to see the
non-universal effects that bring about the breaking of high-energy factorization.
Basically, the breaking of universality can be tracked back to three factors.
\begin{itemize}
 \item The appearance of the operator $({\bf O}_{s - u})^2$, which, acting on $H^{(0)}$, 
 gives a non-vanishing octet component, in contrast to ${\bf O}_{s - u}$, ${\bf O}_{t,s}$, 
 and in general ${\bf O}_{t, \ldots, t, s}$: these operators, when applied to a pure color octet 
 amplitude, give a vector with a vanishing octet component. To be more explicit, one has 
\beq
  \Big[  \left( {\bf O}_{s - u} \right)^2 H^{(0)} \Big]^{[8]} \, = \,
  \sum_{[i]}  \left[ {\bf O}_{s - u} \right]^{[8],[i]} \left[ {\bf O}_{s-u} \right]^{[i],[8]} 
  H^{(0),[8]} \neq 0 \, .
\eeq 
\item A mismatch between the Regge factorization formula and
 the high-energy limit of the infrared factorization formula
 in the octet channel itself: this can be easily seen by noting that 
\beq
  - \frac{\pi^2}{8} \left( K^{(1)} \right)^2 \left[ {\bf O}_t^2 \, \left(1 + \kappa_{\, {\rm r s}} 
  \right)^2 H^{(0)} \right]^{[8]} \, \neq \,
  - \, \frac{\pi^2}{4} \, \left(1 + \kappa_{\, {\rm r s}} \right) \left( \alpha^{(1)} \right)^2 
  H^{(0),[8]}_{\, {\rm r s}} \, . 
\label{crosster}
\eeq
\item The contributions of the other color components of the amplitude. Starting at 
two loops, one needs to take into account the effect of the operators  ${\bf O}_{s - u}$, 
${\bf O}_{t,s}$, and in general ${\bf O}_{t, \ldots, t, s}$ on the non-octet components of 
the amplitude, which are zero at tree level, but have contributions starting at one 
loop. For instance, in \eq{eq:coeffim20} one sees that the color octet amplitude 
receives a contribution proportional to the one-loop imaginary part of the sub-leading
color amplitudes, through  
\beq
  \bigg[ {\bf O}_{s-u} \, \IM \Big[ H^{(1),0}_{\, {\rm rs}} \Big] \bigg]^{[8]} \, \neq \, 0 \, .
\label{exosmu}
\eeq
As we will see below, this last effect is subtle, because it involves terms which are at 
least ${\cal O}(\eps)$ from $\IM \Big[H^{(1),0}_{\, {\rm rs}} \Big]$, and it can be made to
disappear by expanding the amplitude in powers of $\tilde \alpha_s = c_{\Gamma} \alpha_s$.
\end{itemize}
In general, ${\bf O}_{s - u}$, ${\bf O}_{t,s}$ and the factor $\kappa_{\rm rs}$ depend on 
the scattering process through color correlations and through the species of the incoming 
partons, so we expect that such terms will break the universality of high-energy factorization.
Our proposal is to identify all universality-breaking terms and include them in the definition 
of the remainder function $R^{(2),0,[8]}$. The analysis of infrared singularities is crucial to 
pinpoint the non-universal contribution. Indeed, if we replace the one loop impact factor 
and Regge trajectory \Eqns{imp1}{alpha1} in \eq{ReggeCoeff2}, and we compare it with 
the real part of $M^{(2),0,[8]}_{\rm rr}$ obtained from \eq{AmpCoeff2}, after using the 
identities in \eq{ottott} and \eq{H10}, we get an equation for the sum of impact 
factors and remainder functions, of the form
\beqa
  \left( 2 \, C^{(2)}_{\, {\rm rr}} + \frac{R^{(2),0,[8]}_{\, {\rm rr}}}{2} \right) 
  H^{(0),[8]}_{\, {\rm rr}} & = & \bigg[ Z^{(2)}_{1,{\bf R}, {\rm rr}} - 
  \frac{1}{4} \left(Z^{(1)}_{1, {\bf R}, {\rm rr}} \right)^2
  + \frac{1}{2} Z^{(1)}_{1,{\bf R},{\rm rr}} \RE \Big[\widehat{H}^{(1),0,[8]}_{\rm rr}\Big] 
  \nonumber \\
  & & \hspace{0.5cm} + \, \RE \Big[ \widehat{H}^{(2),0,[8]}_{\rm rr} \Big]
  - \frac{1}{4} \left( \RE \Big[ \widehat{H}^{(1),0,[8]}_{\rm rr} \Big] \right)^2 \bigg]
  H^{(0),[8]}_{\rm rr} \nonumber \\
  & & - \, \frac{\pi}{2} K^{(1)} \bigg\{ \frac{\pi K^{(1)}}{4} \bigg[ \big( {\bf O}_{s-u}^2
  - {\bf O}_t^2 \, (1 - \kappa_{\rm rr}^2) \big) H^{(0)}_{\rm rr} \bigg]^{[8]} 
  \label{C+R} \\
  & & \hspace{0.5cm} + \, \bigg[ {\bf O}_{s-u} \IM \Big[H^{(1),0}_{\rm rr} \Big] \bigg]^{[8]} \!
  - \frac{\pi N_c}{2} \, (1 - \kappa_{\rm rr}^2) \, \RE \Big[H^{(1),1,[8]}_{\rm rr} \Big] \bigg\} \, .
  \nonumber 
\eeqa
According to the considerations above, we assign all terms containing the operator 
${\bf O}_{s - u}$ and the factor $\kappa_{\rm rs}$ to the remainder function, 
while all the other contributions will define the impact factors. We write then
\beqa
  R^{(2),0,[8]}_{\rm rs} & = &  - \frac{\pi^2}{4} \left( K^{(1)} \right)^2 \, 
  \frac{1}{H^{(0),[8]}_{\rm rs}} \left[ \Big( {\bf O}_{s - u}^2 - {\bf O}_t^2 
  \left(1 - \kappa_{\, {\rm r s}}^2 \right) \Big) H^{(0)}_{\rm rs} \right]^{[8]}
  \label{rest2R} \\
  & & - \, \frac{\pi K^{(1)}}{H^{(0),[8]}_{\rm rs}} \, \bigg[{\bf O}_{s-u} \, 
  \IM \Big[ H^{(1),0}_{\rm rs} \Big] \bigg]^{[8]} + \frac{\pi^2}{2} K^{(1)} N_c \,
  \left( 1 - \kappa_{\rm rs}^2 \right) \, \RE \Big[ \widehat{H}^{(1),1,[8]}_{\rm rs} \Big] \, , 
  \nonumber 
\eeqa
and
\beqa
  C_{\rm r}^{(2)} & = & - \frac{1}{8} \left( Z^{(1)}_{1, {\bf R}, {\rm rr}} \right)^2 
  + \frac{1}{2} Z^{(2)}_{1, {\bf R}, {\rm rr}}
  + \frac{1}{4} Z^{(1)}_{1, {\bf R}, {\rm r r} } \, {\rm Re} 
  \left[ \widehat{H}^{(1), 0, [8]}_{{\rm r r} } \right]  \nonumber  \\
  & & - \, \frac{1}{8} \left({\rm Re} 
  \left[\widehat{H}^{(1),0,[8]}_{{\rm r r} }\right] \right)^2 
  + \frac{1}{2} {\rm Re} \left[ \widehat{H}^{(2),0,[8]}_{{\rm r r} } \right] \, , 
\label{newC}
\eeqa
with ${\rm r} = q,g$. 
We use this definition to compute quark and gluon impact factors 
at two loops and report their singularities
\beqa
\label{2loopimp}
  C_{q}^{(2)} & = & \frac{1}{8 \eps^4} \, C_F^2
  + \frac{1}{\eps^3} \left( \frac{17}{64} N_c^2 - \frac{23}{64}
  - \frac{1}{16} \, n_f C_F + \frac{3}{32}\frac{1}{N_c^2} \right) \nonumber \\
  & & + \, \frac{1}{\eps^2} \bigg[ N_c^2 \left( \frac{31}{384} - \frac{3}{16} \zeta(2) \right) +
  \frac{7}{32} \zeta(2) - \frac{77}{192} + \frac{1}{24} n_f C_F 
  + \frac{1}{N_c^2} \left( \frac{41}{128} - \frac{1}{32} \zeta(2) \right) \bigg] \nonumber \\
  & & + \,  \frac{1}{\eps} \bigg[ N_c^2 \left( - \frac{1037}{2304} - \frac{19}{48} \zeta(2) + 
  \frac{1}{96} \zeta(3) \right) - \frac{119}{288} + \frac{19}{48} \zeta(2) + 
  \frac{31}{96} \zeta(3)   \nonumber \\
  & & \hspace{1cm} + \, \left( \frac{1}{24} \zeta(2) + \frac{89}{288} \right) n_f C_F
  + \frac{1}{N_c^2} \left( \frac{221}{256} - \frac{1}{3} \zeta(3) \right) \bigg] \, 
  + \, \ord{ \left( \eps^0 \right) } \, , \\
  C_g^{(2)} & = & \frac{1}{8 \eps^4} \, N_c^2 + \frac{7}{32 \eps^3} b_0 N_c
  + \frac{1}{\eps^2} \bigg[ N_c^2 \left( \frac{103}{96} - \frac{7}{16} \zeta(2) \right)
   - \frac{49}{144} \, n_f N_c + \frac{1}{36} \,n_f^2 \bigg] \nonumber \\
  & & + \, \frac{1}{\eps} \bigg[ N_c^2 \left( \frac{853}{864} - \frac{11}{12} \zeta(2) - 
  \frac{31}{48} \zeta(3) \right) + n_f N_c \left( - \frac{67}{288} + \frac{1}{6} \zeta(2)
  \right) \nonumber \\
  & & \hspace{1 cm} + \, \frac{5}{216} \, n_f^2 - \frac{1}{32} \, \frac{n_f}{N_c} \bigg] 
  + \, \ord{ \left( \eps^0 \right) } \, , \nonumber
\eeqa
where we have adopted for simplicity a mixed notation, writing explicitly $C_F$ instead
of $(N_c^2 - 1)/(2 N_c)$ and $b_0$ instead of  $(11 N_c - 2 n_f)/3$ whenever such factors 
appear explicitly in the calculation. 

Similarly, the remainder functions $R^{(2),0,[8]}_{\rm rs}$ are written somewhat formally
in \eq{rest2R}, but they can be easily made explicit, for each parton species, upon 
picking specific color bases for the various amplitudes. Working in the orthonormal 
bases described in detail in Appendix \ref{AppColor} and in Refs.~\cite{Beneke:2009rj,
Beenakker:2013mva}, we get, for the octet components of quark and gluon amplitudes, 
\beqa
  R^{(2), 0, [8]}_{qq} & = & \frac{\pi^2}{4 \epsilon^2}
  \left(1 - \frac{3}{N_c^2} \right) \Big(1 - \epsilon^2 \zeta(2) \Big) 
  \, , \nonumber \\
  R^{(2), 0, [8]}_{gg} & = & - \, \frac{ 3 \pi^2}{2 \epsilon^2} 
  \Big(1 - \epsilon^2 \zeta(2) \Big) \, ,
  \label{vecrest} \\
  R^{(2), 0, [8]}_{qg} & = & - \, \frac{\pi^2}{4 \epsilon^2}
  \Big(1 - \epsilon^2 \zeta(2) \Big) \, .
  \nonumber
\eeqa
Notice that our remainder at this order has no contributions of order $\eps^{-1}$, as a 
consequence of the structure of infrared factorization, while the factor $\left(1 - \epsilon^2
\zeta(2) \right)$ can be absorbed in the constant $c_{\Gamma}^2$ by performing 
the expansion in terms of $\tilde{\alpha}_s = \alpha_s c_{\Gamma}$, instead of using
$\alpha_s$. This explains, as we will see shortly, the result of Ref.~\cite{DelDuca:2001gu}, 
where a violation of universality with only a double-pole contribution was discovered at 
the two-loop level. 

We finally consider our proposed expression for the impact factors, \eq{newC}. It contains 
terms which are manifestly universal and consistent with the interpretation of $C_{\rm r}$: 
for example, the first line of \eq{newC} naturally arises from the action of the exponential 
jet factors $Z_{1, {\bf R}, {\rm r r}}$, as defined in \eq{jetfactors}, on the hard factors, and 
can be unambiguously assigned to the external legs of the amplitude. Armed with these 
results and definitions, we can now check that our corrected high-energy factorization 
formula, \eq{ReggeFact}, works, by comparing the exact $qg\to qg$ amplitude at two 
loops, taken from Ref.~\cite{Bern:2003ck}, with the one constructed using \eq{ReggeFact}, 
with the impact factors, the Regge trajectory, and the remainder functions defined above. 
We find, as expected, that $R^{(2),0,[8]}_{qg}$, together with the impact factors defined 
in \eq{newC}, accounts for the complete two-loop quark-gluon scattering amplitude in 
the high-energy limit, including finite parts. Note that the same consistency check was 
performed in \secn{1loopme} on the $qg$ amplitude at one loop: in that case, universality 
was recovered by means of both \eq{qgrelRE1}, derived from the definition of ${\cal Z}_{1,
{\bf R},{\rm rr}}$, and \eq{qgrelRE2}, which is of the same form but is required by high-energy
factorization. Similarly, an important ingredient for universality of \eq{ReggeFact} at two 
loops is the two-loop jet factor identity
\beq
  Z^{(2)}_{1, {\bf R}, qg} \, = \,  \frac{1}{8} \bigg[ 4 Z^{(2)}_{1, {\bf R}, qq}
  + 4 Z^{(2)}_{1, {\bf R}, gg} + 2 Z^{(1)}_{1, {\bf R}, qq} Z^{(1)}_{1, {\bf R}, gg} 
  - \left( Z^{(1)}_{1, {\bf R}, qq} \right)^2 - \left(Z^{(1)}_{1, {\bf R}, gg} \right)^2 \bigg] \, , 
\label{jetfatwol}
\eeq
which is a simple consequence of \eq{jetfactors} and of the exponential form of the color 
singlet functions ${\cal Z}_{1, {\bf R}}$. On the other hand, the consistency check on the 
$qg\to qg$ scattering amplitude implies that an identity of the same form must hold for 
the finite parts too. This can be verified directly using the results of Ref.~\cite{Bern:2003ck}.
One finds that
\beqa
  \RE \Big[ \widehat{H}^{(2),0,[8]}_{qg} \Big] & = & \frac{1}{8} 
  \bigg[ 4 \, \RE \Big[ \widehat{H}^{(2),0,[8]}_{qq} \Big]
  + 4 \, \RE \Big[ \widehat{H}^{(2),0,[8]}_{gg} \Big] 
  + 2 \, \RE \Big[\widehat{H}^{(1),0,[8]}_{qq} \Big] 
  \RE \Big[ \widehat{H}^{(1),0,[8]}_{gg} \Big] \nonumber \\
  & & - \, \left( \RE \Big[ \widehat{H}^{(1),0,[8]}_{qq} \Big] \right)^2 
  - \left( \RE \Big[ \widehat{H}^{(1),0,[8]}_{gg} \Big] \right)^2 \bigg] \, .
\label{h208id}
\eeqa
The structure of \eq{h208id}, as well as that of \eq{qgrelRE2}, suggest a simple 
exponential ansatz for the impact factors, involving the jet factors of \eq{jetfactors}
and the non-logarithmic terms of the hard functions. To see it, we define the functions
\beq
  \widehat{H}_{{\bf R}, {\rm rs}}^{0,[8]} \, = \, \sum_{n = 0}^\infty \left( \frac{\alpha_s}{\pi} 
  \right)^n \, \RE \Big[ \widehat{H}^{(n),0,[8]}_{\rm rs} \Big] \, ,
\label{alloH0}
\eeq
and we simply assume that they exponentiate just like the jet factors $Z_{1,{\bf R},{\rm rs}}$ 
in \eq{jetfactors}. We can then write
\beq
  Z_{1,{\bf R},{\rm rs}} \, = \, \exp \Big[ \zeta_{\rm r} \Big] \times \exp \Big[\zeta_{\rm s} 
  \Big] \, , \qquad \,\,\,
  \widehat{H}_{{\bf R},{\rm rs}}^{0,[8]} \,  = \, \exp \Big[ h_{\rm r} \Big]
  \times \exp \Big[ h_{\rm s} \Big] \, ,
\label{exph}
\eeq
where the functions $h$ and $\zeta$ have perturbative expansions 
\beq
  h_{\rm r} (\alpha_s) \, = \, \frac{\alpha_s}{\pi} \, h_{\rm r}^{(1)} + 
  \left( \frac{\alpha_{s}}{\pi} \right)^2 h_{\rm r}^{(2)} + \ldots \, , \qquad 
  \zeta_{\rm r} (\alpha_s) \, = \, \frac{\alpha_s}{\pi} \, \zeta_{\rm r}^{(1)} + 
  \left( \frac{\alpha_{s}}{\pi} \right)^2 \zeta_{\rm r}^{(2)} + \ldots \, .
\label{pertexp}
\eeq
By using this notation, we can express the coefficients of the perturbative expansions 
of $Z_{1,{\bf R},{\rm rs}}$ and $\widehat{H}^{(n),0,[8]}_{\rm rs}$ simply as
\beqa
  \label{Z_1H_1coef}
  Z^{(1)}_{1, {\bf R}, {\rm rs}} & = & \frac{1}{2} \bigg( \zeta_{\rm r}^{(1)} + 
  \zeta_{\rm s}^{(1)} \bigg) \, , \nonumber \\
  Z^{(2)}_{1, {\bf R}, {\rm rs}} & = & \frac{1}{2} \bigg( \zeta_{\rm r}^{(2)} + 
  \zeta_{\rm s}^{(2)} \bigg) + \frac{1}{2} \bigg( \frac{\zeta^{(1)}_{\rm r} + 
  \zeta^{(1)}_{\rm s}}{2} \bigg)^2 \, , \nonumber \\
  \RE \Big[ \widehat{H}^{(1),0,[8]}_{\rm rs} \Big] & = & \frac{1}{2} 
  \bigg( h_{\rm r}^{(1)} + h_{\rm s}^{(1)} \bigg) \, , \\
  \RE \Big[ \widehat{H}^{(2),0,[8]}_{\rm rs} \Big] & = & \frac{1}{2}
  \bigg( h_{\rm r}^{(2)} + h_{\rm s}^{(2)} \bigg) + \frac{1}{2}
  \bigg( \frac{h^{(1)}_{\rm r} + h^{(1)}_{\rm s}}{2} \bigg)^2 \, . \nonumber
\eeqa
We are now in a position to rewrite the definitions of the impact factors at one and two 
loops, \eq{imp1} and \eq{newC}, using \eq{Z_1H_1coef}. We obtain simply
\beqa
  C^{(1)}_{\rm r} & = & \frac{\zeta_{\rm r}^{(1)} + h_{\rm r}^{(1)}}{2} \, , \nonumber  \\
  C^{(2)}_{\rm r} & = & \frac{\zeta_{\rm r}^{(2)} + h_{\rm r}^{(2)}}{2} + 
  \frac{1}{2} \bigg( \frac{\zeta_{\rm r}^{(1)} + h_{\rm r}^{(1)}}{2} \bigg)^2 \, .
\eeqa
This suggests a formal definition of impact factors to all orders, based on the information 
we get from infrared factorization, and on the properties of the hard functions up to two 
loops. We write
\beq
  C_{\rm r} \, = \, \exp \left[ \frac{\zeta_{\rm r} + h_{\rm r}}{2} \right] \, ,
\label{expans}
\eeq
which is exact at two loops with our definition of impact factor, and can be conjectured 
to provide a consistent definition to all orders. Intriguingly, \eq{expans} involves the
exponentiation of non-logarithmic, finite contributions to the amplitude: similar effects
have been known for a long time~\cite{Parisi:1979xd,Sterman:1986aj,Magnea:1990zb,
Eynck:2003fn,Ahrens:2008qu} for form factors and cross sections that are electroweak 
at tree level: \eq{expans} provides a hint that this kind of exponentiation might extend
to multi-particle amplitudes, at least in the high-energy limit.

We conclude the discussion at the two-loop level by noting that we are now in a position
to recover the violation of universality first diagnosed in Ref.~\cite{DelDuca:2001gu}, where 
the authors were assuming that high-energy factorization would work without a remainder 
function. Under that assumption, one finds a discrepancy between the exact two-loop 
quark-gluon scattering amplitude and the one predicted by the high-energy factorization 
formula, \eq{ReggeFact}, in the absence of the remainder $R$. That mismatch may be 
quantified by the function~\cite{DelDuca:2013ara}
\beqa
  \Delta_{(2),0,[8]} & = & \frac{M^{(2),0}_{qg}}{H^{(0),[8]}_{qg}} - 
  \bigg[C^{(2)}_q + C^{(2)}_g + C^{(1)}_q C^{(1)}_g - \frac{\pi^2}{4} 
  \left(1 + \kappa \right) \left(\alpha^{(1)} \right)^2 \bigg] \nonumber \\
  & = & \frac{1}{2}\bigg[R^{(2), 0, [8]}_{qg} - \frac{1}{2} 
  \left(R^{(2), 0, [8]}_{qq} + R^{(2), 0, [8]}_{gg} \right) \bigg] \, .
\label{delta}
\eeqa
Using \eq{rest2R} and \eq{vecrest}, we may evaluate explicitly \eq{delta}, finding
\beqa
  \Delta_{(2),0,[8]} & = & \frac{3}{2} \, \pi^2 \, \left( K^{(1)} \right)^2  
  \left(\frac{N_c^2 + 1}{N_c^2} \right) \Big( 1 - \epsilon^2 \zeta(2) \Big) \nonumber \\
  & = & \frac{\pi^2}{\eps^2} \, \frac{3}{16} 
  \left(\frac{N_c^2 + 1}{N_c^2} \right) \Big( 1 - \epsilon^2 \zeta(2) \Big) \, .
\label{findelta}
\eeqa
Up to our different normalization, already discussed above \eq{KNNLO}, \eq{findelta} 
is in complete agreement with Ref.~\cite{DelDuca:2001gu}, and explains the origin of 
the problem, as arising from the mixing of color representations induced by infrared 
factorization.


\subsection{Three-loop matrix elements}
\label{3loopme}

Proceeding to three-loop order, we expect that matching the single-logarithmic terms in
 Eqs.~(\ref{ReggeFact}) and (\ref{AmpExpansion}) will lead to a breaking of universality
similar to that observed for non-logarithmic terms at two loops. Indeed, as predicted in 
Refs.~\cite{DelDuca:2011ae,Bret:2011xm}, a direct comparison yields a non-universal 
result. As before, we begin by expanding \eq{AmpExpansion} to third order in $\as$. 
We obtain
\beqa 
\label{AmpCoeff30}
  \nonumber
  M^{(3),0}&=&
  \bigg[\, {\rm i} \, \frac{\pi^3}{48} \, \left( K^{(1)} \right)^3 \Big( - {\bf O}_{s - u}^3 + 
       8 \, {\bf O}_{s,t,s} (1 + \kappa_{\, {\rm r s}} ) - 
       6 \, {\bf O}_{s - u} \, {\bf O}_{t,s} (1 + \kappa_{\, {\rm r s}} ) \\ \nonumber
  & & \hspace{2.0cm} + \,
       2 \, {\bf O}_{t,t,s} (1 + \kappa_{\, {\rm r s}} )^2 - 
       3 \, {\bf O}_t^2 \, {\bf O}_{s-u} (1 + \kappa_{\, {\rm r s}} )^2  \\ \nonumber
  & & \hspace{2.0cm} + \,
       {\bf O}_t^3 (1 + \kappa_{\, {\rm r s}} )^3 + 
       3 \,{\bf O}_t \, {\bf O}_{s-u}^2 (1 + \kappa_{\, {\rm r s}} ) + 
       6 \, {\bf O}_t \, {\bf O}_{t,s} (1 + \kappa_{\, {\rm r s}} )^2 \Big)  \\ \nonumber
  & & + \, 
     \frac{\pi^2}{8} K^{(1)} \left(K^{(1)} Z_{1, {\bf R}}^{(1)} + 2 \, K^{(2)} \right) 
     \Big( - {\bf O}_{s - u}^2 - 
       2 \, {\bf O}_{t,s} (1 + \kappa_{\, {\rm r s}} )  \\ \nonumber
  & & \hspace{3.5cm} + \, 
       2 \, {\bf O}_t \, {\bf O}_{s - u} (1 + \kappa_{\, {\rm r s}} ) - 
        {\bf O}_t^2 (1 + \kappa_{\, {\rm r s}} )^2 \Big) \\ \nonumber
  & & + \, 
        {\rm i} \, \frac{\pi}{2} \left(K^{(1)} Z_{1, {\bf R}}^{(2)} + K^{(2)} Z_{1,{\bf R}}^{(1)} + 
        K^{(3)} \right) \Big( {\bf O}_{s-u} - {\bf O}_t (1 + \kappa_{\, {\rm r s}} ) \Big) + 
      Z_{1,{\bf R}}^{(3)}\bigg] H^{(0)}  \\ \nonumber
  & + & \bigg[ \frac{\pi^2}{8} \left( K^{(1)} \right)^2 \Big( - {\bf O}_{s - u}^2 - 
       2 \, {\bf O}_{t,s} (1 + \kappa_{\, {\rm r s}} )  \\ \nonumber
  & & \hspace{2.0cm} + \,  
       2 \, {\bf O}_t \, {\bf O}_{s - u} (1 + \kappa_{\, {\rm r s}} ) - 
       {\bf O}_t^2 (1 + \kappa_{\, {\rm r s}} )^2 \Big)  \\ \nonumber
  & & + \, 
  {\rm i} \, \frac{\pi}{2} \, \left(K^{(1)} Z_{1,{\bf R}}^{(1)} + K^{(2)} \right) 
   \Big( {\bf O}_{s - u} - {\bf O}_t (1 + \kappa_{\, {\rm r s}} ) \Big)  + 
      Z_{1,{\bf R}}^{(2)} \bigg] H^{(1), 0}  \\
  & + &  \bigg[ {\rm i} \, \frac{\pi}{2} \, K^{(1)} \Big({\bf O}_{s - u} - 
       {\bf O}_t (1 + \kappa_{\, {\rm r s}} )\Big) + Z_{1, {\bf R}}^{(1)} \bigg] H^{(2), 0} +
  H^{(3), 0} \, ,
\eeqa 
\beqa
\label{AmpCoeff31}
  \nonumber
  M^{(3),1} & = &  \bigg[ \frac{\pi^2}{24} \, 
    \left( K^{(1)} \right)^3 \Big( - 8 \, {\bf O}_{s,t,s} + 6 \, {\bf O}_{s - u} \, {\bf O}_{t,s} + 
       6 \, {\bf O}_t^2 \, {\bf O}_{s - u} (1 + \kappa_{\, {\rm r s}} ) \\ \nonumber
  & & \hspace{2.0cm} - \, 
       3 \, {\bf O}_t^3 (1 + \kappa_{\, {\rm r s}} )^2  
       - 3 \, {\bf O}_t \, {\bf O}_{s - u}^2 
      - 12 \, {\bf O}_t \, {\bf O}_{t,s} (1 + \kappa_{\, {\rm r s}} ) \Big)  \\ \nn
  & & + \, {\rm i}  \, \frac{\pi}{2} \, K^{(1)} \left(K^{(1)} Z_{1,{\bf R}}^{(1)} + 2 K^{(2)} \right) 
  \Big( - \, {\bf O}_{t,s} + {\bf O}_t \, {\bf O}_{s - u} - 
      {\bf O}_t^2 (1 + \kappa_{\, {\rm r s}} ) \Big) \\ \nonumber
  & & + \, \Big( K^{(3)} + K^{(2)} Z_{1,{\bf R}}^{(1)}  + 
    K^{(1)} Z_{1,{\bf R}}^{(2)} \Big) \, {\bf O}_t \bigg] H^{(0)} \\ \nonumber
  & & \hspace{-5mm} + \, \bigg[ 
  {\rm i} \, \frac{\pi}{2} \, \left( K^{(1)} \right)^2 \Big( - {\bf O}_{t,s} + {\bf O}_t \, {\bf O}_{s - u} 
  - {\bf O}_t^2 (1 + \kappa_{\, {\rm r s}} ) \Big) 
  + \left(K^{(1)} Z_{1,{\bf R}}^{(1)} + K^{(2)} \right) {\bf O}_t \bigg] H^{(1), 0} \\ \nonumber
  & &  \hspace{-5mm} + \, \bigg[  \frac{\pi^2}{8} \,  \left( K^{(1)} \right)^2 
       \Big( - {\bf O}_{s - u}^2  - 2 \, {\bf O}_{t,s} (1 + \kappa_{\, {\rm r s}} )  \\ \nonumber
  & & \hspace{2.0cm} + \, 
       2 \, {\bf O}_t \, {\bf O}_{s - u}  (1 + \kappa_{\, {\rm r s}} ) - 
       {\bf O}_t^2  (1 + \kappa_{\, {\rm r s}} )^2 \Big) \\ \nonumber
  & & + \, {\rm i} \, \frac{\pi}{2} \, \left( K^{(1)} Z_{1,{\bf R}}^{(1)} + K^{(2)} \right) 
  \Big( {\bf O}_{s - u} - {\bf O}_t (1 + \kappa_{\, {\rm r s}} ) \Big) + 
      Z_{1,{\bf R}}^{(2)} \bigg] H^{(1), 1} \\
  & + & K^{(1)} {\bf O}_t \, H^{(2), 0} 
  + \bigg[ {\rm i} \, \frac{\pi}{2} \, K^{(1)} \Big( {\bf O}_{s - u} - 
  {\bf O}_t (1 + \kappa_{\, {\rm r s}} ) \Big) + Z_{1,{\bf R}}^{(1)} \bigg] H^{(2), 1}
  + H^{(3), 1} \, ,
\eeqa
\beqa
\label{AmpCoeff32}
  \nonumber
  M^{(3),2} & = & \bigg[ \,
    {\rm i} \, \frac{\pi}{12} \, \left( K^{(1)} \right)^3 \Big( 2 \, {\bf O}_{t,t,s} - 6 \, {\bf O}_t  \, 
    {\bf O}_{t,s} + 3 \, {\bf O}_t^2  {\bf O}_{s - u}  - 
    3 \, {\bf O}_t^3  (1 + \kappa_{\, {\rm r s}} ) \Big) \\ \nonumber 
  & & \hspace{2.0cm} + \,
    \frac{1}{2} \, K^{(1)} \left( K^{(1)}  Z_{1, {\bf R}}^{(1)} + 2 K^{(2)} \right) {\bf O}_t^2 \,
    \bigg] H^{(0)} + \frac{1}{2} \, \left( K^{(1)} \right)^2 {\bf O}_t^2  \, H^{(1),0} \\ \nn
  & + & \bigg[ \, {\rm i} \, \frac{\pi}{2} \,  \left( K^{(1)} \right)^2 
  \Big( - {\bf O}_{t,s} + {\bf O}_t \, {\bf O}_{s - u} - {\bf O}_t^2 (1 + \kappa_{\, {\rm r s}} ) \Big) 
  \\ \nonumber
  & & \hspace{2.0cm} + \, \left( K^{(2)}  + K^{(1)} Z_{1, {\bf R}}^{(1)} \right) {\bf O}_t \bigg]  
  H^{(1),1}  + K^{(1)} \, {\bf O}_t \, H^{(2),1} \\
  & + &  \bigg[ \, {\rm i} \, \frac{\pi}{2} \, K^{(1)} \Big( {\bf O}_{s - u} - 
  {\bf O}_t ( 1 + \kappa_{\, {\rm r s}} ) \Big) + Z_{1, {\bf R}}^{(1)} \bigg] H^{(2),2}  
  + H^{(3),2} \, , 
\eeqa
\beq
\label{AmpCoeff33}
  M^{(3),3} \, = \, 
  \frac{\left( K^{(1)} \right)^3}{6} \, {\bf O}_t^2 \, H^{(0)}
  + \frac{ \left( K^{(1)} \right)^2}{2} \, {\bf O}_t^2 \, H^{(1),1} + K^{(1)} \, {\bf O}_t \, H^{(2),2} 
  + H^{(3),3} \, ,
\eeq
where each of \Eqnss{AmpCoeff30}{AmpCoeff33} is a vector in color space. For the octet 
component, expanding \eq{ReggeFact} to third order in $\as$ we find
\beqa
\label{ReggeCoef3}
  \nonumber
  M^{(3),0,[8]}_{\, {\rm r s}} & = & \Bigg\{ C_{\rm r}^{(3)} + C_{\rm s}^{(3)} + 
  C_{\rm r}^{(1)} C_{\rm s}^{(2)} + C_{\rm r}^{(2)}C_{\rm s}^{(1)} \\ \nonumber
  & & - \, \frac{\pi^2}{4} \left( \alpha^{(1)} \right)^2 \left(C_{\rm r}^{(1)} + C_{\rm s}^{(1)} \right) 
  (1 + \kappa_{\, {\rm r s}} ) - \frac{\pi^2}{2} \, \alpha^{(1)} \alpha^{(2)} 
  (1 + \kappa_{\, {\rm r s}} ) \\
  & & + \, {\rm i} \, \pi \Bigg[ \bigg( \frac{\pi^2}{12} \left( \alpha^{(1)} \right)^3
  - \frac{\alpha^{(2)}}{2} \left( C_{\rm r}^{(1)} + C_{\rm s}^{(1)} \right) - 
  \frac{\alpha^{(3)}}{2} \bigg) (1 + \kappa_{\, {\rm r s}} ) \nonumber \\
  & & \qquad - \frac{\alpha^{(1)}}{2} \, (1 + \kappa_{\, {\rm r s}}) 
  \left( C_{\rm r}^{(2)} + C_{\rm s}^{(2)} + C_{\rm r}^{(1)} C_{\rm s}^{(1)} \right)  
  \Bigg] \Bigg\} \, H^{(0),[8]}_{\rm r s} \, + \,  \frac{R^{(3),0,[8]}}{2} \, H^{(0)}_{\rm r s} 
  \, , \nonumber \\
  M^{(3),1,[8]}_{\, {\rm r s}} & = & \Bigg[ \alpha^{(3)} + \alpha^{(2)} 
  \left( C_{\rm r}^{(1)} + C_{\rm s}^{(1)} \right) + \alpha^{(1)} 
  \left( C_{\rm r}^{(1)} C_{\rm s}^{(1)} +  C_{\rm r}^{(2)} + C_{\rm s}^{(2)} \right) \nonumber \\
  & & - \frac{\pi^2}{4} \left( \alpha^{(1)} \right)^3 (1 + \kappa_{\, {\rm r s}} ) \nonumber \\
  & & - \,  {\rm i} \, \pi \, ( 1 + \kappa_{\, {\rm r s}} ) \, 
  \bigg( \frac{ \left( \alpha^{(1)} \right)^2}{2} \left(C_{\rm r}^{(1)} + C_{\rm s}^{(1)} \right)
  + \alpha^{(1)} \alpha^{(2)} \bigg) \Bigg] \, H^{(0),[8]}_{\rm r s} \, + \, 
  \frac{R^{(3),1,[8]}}{2} \, H^{(0)}_{\rm r s} \, , \nonumber \\
  M^{(3),2,[8]}_{\, {\rm r s}} & = & \left[ \frac{ \left( \alpha^{(1)} \right)^2}{2} 
  \left( C_{\rm r}^{(1)} + C_{\rm s}^{(1)} \right)
  + \alpha^{(1)} \alpha^{(2)} - {\rm i} \, \pi \, \frac{ \left( \alpha^{(1)} \right)^3}{4} 
  (1 + \kappa_{\, {\rm r s}} ) \right] H^{(0),[8]}_{\rm r s} \, , \nonumber \\
  M^{(3),3,[8]}_{\, {\rm r s}} & = & \frac{ \left( \alpha^{(1)} \right)^3}{6} 
  \, H^{(0),[8]}_{\rm r s} \, .
\eeqa
where in the first two equations we allow for a non-factorizing remainder, as in 
\eq{Rexp}. Notice that the Reggeization of next-to-leading logarithms was proven 
in Ref.~\cite{Fadin:2006bj} only for the real part of the scattering amplitude, therefore 
in principle we should allow for a non-vanishing purely imaginary remainder $R^{(3),2,[8]}$.
We have seen at two loops, however, that \eq{ReggeFact} yields the correct result for the 
imaginary part of the octet amplitude not only at NLL level, but in fact even at NNLL. 
Furthermore we note that IR factorization, as seen for example in \eq{Zfact3}, does 
not generate any contribution at NLL for the octet component of the amplitude, thanks 
to the identity in \eq{ottott}. We conjecture therefore that \eq{ReggeFact} yields the 
exact result for the octet component of the amplitude at NLL level, both for the real 
and for the imaginary part, and we set $R^{(n),n - 1,[8]} = 0$.

With this premise, we can proceed as we did at two loops. We start by comparing the 
coefficients of the leading logarithms, that is \eq{AmpCoeff33} and the bottom expression 
of \eq{ReggeCoef3}, and we verify that the coefficient of the highest power of the energy 
logarithm is determined by the one-loop result. In particular, the imaginary part of the 
hard matrix element vanishes, while the real part is of higher order in $\eps$,
\beqa
\label{H33}
  \IM \Big[ H^{(3),3,[8]}_{\, {\rm r s}} \Big] & = & 0 \, , \nonumber \\
  \RE \Big[ H^{(3),3,[8]}_{\, {\rm r s}} \Big]
  & = & \frac{1}{6} \frac{{\RE \Big[H^{(1),1,[8]}_{\, {\rm r s}} \Big]}^3}{\left(H^{(0),[8]}_{\, {\rm r s}}\right)^2 } 
  \, = \, \ord(\eps^3) \, .
\eeqa
In order to inspect the NLL terms, we compare \eq{AmpCoeff32} with the next-to-last 
expression in \eq{ReggeCoef3}. Considering first the imaginary part, and using our 
assumption that \eq{ReggeFact} works up to NLL for  the octet amplitude, we expect 
to find relations allowing us to express the NLL finite parts $H^{(3),2,[8]}_{\, {\rm r s}}$ 
in terms of lower-loop amplitudes. A direct comparison yields a somewhat unwieldy 
expression
\beqa
\label{IMH32}
  \IM \Big[ H^{(3),2,[8]}_{\, {\rm r s}} \Big] & = & 
  -  \frac{\pi}{12} \left( K^{(1)} \right)^3 \left[ \Big( 2 \, {\bf O}_{t,t,s} - 6 \, {\bf O}_t  {\bf O}_{t,s} + 
  3 \, {\bf O}_t^2  {\bf O}_{s - u}  - 3 \, {\bf O}_t^3  (1 + \kappa_{\, {\rm r s}} ) \Big) 
  H^{(0)}_{\rm r s} \right]^{[8]} \nonumber \\
  & & - \,  \frac{\pi}{2} \left( K^{(1)} \right)^2 \left[ \Big( - {\bf O}_{t,s} + {\bf O}_t {\bf O}_{s - u}  - 
  {\bf O}_t^2 (1 + \kappa_{\, {\rm r s}} ) \Big) \, \RE \left( H^{(1),1}_{\rm r s} \right) 
  \right]^{[8]}  \nonumber \\
  & & - \, \left( K^{(2)}  + K^{(1)} Z_{1,{\bf R}}^{(1)} \right) \left[ {\bf O}_t \, 
  \IM \left( H^{(1),1}_{\rm r s} \right) \right]^{[8]} - K^{(1)} \left[ {\bf O}_t  \, 
  \IM \left( H^{(2),1}_{\rm r s} \right) \right]^{[8]} \nonumber \\
  & & - \, \frac{\pi}{2} K^{(1)} \left[ \Big( {\bf O}_{s - u} - 
  {\bf O}_t (1 + \kappa_{\, {\rm r s}} ) \Big) \, \RE \left( H^{(2),2}_{\rm r s} \right) \right]^{[8]} 
  + Z_{1, {\bf R}}^{(1)} \, \IM \left[ H^{(2),2, [8]}_{\rm r s} \right] \nonumber \\ 
  & & - \, \frac{1}{2} \left( K^{(1)} \right)^2 \left[ {\bf O}_t^2 \, 
  \IM \left( H^{(1),0}_{\rm r s} \right) \right]^{[8]}
  - \frac{\pi}{4} \left(\alpha^{(1)}\right)^3 (1 + \kappa_{\, {\rm r s}} ) \, H^{(0),[8]}_{\rm r s} \, .
\eeqa
The right-hand side of \eq{IMH32} can however be simplified considerably, by making 
use of the fact that all color-mixing operators appearing in \eq{IMH32} give zero when 
applied either on the tree level amplitude $H^{(0),[8]}$, or on the leading-logarithmic finite 
parts $\RE \big[ H^{(1),1} \big]$, $\RE \big[ H^{(2),2} \big]$, whose only non-vanishing 
component is the octet amplitude. Furthermore, one can use the identities in Eqs.~(\ref{H22}),  
(\ref{H10}), (\ref{eq:h118}) and (\ref{reh118}) to express the corresponding finite parts 
of the amplitude in terms of lower order quantities, or set them to zero. In this way, 
we find that
\beq
\label{IMH32fin}
  \IM \Big[ H^{(3),2,[8]}_{\, {\rm r s}} \Big] \, = \, 
  -  \, \frac{\pi}{4} \, (1 + \kappa_{\, {\rm r s}} ) \left[ \widehat{H}^{(1),1,[8]}_{\, {\rm r s}} 
  \right]^3 \, H^{(0),[8]}_{\, {\rm r s}} \, = \, O \left( \eps^3 \right) \, .
\eeq
Performing the same procedure on the real part, we find
\beq
\label{REH32fin}
  \RE \Big[ H^{(3),2,[8]}_{\, {\rm r s}} \Big] \, = \, \RE \Big[ H^{(2),1,[8]}_{\, {\rm r s}} \Big] \, 
  \widehat{H}^{(1),1,[8]}_{\, {\rm r s}} - \frac{1}{2} \RE \Big[ H^{(1),0,[8]}_{\, {\rm r s}} \Big]
  \left( \widehat{H}^{(1),1,[8]}_{\, {\rm r s}} \right)^2 \, = \, {\cal O} \left( \epsilon \right) \, .
\eeq
As was the case at two loops, we predict that octet hard parts actually vanish in $d  = 4$
up to NLL.

Proceeding to the NNLL terms, it is interesting to inspect first the imaginary part of 
$M^{(3),1,[8]}_{\, {\rm r s}}$, and briefly explore the possibility that also at three loops
the imaginary part of the NNLL octet remainder function might vanish, as it does at two 
loops. In that case, equating the second expressions in Eqs.~(\ref{AmpCoeff33}) and  
(\ref{ReggeCoef3}), we obtain a lengthy expression which can be drastically simplified
using the same techniques that led to \eq{IMH32fin}. The result is
\beqa
\label{IMH31b}
  \IM  \Big[ H^{(3),1,[8]}_{\, {\rm r s}} \Big] & = &  
  \frac{\pi}{2} \left( K^{(1)} \right)^2 \Big( {\bf O}_{t,s} - {\bf O}_t {\bf O}_{s - u} \Big) 
  \RE \Big[ H^{(1), 0}_{\rm r s} \Big] - \frac{\pi}{2} \, K^{(1)} \, {\bf O}_{s - u} \, 
  \RE \Big[ H^{(2), 1}_{\rm r s} \Big] \nonumber \\
  & - & \pi (1 + \kappa_{\, {\rm r s}} ) \, \RE \Big[ H^{(3),2,[8]}_{\, {\rm r s}} \Big] \, .
\eeqa
The first line involves color-mixing operators acting on $\RE \big[ H^{(1), 0} \big]$, 
$\RE \big[ H^{(2), 1} \big]$. The effect of these operators is to generate contributions
to the non-octet components of the amplitude. The information available 
about those terms can be extracted from the calculation of the two-loop 
amplitudes for parton-parton scattering of Refs.~\cite{Bern:2002tk,Bern:2003ck,
DeFreitas:2004tk}, whose high-energy limit can be found in appendix \ref{AppHard}. 
Both terms vanish to the highest available order in $\epsilon$, {\it i.e.}, $\RE \big[ 
H^{(1),0,[k]} \big] = 0$ through $O (\epsilon^2)$, and $\RE \big[ H^{(2),1,[k]}_{\, {\rm r s}} 
\big] = 0$ through $O(\epsilon^0)$ for $k\neq 8$. As a consequence, we can conclude 
that the first term in the first line of \eq{IMH31b} is at least ${\cal O}(\eps)$, while the 
second is ${\cal O}(\eps^0)$. It is interesting to note that, if the uncalculated ${\cal O} 
(\epsilon)$ terms in $\RE \big[ H^{(2),1,[k]}_{\, {\rm r s}} \big] = 0$ turn out to vanish, 
then, at least in the $\epsilon \to 0$ limit, one would find the simplified expression
\beq
\label{IMH31ba}
  \IM \Big[ H^{(3),1,[8]}_{\, {\rm r s}} \Big] \, = \, - \, \pi \, (1 + \kappa_{\, {\rm r s}} ) \, 
  \RE \Big[ H^{(3),2,[8]}_{\, {\rm r s}} \Big] \, ,
\eeq
which is strongly reminiscent of the NNLL imaginary part at two loops, \eq{eq:coeffim20fin}.

We finally proceed to the real part of $M^{(3),1,[8]}_{\, {\rm r s}} $ at NNLL accuracy, which
is the level at which the three-loop Regge trajectory shows up. Inspecting \eq{AmpCoeff31}, 
we can easily single out potential contributions to the remainder function $R^{(3),1,[8]}$, by 
looking for the color mixing operators which give a non-zero result when acting on the color 
octet amplitude. It is however clear that at three loops we will not be able to give a complete 
expression for $R^{(3),1 [8]}$, since single-logarithmic finite contributions can arise directly 
in $H^{(3)}$, which is unknown. Singular single-logarithmic terms  at three loops are however 
completely predicted, under our assumption that the dipole formula applies at this order, 
since all relevant anomalous dimensions are known, and finite contributions to the amplitude
are known up to two loops. 

With this in mind, we start our inspection of $M^{(3),1}$ in \eq{AmpCoeff31} by identifying 
the terms involving the operators ${\bf O}_{s,t,s}$, ${\bf O}_{s - u} {\bf O}_{t,s}$ and ${\bf O}_t 
{\bf O}_{s - u}^2$ in the first two lines as contributions to the octet remainder. These terms 
multiply $(K^{(1)})^3 H^{(0)}$, and thus are ${\cal O}(\eps^{-3})$, which is consistent with 
the fact that the NNLL octet remainder at two loops is ${\cal O}(\eps^{-2})$. Additional 
contributions arise from the operator ${\bf O}_{s - u}^2$ applied to $H^{(1),1}$: $H^{(1),1}$ 
is ${\cal O} (\eps)$, but it is multiplied by $(K^{(1)})^2$, so this term is ${\cal O}(\eps^{-1})$. 
Proceeding further, we see that terms which involve $H^{(0)}$ or $H^{(1),1}$ and are linear 
in operators like ${\bf O}_{s - u}$ and ${\bf O}_{t,s}$ do not contribute to the octet 
remainder: in fact, $H^{(0)}$ and $H^{(1),1}$ are pure octets, and therefore they give 
vectors with a vanishing octet component when acted upon by these color-mixing 
operators. This reasoning however does not work when we consider terms involving 
$H^{(1),0}$: in fact, in this case $\IM \big[ H^{(1),0,[k]} \big]$ with $k \neq 8$ is different 
from zero; thus, the operators ${\bf O}_{t,s}$ and ${\bf O}_t{\bf O}_{s - u}$, acting on 
$H^{(1),0}$, will contribute to the remainder. Likewise, contributions could arise from the 
operator ${\bf O}_{s - u}$ acting on $\IM \big[ H^{(2),1,[k]} \big]$, with $k \neq 8$, in the 
last line of \eq{AmpCoeff31}. These contributions however will not contribute any poles, 
since $\IM(H^{(2),1,[k]}) = 0$, for any $k \neq 8$, as can be seen by using the explicit 
results given in appendix \ref{AppHard}. 

Further contributions to the octet remainder arise from terms involving the color diagonal 
operator ${\bf O}_t = {\bf T}_t^2$, because of  a mismatch between the structure predicted 
by \eq{ReggeFact} and the terms originating from ${\bf O}_t$ in \eq{Zfact3}, similarly to 
what we observed at two loops. In the case of $M^{(3),1}$, these are the terms involving 
the operators ${\bf O}_t^3 (1+\kappa_{\, {\rm r s}} )^2$ and ${\bf O}_t^2 (1 + \kappa_{\, 
{\rm r s}} )^2$. The mismatch occurs because the factor $(1+\kappa_{\, {\rm r s}} )$ appears 
only linearly in the corresponding terms of $M^{(3),1}$ in \eq{ReggeCoef3}, after the explicit 
values of the Regge trajectory and the impact factors are inserted. The two terms above are  
${\cal O}(\eps^{-3})$ and ${\cal O}(\eps^{-1})$ respectively. Proceeding further, one could pin 
down further contributions to the octet remainder at ${\cal O}(\eps^0)$, but those are beyond 
the reach of the present analysis, as explained previously.

We are now in a position to give an expression for the octet remainder at three loops, 
$R^{(3),1, [8]}$, where we collect all terms through ${\cal O}(\eps^{-1})$. We find
\beqa
\label{3loopR}
  R^{(3),1,[8]}_{\rm r s} & = &   \frac{\pi^2}{4}\left(K^{(1)}\right)^3 \bigg[ \left(
  - \frac{8}{3} \, {\bf O}_{s,t,s} + 2 \, {\bf O}_{s - u} {\bf O}_{t,s} 
  - {\bf O}_t {\bf O}_{s - u}^2 + {\bf O}_t^3 \, 
  (1 - \kappa_{\, {\rm r s}}^2) \right) \widehat{H}^{(0)}_{\rm r s} \bigg]^{[8]} \nonumber \\
  & + & \left( K^{(1)} \right)^2 \bigg[ \pi \, \Big( {\bf O}_{t,s} 
  - {\bf O}_t {\bf O}_{s - u} \Big) \, \IM \Big[ \widehat{H}^{(1),0}_{\rm r s} \Big]
  - \frac{\pi^2}{4} \, {\bf O}_{s - u}^2 \, \RE \Big[ \widehat{H}^{(1),1}_{\rm r s} \Big] \nonumber \\
  & & \hspace{1.0cm} + \, \frac{3}{4} \pi^2 \, {\bf O}_t^2 \, (1 - \kappa_{\, {\rm r s}}^2) \, 
  \RE \Big[ \widehat{H}^{(1),1}_{\rm r s} \Big] \bigg]^{[8]} + {\cal O}(\epsilon^{0}) \, .
\eeqa
With this definition, it is easy to verify that the divergent part of the three-loop Regge 
trajectory retains a universal form. One finds, as expected
\beq
 \alpha^{(3)} \, = \, K^{(3)} N_c + {\cal O} \left( \epsilon^0 \right) \, . 
\label{alpha3}
\eeq
Introducing in \eq{3loopR} the appropriate color factors and hard functions, and working in 
the color bases discussed in the appendix \ref{AppColor}, we obtain the explicit results
\beqa
  R^{(3), 1, [8]}_{qq} & = & \left(\frac{\as}{\pi} \right)^3 \frac{\pi^2}{\epsilon^3} \, 
  \frac{2 N_c^2 - 5}{12 N_c} \left(1 - \frac{3}{2} \eps^2 \zeta(2) \right) 
  + \, \ord{ \left( \eps^0 \right) }  \, , \nonumber \\
  R^{(3), 1, [8]}_{gg} & = & - \, \left(\frac{\as}{\pi} \right)^3 \frac{\pi^2}{\epsilon^3} \,
  \frac{2}{3} \, N_c  \left(1 - \frac{3}{2} \eps^2 \zeta(2) \right)
  + \, \ord{ \left( \eps^0 \right) }  \, , \\
  R^{(3), 1, [8]}_{qg} & = & - \, \left(\frac{\as}{\pi} \right)^3 \frac{\pi^2}{\epsilon^3} \,
  \frac{N_c}{24} \, \left(1 - \frac{3}{2} \eps^2 \zeta(2) \right)
  + \, \ord{ \left( \eps^0 \right) }  \, , \nonumber
\label{explR3}
\eeqa
which can be consistently used in \eq{ReggeFact}, provided one substitutes the impact 
factors and the Regge trajectory as defined in Eqs.~(\ref{newC}) and (\ref{alpha3}). Once
again, remarkably, we find that the singular parts of the remainders originate from a 
high-order pole (here $\epsilon^{-3}$), with lower order poles arising exclusively from 
the expansion of the constant $c_\Gamma^3$.

Comparing $M^{(3),0}$ in Eqs.~(\ref{AmpCoeff33}) and (\ref{ReggeCoef3}), one could
single out contributions to the octet remainder $R^{(3),0,[8]}$, which would be necessary 
to obtain a consistent definition of the impact factors $C_{\, {\rm r s}} ^{(3)}$ to N$^3$LL 
accuracy, through ${\cal O}(\eps^{-1})$. That analysis is straightforward but lengthy, and 
since it does not provide additional insight in the mechanism of breaking of \eq{ReggeFact}, 
we will not perform it in this work. We conclude by noting that, if a non-vanishing quadrupole
contribution to the soft anomalous dimension were to be discovered, it would affect \eq{3loopR}
and \eq{explR3} at the level of single poles


\subsection{Beyond three-loops}
\label{allloopme}

As we have seen, by using the information provided by infrared factorization, we are 
able to pin down the origin of the breakdown of \eq{ReggeFact} at NNLL accuracy, 
and define a remainder function which collects non-universal terms. Since infrared 
factorization correctly reproduces the infrared poles of an amplitude, but gives no 
prediction for its finite parts, we are able to extend this procedure only up to terms 
which involve finite parts which are known through explicit calculations. On the other 
hand, the high-energy factorization embodied in \eq{ReggeFact} is exact up to NLL 
accuracy for real parts of amplitudes~\footnote{Specifically, we note that we are assuming 
here that high-energy factorization holds at NLL also for $\ord(\eps)$ terms, which are 
not known at two loops.}, and, we assume, for octet imaginary parts as well. This enables 
us to give NLL predictions concerning finite parts of amplitudes as well, to all orders 
in perturbation theory. 

Let us start by inspecting leading logarithmic terms. First we note that, at LL accuracy,
the infrared factorization formula, \eq{IRfact}, can be written as 
\beqa
\label{MirLL}
  & & {\cal M}_{\rm LL}^{[8]} \left(\frac{s}{\mu^2}, \frac{t}{\mu^2}, \as \right) \, = \,  
  \left[ \left( \frac{s}{-t} \right)^{K (\alpha_s) \, {\bf O}_t} {\cal H}_{\rm LL} \right]^{[8]} 
  \\ \nonumber 
  & & \hspace{1cm} = \, 4 \pi \alpha_s \, H^{(0),[8]} 
  \sum_{n = 0}^{\infty} \left( \frac{\alpha_s}{\pi} \right)^n 
  \log^n\left(\frac{s}{-t}\right) \left( \, \sum_{k = 0}^n \frac{N_c^k}{k!} \left(K^{(1)}\right)^k
  \RE \Big[ \widehat{H}^{(n - k),n - k,[8]} \Big] \right) \, ,
\eeqa
while in case of the Regge factorization formula we can write
\beqa
\label{ReggeFactLL}
  {\cal M}_{\rm r s, LL}^{[8]} \left( \frac{s}{\mu^2}, \frac{t}{\mu^2}, \as \right)
  & = & 2 \pi \alpha_s \, H^{(0),[8]}_{\, {\rm r s}} \left[
  \left( \frac{s}{-t} \right)^{\alpha(t)} 
  \bigg[ 1 + {\rm e}^{- {\rm i} \pi \alpha(t)} + \kappa_{\, {\rm r s}}  
  \left(1 - {\rm e}^{- {\rm i} \pi \alpha(t)} \right)
  \bigg] \right]_{\rm LL} \nonumber \\ 
  && = \, 4 \pi \alpha_s \, H^{(0),[8]}_{\, {\rm r s}} \sum_{n=0}^{\infty} 
  \left( \frac{\alpha_s}{\pi} \right)^n 
  \frac{(\alpha^{(1)}(t))^n}{n!} \log^n \left( \frac{s}{-t} \right) \, . 
 \eeqa
Using the explicit result for $\alpha^{(1)}(t)$, and comparing term by term \eq{MirLL} with 
\eq{ReggeFactLL}, it is easy to find that
\beqa
\label{HLL}
  \IM \Big[ H^{(n),n} \Big] & = & 0 \, , \nonumber \\ 
  \RE \Big[ H^{(n),n,[8]} \Big] & = & \frac{1}{n!} \left( \RE \big[ H^{(1),1,[8]} \big] \right)^n 
  \, = \, {\cal O} (\epsilon ^{n}) \, .
\eeqa
Interestingly, and extending to all orders the results obtained up to three loops, LL hard 
parts vanish in $d = 4$, as a consequence of the fact that the one-loop Regge trajectory
is essentially a pure pole in dimensional regularization. A finite contribution to $\alpha^{(1)}$
would in fact spoil \eq{HLL}.

With a little more work, this result generalizes to NLL. At this accuracy, the infrared 
factorization formula in \eq{IRfact} yields
\beqa
\label{MirNLL1}
  {\cal M}_{\rm NLL}^{[8]} \left(\frac{s}{\mu^2}, \frac{t}{\mu^2}, \as \right) & = &  
  \Bigg[ {\cal Z}_{\rm NLL} \left( \frac{s}{\mu^2}, \frac{t}{\mu^2}, \as \right)
  {\cal H}_{\rm LL} \left( \frac{s}{\mu^2}, \frac{t}{\mu^2}, \as \right) \nonumber \\
  & & \hspace{1cm} + \, {\cal Z}_{\rm R, LL} \left(\frac{s}{t}, \as \right)
  {\cal H}_{\rm NLL} \left(\frac{s}{\mu^2}, \frac{t}{\mu^2}, \as \right) \Bigg]^{[8]} \, .
\eeqa
The second term can easily be written down explicitly. It is given by
\beqa
\label{ZLL-HNLL}
  \left[{\cal Z}_{\rm R,LL} \left(\frac{s}{t}, \as \right){\cal H}_{\rm NLL} 
  \left( \frac{s}{\mu^2}, \frac{t}{\mu^2}, \as \right) \right]^{[8]} & = & 4 \pi \as \, 
  H^{(0),[8]}_{\rm rs} \sum_{n = 1}^{\infty} \left( \frac{\alpha_s}{\pi} \right)^{n}
  \log^{n - 1} \left(\frac{s}{- t} \right) \nonumber \\
  & & \sum_{\sigma = 0}^{n - 1} \frac{ \left(K^{(1)} N_c \right)^{n - \sigma - 1}}{(n - \sigma - 1)!}
  \, \widehat{H}^{(\sigma + 1), \sigma,[8]}_{\rm rs} \, .
\eeqa
The first term in \eq{MirNLL1}, on the other hand, can be significantly simplified
by noting that the only non-vanishing component of the vector ${\cal H}_{\rm LL} $ is 
the color octet, and therefore the color octet component of the result is annihilated 
by the operators ${\bf O}_{s - u}$ and ${\bf O}_{t,\ldots,t,s}$ appearing in ${\cal Z}_{\rm 
NLL}$. One obtains then
\beqa
\label{ZfactNLL}
  & & \left[ {\cal Z}_{\rm NLL} \left( \frac{s}{\mu^2}, \frac{t}{\mu^2}, \as \right)
  {\cal H}_{\rm LL} \left( \frac{s}{\mu^2}, \frac{t}{\mu^2}, \as \right) \right]^{[8]} \, = \, 
  \sum_{n = 1}^{\infty} \left( \frac{\as}{\pi} \right)^{n} \log^{n - 1} \left( \frac{s}{- t} \right)
  \nonumber \\
  & & \hspace{1cm} \times \, \Bigg\{ \sum_{\sigma = 0}^{n - 1} 
  \frac{\left(K^{(1)} N_c \right)^{n - \sigma - 1}}{(n - \sigma - 1)!} 
  \bigg[ {\cal Z}_{\bf R}^{(1)} - {\rm i} \, \pi K^{(1)} \frac{ \left(1 + \kappa_{\rm rs} \right)}{2}
  N_c \bigg] \widehat{H}^{(\sigma), \sigma,[8]}_{\rm rs} \nonumber \\
  & & \hspace{1cm} + \, \sum_{\sigma = 0}^{n - 2} 
  \frac{\left(K^{(1)} N_c\right)^{n - \sigma - 2}}{(n - \sigma - 2)!} K^{(2)} N_c \,
  \widehat{H}^{(\sigma), \sigma,[8]}_{\rm rs} \Bigg\} \, .
\eeqa
The NLL prediction from the Regge factorization formula, \eq{ReggeFact}, gives instead
\beqa
\label{ReggeFactNLL}
  {\cal M}_{\rm rs, NLL}^{[8]} & = & 4 \pi \alpha_s \, H^{(0),[8]}_{\, {\rm r s}}  
  \sum_{n = 1}^{\infty} \left( \frac{\alpha_s}{\pi} \right)^n 
  \Bigg[ \left( C^{(1)}_{\rm r} + C^{(1)}_{\rm s} \right) \frac{\left( \alpha^{(1)} \right)^{n - 1}}{(n-1)!} 
  + \alpha^{(2)} \, \frac{ \left( \alpha^{(1)} \right)^{n - 2}}{(n - 2)!}  \nonumber \\ 
  && \hspace{4cm} 
  - {\rm i} \, \frac{\pi}{2} \, ( 1 + \kappa_{\, {\rm r s}} ) \, \frac{ \left( \alpha^{(1)} \right)^n}{n!}  
  \Bigg]  \log^{n - 1}\left( \frac{s}{-t} \right) \, .
 \eeqa
Under our assumption that \eq{ReggeFact} is correct up to NLL also for the imaginary 
part of the octet component of the amplitude, we can use the fact that the Regge 
trajectory and the impact factors are real. We can then substitute their explicit values, 
as obtained in Eqs.~(\ref{alpha1qqqggg}) and (\ref{imp1}), and proceed to compare 
\eq{ReggeFactNLL} with the sum of Eqs.~(\ref{ZLL-HNLL}) and (\ref{ZfactNLL}). 
We get 
\beqa
\label{HNLL}
  {\rm Im} \Big[ \widehat{H}^{(n), n - 1, [8]}_{\rm r s} \Big] & = & - \, \pi \, 
  \frac{1 + \kappa_{\rm r s}}{2} \, n \, \widehat{H}^{(n), n, [8]}
  \, = \, O(\epsilon^n) \, , \nonumber \\
  \RE \Big[ \widehat{H}^{(n), n - 1, [8]} \Big] & = & \RE \Big[ 
  \widehat{H}^{(2), 1, [8]} \Big] \, \widehat{H}^{(n - 2), n - 2, [8]} + (2 - n) \, 
  \RE \Big[ \widehat{H}^{(1), 0, [8]} \Big] \widehat{H}^{(n - 1), n - 1, [8]} 
  \nonumber \\ & = & {\cal O} (\epsilon ^{n - 2}) \, . 
\eeqa
Also at NLL, we find that hard coefficients vanish in $d = 4$. In essence, \eq{HNLL} 
reinforces the idea that high-energy logarithms are infrared in nature: indeed, leading and 
next-to-leading logarithmic contributions to hard scattering coefficients are forced to vanish 
with increasing powers of the regulator $\eps$. This means that infrared-finite high-energy 
logarithms must come exclusively from the interference of soft and collinear functions with 
lower-order contributions subleading in $\eps$.


\section{On subleading color amplitudes}
\label{even-even}

The central idea at the basis of our analysis in \secn{coirf} is that the interplay of 
high-energy and infrared factorization allows one to obtain more information 
about the leading color-octet amplitude than would be allowed by inspection of the 
two factorization formulas separately. Taken individually, the two factorizations have
different limitations: infrared factorization predicts all infrared poles of the amplitude, 
but does not control finite parts. High-energy factorization, on the other hand, predicts 
both poles and finite parts of the color octet amplitude, once the Regge trajectory and 
impact factors are known, but the predictions have a limited logarithmic accuracy. 
Combining the two factorizations, on the one hand one can use infrared information 
to determine the poles of the remainder function at NNLL and beyond: this results 
in consistent definitions of the two-loop impact factors, \eq{newC}, and of the three-loop 
Regge trajectory, \eq{alpha3}. On the other hand, high-energy factorization allows one 
to derive, to all orders, the NLL part of the color-octet hard functions appearing in the 
infrared factorization formula. The knowledge of these terms, in turn, gives access to 
a set of higher-order contributions to the amplitude, not only for color octet exchange, 
but also for other representations contributing to the scattering process. In this section 
we briefly explore the predictions that can be obtained in this way.

In order to be more explicit, let us start with an example. The state of the art QCD 
computations provide us with one-loop amplitudes to all order in $\eps$, and two-loop 
amplitudes up to ${\cal O}(\eps^0)$, for $2 \to 2$ processes in all partonic channels, 
$qq \to qq$, $gg \to gg$ and $qg \to qg$. Organising this knowledge in terms of the 
dipole formula allows us to extract the one-loop and two-loop hard functions, whose 
high-energy limit is given in appendix \ref{AppHard}, respectively up to ${\cal O}
(\eps^2)$ and ${\cal O}(\eps^0)$. With this knowledge at hand, one can 
generically predict the corresponding amplitudes at three loops up to ${\cal O}
(\eps^{-2})$, at four loops up to ${\cal O}(\eps^{-4})$, and so on. If, however,
one inputs information from high-energy factorization, such as \eq{HNLL}, we can 
extend the prediction to lower-order poles in $\eps$, for the coefficients of leading
and next-to-leading logarithms. 
The pole structure of the leading logarithms is simple: for gluon-gluon scattering we have
\beqa
  M^{(3),3,[1]}_{gg} & = & M^{(3),3,[8_s]}_{gg} \, = \, M^{(3),3,[10+\overline{10}]}_{gg} \, 
  = \, M^{(3),3,[27]}_{gg} \, = \, M^{(3),3,[0]}_{gg} \, = \, 0 \, , \nonumber \\
  M^{(3),3,[8_a]}_{gg} & = & - \frac{N_c^4 \sqrt{N_c^2 - 1}}{24 \epsilon^3} \, \frac{s}{t}
  \left(1 - \frac{3}{2} \, \eps^2 \zeta(2) \right) + {\cal O} \left( \epsilon^0 \right) \, ,
\label{M331gg}
\eeqa
where the terms in bracket recover the expansion of $c_{\Gamma}$. 
The quark-quark amplitude gives
\beqa
  M^{(3),3,[1]}_{qq} & = & 0 \, , \nonumber \\
  M^{(3),3,[8]}_{qq} & = & \frac{N_c^3 \sqrt{N_c^2 - 1}}{24 \epsilon^3} \, \frac{s}{t}
  \left(1 - \frac{3}{2} \, \eps^2 \zeta(2) \right) + {\cal O} \left(\epsilon^0\right) \, .
\label{M331qq}
\eeqa
Finally, for quark-gluon scattering,
\beqa
  M^{(3),3,[1]}_{qg} & = & M^{(3),3,[8_s]}_{qg} \, = \, 0 \, , \nonumber \\
  M^{(3),3,[8_a]}_{qg} & = & \frac{N_c^3 \sqrt{N_c (N_c^2 - 1)}}{24 \sqrt{2} \epsilon^3} \,
  \frac{s}{t} \left(1 - \frac{3}{2} \, \eps^2 \zeta(2) \right) + {\cal O} \left(\epsilon^0\right) \, .
\label{M331qg}
\eeqa
We can similarly obtain the poles of NLL amplitudes. For instance, using \eq{AmpCoeff32}, 
we realize that, since $H^{(2),2} = {\cal O}(\eps^2)$ and $H^{(3),3} = {\cal O}(\eps^3)$, 
$M^{(3),3,[8]}$ can be predicted up to ${\cal O}(\eps^0)$, and $M^{(3),2,[8]}$ can be 
predicted up to ${\cal O}(\eps^{-1})$. We find
\beqa
  M_{gg}^{(3),2,[8]} & = & N_c^3 \sqrt{N_c^2 - 1} \, \frac{s}{t}
  \bigg\{ \frac{1}{4 \epsilon^4} + \frac{11 N_c - 2 n_f + {\rm i} \pi N_c}{16 \epsilon^3} - 
  \frac{5}{8} N_c \, \zeta(2) \frac{1}{\epsilon^2} \nonumber \\
  & + & \frac{1}{\epsilon} \bigg[ - \frac{3}{16} b_0 \, \zeta(2) - \frac{7}{4} N_c \, \zeta(3) - 
  {\rm i} \pi \frac{3}{32} N_c \, \zeta(2) \bigg] \bigg\} + {\cal O} \left( \epsilon^0 \right) \, .
\label{gg328}
\eeqa
Regarding quark-quark and quark-gluon scattering processes, we find respectively
\beqa
\label{M32qq}
  M_{qq}^{(3),2,[8]} & = & N_c \sqrt{N_c^2 - 1} \, \frac{s}{t} \, \Bigg\{
  \frac{1}{\epsilon^4} \left( - \frac{N_c^2 - 1}{16} \right)+\frac{1}{\epsilon^3} \bigg(
  - \frac{5}{24} N_c^2 + \frac{1}{48} n_f N_c + \frac{3}{32} - \frac{{\rm i} \pi}{8} \bigg)
  \nonumber \\
  & & + \, \frac{1}{\epsilon^2} \bigg[ N_c^2 \bigg(\frac{7}{32} \zeta(2) + \frac{5}{18} \bigg)
  - \frac{5}{72} n_f N_c + \frac{1}{4} - \frac{3}{32} \zeta(2) \bigg] \nonumber \\
  & & + \, \, \frac{1}{\epsilon} \, \bigg[ N_c^2 \bigg( \frac{121}{216} + \frac{9}{64} \zeta(2) + 
  \frac{7}{16} \zeta(3) \bigg) \nonumber \\
  & & \hspace{1cm} - \, \frac{7}{54} n_f N_c + \frac{1}{2} - \frac{9}{64} \zeta(2) - 
  \frac{7}{16} \zeta(3) + {\rm i} \pi \frac{3}{16} \zeta(2) \bigg) \bigg] \Bigg\} + 
  {\cal O} \left( \epsilon^0 \right) \, ,
\eeqa
and
\beqa
\label{M32qg}
  M_{qg}^{(3),2,[8]}  & = & \frac{\sqrt{N_c^3 (N_c^2 - 1)}}{\sqrt{2}} \, \frac{s}{t} \, 
  \Bigg\{ \frac{1}{\epsilon^4} \left(\frac{1 - 3 N_c^2}{16} \right) + \frac{1}{\epsilon^3}
  \left(- \frac{53}{96} N_c^2 + \frac{n_f N_c}{12} + \frac{3}{32} - {\rm i} \pi 
  \frac{N_c^2}{16} \right) \nonumber \\
  & & + \, \frac{1}{\epsilon^2} \bigg[ N_c^2 \bigg( \frac{5}{18} + \frac{17}{32} \zeta(2) 
  \bigg) - \frac{5}{72} \, n_f N_c + \frac{1}{4} - \frac{3}{32} \zeta(2) \bigg] \nonumber \\
  & & + \, \, \frac{1}{\epsilon} \, \bigg[ N_c^2 \, \bigg( \frac{121}{216} + \frac{31}{64} \zeta(2) + 
  \frac{21}{16} \zeta(3) + {\rm i} \pi \, \frac{3}{32} \zeta(2) \bigg) \nonumber \\
  & & \hspace{1cm} - \,  n_f N_c \bigg( \frac{7}{54} + \frac{1}{16} \zeta(2) \bigg) 
  + \frac{1}{2} - \frac{9}{64} \zeta(2) - \frac{7}{16} \zeta(3) \bigg) \bigg] \Bigg\} + 
  {\cal O} \left( \epsilon^0 \right) \, .
\eeqa
We next ask to what extent we can predict other color components of the amplitudes,
which are subleading in the high-energy limit. The dipole formula applies to the vector 
amplitude in color space, therefore of course we are able to obtain predictions for the 
infrared poles of subleading color amplitudes as well. These amplitudes however do 
not admit a high-energy factorization formula of the form of \eq{ReggeFact}. This can 
be easily understood inspecting \eq{Zfact2}: non-octet amplitudes vanish at tree level 
in the high-energy limit, and are generated at one loop because of the effect of the operator 
${\bf O}_{s - u}$, acting on the octet component. In the language of Regge theory, these 
contributions are associated with cuts in the complex angular momentum plane, as 
opposed to the leading color-octet amplitude, which can be described at least up to NLL 
in terms of angular momentum poles only. Contributions arising from Regge cuts are 
expected to obey their own form of Regge factorization, different from \eq{ReggeFact}: 
a proposal in this direction was put forward in Ref.~\cite{Caron-Huot:2013fea}, where 
a formula for the even-even color-subleading amplitudes at NLL was provided, leading 
in particular to the prediction that the dipole formula must receive corrections at NLL 
and at  the four-loop order. For comparison and future reference, we provide here a 
set of predictions at three and four loops for the poles of color-subleading amplitudes 
associated with leading and next-to-leading logarithms, which we derive from the dipole 
formula in the absence of corrections. The starting point is the vanishing of the one- 
and two-loop LL and NLL hard parts for the non-octet amplitudes, which can be 
seen from the explicit result in appedix \ref{AppHard}, and is in agreement with 
Ref.~\cite{Caron-Huot:2013fea}.
Specifically, we find that
\beqa
\label{0even}
  H_{\rm r s}^{(n),n,[i \neq 8]} \, = \, 0 \, \quad & & \quad (n = 1,2) \, , \nonumber \\ 
  \RE \Big[ H_{\rm r s}^{(1),0,[i \neq 8]} \Big] \, = \, 0 \, ,  \quad
  \IM \Big[ H_{\rm r s}^{(1),0,[i \neq 8]} \Big] & = & {\cal O}(\eps), \quad 
  H_{\rm r s}^{(2),1,[i \neq 8]} \, = \, {\cal O}(\eps) \, .
\eeqa
Using this information, and inspecting \eq{AmpCoeff32}, we see that all components 
of four-parton amplitudes in the high-energy limit can be fully predicted at NLL, up to 
${\cal O}(\eps^{-1})$. The results are, for the three-loop gluon-gluon amplitude, 
\beqa
\label{ggnonoct}
   M_{gg}^{(3),2,[1]} & = & {\rm i} \pi \, \frac{N_c^4}{12 \epsilon^3} \, \frac{s}{t} \, 
   \left( 1 - \frac{3}{2} \epsilon^2 \, \zeta(2) \right) + {\cal O} \left( \epsilon^0 \right) \, ,
   \nonumber \\
   M_{gg}^{(3),2,[8_s]} & = & {\rm i} \pi \, \frac{N_c^4 \sqrt{N_c^2 - 1}}{16 \epsilon^3}
   \, \frac{s}{t} \, \left( 1 - \frac{3}{2} \epsilon^2 \, \zeta(2) \right) + 
   {\cal O} \left( \epsilon^0 \right) \, , \nonumber \\
   M_{gg}^{(3),2,[10+\overline{10}]} & = & {\cal O} \left( \epsilon^0 \right) \, , \\
   M_{gg}^{(3),2,[27]} & = & {\rm i} \pi \, \frac{N_c \sqrt{(N_c + 3)(N_c - 1)}}{24 \epsilon^3} \, 
   \frac{s}{t} \, \big( 7 N_c^2 + 10 N_c + 4 \big) \, \left( 1 - \frac{3}{2} \epsilon^2 \, \zeta(2) \right)
   + {\cal O} \left( \epsilon^0 \right) \, , \nonumber \\
   M_{gg}^{(3),2,[0]} & = & {\rm i} \pi \, \frac{N_c \sqrt{(N_c - 3)(N_c + 1)}}{24 \epsilon^3} \, 
   \frac{s}{t} \, \big( 7 N_c^2 - 10 N_c + 4 \big) \, \left( 1 - \frac{3}{2} \epsilon^2 \, \zeta(2) \right)
   + {\cal O} \left( \epsilon^0 \right) \, . \nonumber
\eeqa
For the three-loop quark-quark amplitude we find
\beq
\label{qqnonoct}
  M_{qq}^{(3),2,[1]} \, = \, {\rm i} \pi \, \frac{N_c (N_c^2 - 1)}{48 \epsilon^3} \, 
  \frac{s}{t} \, \left( 1 - \frac{3}{2} \epsilon^2 \, \zeta(2) \right) + {\cal O} \left( \epsilon^0 \right) \, .
\eeq
Finally, for the quark gluon amplitude we find
\beqa
\label{qgnonoct}
  M_{qg}^{(3),2,[1]} & = & - \, {\rm i} \pi \, \frac{N_c^2 \sqrt{N_c (N_c^2 - 1)}}{24 \epsilon^3}
  \, \frac{s}{t} \, \left( 1 - \frac{3}{2} \epsilon^2 \, \zeta(2) \right) + 
  {\cal O} \left( \epsilon^0 \right) \, , \\
  M_{qg}^{(3),2,[8_s]} & = & - \, {\rm i} \pi \, \frac{N_c^2 
  \sqrt{N_c (N_c^2 - 1)(N_c^2 - 4)}}{16 \sqrt{2} \epsilon^3} \, \frac{s}{t} \, 
  \left( 1 - \frac{3}{2} \epsilon^2 \, \zeta(2) \right) + 
  {\cal O} \left( \epsilon^0 \right) \, . \nonumber
\eeqa
To complete this section, we go one order higher and consider the four-loop expression 
for the amplitude, as predicted by the dipole formula. We provide the amplitude up to NLL 
in $\log(-s/t)$, because we know that constraints from Regge factorization arise only up 
to this order. We find
\beqa
\label{4loops}
  M^{(4),4} & = & \frac{\left( K^{(1)} \right)^4}{24} \, {\bf O}_t^4 \, H^{(0)}  + 
  \frac{\left( K^{(1)} \right)^3}{6} \, {\bf O}_t^3 \, H^{(1), 1} + \frac{\left( K^{(1)} \right)^2}{2} \, 
  {\bf O}_t^2 \, H^{(2), 2} \nonumber \\
  & & + \,  {\bf O}_t \, K^{(1)} \, H^{(3), 3} + H^{(4), 4} \, , \\
  M^{(4),3} & = & \Bigg\{  {\rm i} \, \frac{\pi}{24} \, \left( K^{(1)} \right)^4 \bigg[ \hspace{-2.5pt}
  - {\bf O}_{t,s,t,t} + 4 \, {\bf O}_t \, {\bf O}_{t,t,s} - 6 \, {\bf O}_t^2 \, {\bf O}_{t,s}    
  + 2 \, {\bf O}_t^3 \, {\bf O}_{s - u}  - 2 \, {\bf O}_t^4 \left(1 + \kappa_{\rm rs} \right) 
  \bigg]  \nonumber \\
  & & \hspace{5mm} + \, \frac{ \left( K^{(1)} \right)^3}{6}\,  Z_{1, \bf R}^{(1)} \, {\bf O}_t^3
  + \frac{\left( K^{(1)} \right)^2 K^{(2)} }{2} \, {\bf O}_t^3  \Bigg\} \, H^{(0)} 
  + \frac{\left( K^{(1)} \right)^3}{6} \, {\bf O}_t^3 \, H^{(1),0}  \nonumber \\
  & & + \, \Bigg\{ {\rm i} \, \frac{\pi}{12} \, \left( K^{(1)} \right)^3 \,  \bigg[ 2 \, {\bf O}_{t,t,s}
  - 6 \, {\bf O}_t \, {\bf O}_{t,s} + 3 \, {\bf O}_t^2 \, {\bf O}_{s - u}   
  - 3 \, {\bf O}_t^3 \left( 1 + \kappa_{\rm r s} \right) \bigg] \nonumber \\
  & & \hspace{5mm} + \, K^{(1)} \, K^{(2)} \, {\bf O}_t^2 
  + \frac{ \left( K^{(1)} \right)^2}{2} \, Z_{1, \bf R}^{(1)} \, {\bf O}_t^2 \Bigg\} \, H^{(1),1} 
  + \frac{\left( K^{(1)} \right)^2}{2} \, {\bf O}_t^2 \, H^{(2),1} \nonumber \\
  & & + \, \Bigg\{ {\rm i} \, \frac{\pi}{2} \left( K^{(1)} \right)^2 \bigg[ \hspace{-2pt}
  - {\bf O}_{t,s} 
  + {\bf O}_t \, {\bf O}_{s - u} - {\bf O}_t^2 \left(1 + \kappa_{\rm rs} \right) \bigg]
  + K^{(2)} \, {\bf O}_t + Z_{\bf 1, R}^{(1)} \, {\bf O}_t  \Bigg\} \, H^{(2), 2} \nonumber \\
  & & + \,  K^{(1)} \, {\bf O}_t \, H^{(3), 2} 
  + {\rm i} \, \frac{\pi}{2} \, K^{(1)} \, \bigg[ {\bf O}_{s - u} - {\bf O}_t 
  \left( 1 + \kappa_{\rm r s} \right) \bigg] \, H^{(3), 3} + Z_{\bf 1, R}^{(1)} \, H^{(3), 3}
  + H^{(4), 3} \, . \nonumber 
\eeqa
Taking into account Eqs. (\ref{HNLL}) and (\ref{0even}), as well as the results in Appendix \
\ref{AppAnDim} and Appendix \ref{AppHard}, one can compute the LL amplitudes $M^{(4),4}$ 
up to ${\cal O}(\eps^{-1})$, and the NLL amplitudes $M^{(4),3}$ up to ${\cal O}(\eps^{-2})$. 
Indeed, inspecting \eq{4loops}, we see that one would need the knowledge of $H^{(2),1, 
[i \neq 8]}$ up to ${\cal O}(\eps)$ in order to obtain $M^{(4),3}$ up to ${\cal O}(\eps^{-1})$. 
We will not display here the corresponding lengthy expressions, but they can readily be 
obtained combining \eq{4loops} with the results given in the Appendices.

Here we will focus instead on the color singlet amplitude, which  is a bit special,
because ${\bf O}_t^2 H^{(2),1,[1]} = 0$. This enables us to compute $M^{(4),3,[1]}$ 
up to ${\cal O} (\eps^{-1})$. The result is particularly interesting in light of the recent 
claim \cite{Caron-Huot:2013fea} that this term receives a contribution not predicted by 
the dipole formula. Within our setup we can provide a partial check, in the form an 
independent prediction of the $\eps^{-1}$ poles arising within the infrared factorization 
by using the dipole formula only. 

We start by noting that the terms proportional to ${\bf O}_t \, {\bf O}_{t,s} $, ${\bf O}_t^2 
{\bf O}_{s - u} $ and ${\bf O}_t $ in \eq{4loops} have a vanishing color-singlet component. 
The only source of $\eps^{-1}$ poles in $M^{(4),3,[1]}$ is then the term proportional to 
$\big(K^{(1)} \big)^3 \, {\bf O}_{t,t,s}$. One finds
\beq
  \left. M_{gg}^{(4),3,[1]} \right|_{\eps^{-1}} \, = \, 
  \frac{N_c^5}{12} \, \frac{7}{6\epsilon} \, {\rm i} \, \pi \, \zeta_3 \, \frac{s}{t} \, .
\label{caron}
\eeq
As expected from Ref.~\cite{Caron-Huot:2013fea}, it can be verified that \eq{caron} comes 
entirely from the expansion of the common loop factor $c_{\Gamma}$ in \eq{cgamma}, 
and it can therefore be reabsorbed in the corresponding redefinition of the coupling. We 
confirm then that, when expanded in terms of the high-energy coupling $\tilde{\alpha}_s$, 
the dipole formula does not generate simple poles at NLL for color-singlet $t$-channel 
exchange, and the non-vanishing result found in Ref.~\cite{Caron-Huot:2013fea} must be
considered as a violation of the dipole formula at the four-loop level.


\section{Conclusion}
\label{discu}

The high-energy limit, $s/|t| \to \infty$, of gauge theory amplitudes is of great 
theoretical and phenomenological interest, and has been a major focus of 
investigation for several decades. The range of applications is vast: on the 
one hand, the high-energy limit can be used to study formal properties of 
scattering amplitudes in highly symmetric gauge theories, such as $N=4$ 
Super Yang-Mills theory (see, for example, Refs.~\cite{DelDuca:2009au,
Dixon:2014voa} for recent applications in this direction); on the other hand, 
it is very relevant for phenomenological applications to cross sections of interest 
at colliders such as LHC (see, as recent examples, Refs.~\cite{DelDuca:2013lma,
Andersen:2011hs}).

The main reason beyond this wide range of applications is the simplicity of
high-energy amplitudes. As $s/|t| \to \infty$, amplitudes come to be dominated 
by logarithmic enhancements, which can be studied to all orders in perturbation 
theory with the tool of Gribov-Regge theory, as well as with the more typical tools
of perturbative QCD. In the case of four parton amplitudes, at leading and 
next-to-leading logarithmic accuracy, an all-order
factorization holds, which resums energy logarithms based on the fact that, to this
accuracy, the only relevant singularities in the complex $J$ plane are simple poles.

It is generally understood~\cite{Fadin:1993wh}, however, and in fact it has been 
proven~\cite{Fadin:2006bj}, that this simple Reggeization picture of QCD scattering 
amplitudes cannot be generally applicable beyond NLL accuracy. It is also understood 
that a complete Reggeization picture should include the contributions of Regge cuts 
at sufficiently high orders in perturbation theory, possibly beginning with non-planar 
contributions to scattering amplitudes at the three-loop level~\cite{Collins:1977jy}. 
The details of how those cuts arise, and thus of how the simple Reggeization picture 
breaks down, are however not known.

Building upon the earlier analyses of Refs.~\cite{DelDuca:2011ae,Bret:2011xm,
Korchemsky:1993hr,Korchemskaya:1994qp,Korchemskaya:1996je}, which 
examined the interplay of high-energy factorization and infrared factorization, in 
Ref.~\cite{DelDuca:2013ara} we outlined a roadmap to explore the violations of 
the simple Reggeization picture, and thus the rise of the Regge cuts, by comparing 
the two factorizations, when applied to QCD scattering amplitudes order by order 
in perturbation theory. The immediate outcome was to explain the origin of a 
non-factorizing term, independent of ln$(s/|t|)$, first uncovered in Ref.~\cite{DelDuca:2001gu} 
in amplitudes for parton-parton scattering at the two-loop level. In addition, since it 
was already clear~\cite{DelDuca:2011ae,Bret:2011xm} that the presence of 
non-factorizing terms in ln$(s/|t|)$ at three-loops would invalidate the notion 
of a universal Regge trajectory, in Ref.~\cite{DelDuca:2013ara} we proposed 
a scheme to gather non-universal contributions into a non-factorizing remainder 
function, and we used infrared factorization to re-define the impact factors and 
the Regge trajectory as functions of universal terms only. Those definitions allow 
one in principle to compute the complete three-loop Regge trajectory unambiguously.

In this paper, we have provided the details of the roadmap sketched in 
Ref.~\cite{DelDuca:2013ara}, and we have presented a complete analysis of 
four-parton QCD scattering amplitudes in the high-energy limit, including all 
available results up to three loops, and deriving some all-order relations.
In particular, we have performed a detailed comparison of infrared and 
high-energy factorizations up to three-loop order, for both real and imaginary 
parts of the amplitudes. The cross-fertilization between the two approaches yields 
a number of interesting results. 

First of all, infrared factorization allowed us to identify non-universal terms affecting 
Regge behavior, and therefore to predict the infrared poles of the factorization-breaking 
terms up to three loops. To that accuracy, it is now possible to define unambiguously 
impact factors, the Regge trajectory, and the remainder functions. In addition, we 
analysed the $t$-channel exchange of color representations other than the octet in 
scattering amplitudes up to three-loops, at leading and next-to-leading logarithmic 
accuracy, and, as an example, we computed the infrared pole of a single-logarithmic 
term at four loops, in the singlet component of gluon-gluon scattering, at next-to-leading 
logarithmic accuracy. Our result is consistent with the findings of Ref.~\cite{Caron-Huot:2013fea}.
The four-loop NLL single-pole contribution to singlet exchange arising from the dipole
formula can be reabsorbed in the definition of the coupling: any contribution without 
this property must thus be considered as a violation of the dipole formula at four 
loops. 

On the other hand, high-energy factorization at LL and NLL level provides all-order
constraints on the hard functions defined by infrared factorization. Under mild and 
well-motivated assumptions, that high-energy factorization should extend to the NLL
imaginary part of $t$-channel octet exchange, and to $\ord(\eps)$ contributions to
the amplitude, we have derived a set of all-order identities showing that all hard
functions for four-parton scattering amplitudes in QCD vanish in $d=4$ in the 
high-energy limit, up to NLL accuracy. This result considerably reinforces the idea 
that all high-energy logarithms in QCD originate from infrared enhancements: this 
idea underlies many of the existing approaches to the high-energy limit, and it is 
likely that it will bring further insights in the future.

It is easy to see that the detailed analysis provided here can be extended to scattering 
processes with the production of more than two partons in the final state, as well as to 
quark-gluon scattering with a quark Regge trajectory exchanged in the $t$ channel, 
as outlined in Refs.~\cite{DelDuca:2011ae,Bret:2011xm}. This will hopefully shed further 
light on the interplay of high-energy and infrared factorizations, and will possibly yield
high-order results relevant for high-energy precision phenomenology at colliders.


\vspace{1.5cm}

\noindent {\large{\bf Acknowledgements}}

\vspace{2mm}

\noindent We thank S. Caron-Huot for useful discussions. This work was supported 
by MIUR (Italy), under contracts 2006020509$\_$004 and  2010YJ2NYW$\_$006; 
by the Research Executive Agency (REA) of the European Union, through the Initial 
Training Network LHCPhenoNet under contract PITN-GA-2010-264564; by the 
ERC grant 291377 ``LHCtheory: Theoretical predictions and analyses of LHC physics: 
advancing the precision frontier'', and by the University of Torino and the Compagnia 
di San Paolo under contract ORTO11TPXK. LM thanks CERN, NIKHEF, and the Higgs 
Center for Theoretical Physics at the University of Edinburgh for hospitality and support 
during the completion of this work. LV would like to thank the Mainz Institute for Theoretical 
Physics (MITP) for its hospitality and support during the completion of this work.

\vspace{2cm}


\appendix


\section{A color basis for four-parton amplitudes in QCD}
\label{AppColor}

In this section we provide orthonormal colour bases for each process we considered in the 
text. Most of the considerations we discussed in the paper are independent on the choice 
of basis in colour space, but it is useful to pick explicitly a set of tensors when dealing 
with the actual computation of the amplitudes in the high energy limit. In this kinematic 
regime, scattering amplitudes are organised conveniently by means of orthonormal 
bases diagonalising the operator ${\bf T}_t^2$. To construct them, we followed the approach 
of \cite{Beneke:2009rj,Beenakker:2013mva}, and we used the package {\tt ColorMath}
of Ref.~\cite{Sjodahl:2012nk} to deal with colour algebra.


\subsection{Quark-quark scattering}
\label{qqqq}

The quark-quark scattering amplitude has only two color components. For $N_c = 3$, 
they correspond to the exchange of a singlet or an octet in the $t$ channel, so we label 
the corresponding color tensors as $c^{(1)}_{qq}$ and $c^{(8)}_{qq}$; the expressions
we use are however valid for generic $N_c$. We choose
\beqa
  c^{(1)}_{qq} & = & \frac{1}{N_c} \, {\delta^{\a_4}}_{\a_1} \, {\delta^{\a_3}}_{\a_2} \, , 
  \nonumber \\
  c^{(8)}_{qq} & = & \frac{2}{\sqrt{N_c^2 - 1}} \, \, {\left({\bf T}^a\right)^{\a_4}}_{\a_1} \, 
  {\left({\bf T}_a\right)^{\a_3}}_{\a_2} \, ,
\label{cqqqq}
\eeqa
where $\a_i = 1, \ldots, N_c$ are indices in the fundamental representation of $SU(N_c)$,
while $a = 1, \ldots, N_c^2 - 1$ is in the adjoint representation, and we omit color indices
on the left-hand side for simplicity. Both tensors are normalized to unity with the convention 
$\text{Tr} \left( {\bf T}^{a} {\bf T}^{b} \right) = \frac{1}{2} \delta^{a b} $. For completeness, 
we report also the explicit expressions in this basis of the operators ${\bf T}^2_t$ and 
${\bf T}^2_s$. The matrix ${\bf T}^2_t$ is diagonal by construction, while ${\bf T}^2_s$ 
mixes the different components of the amplitude in colour space. We find 
\beq
  {\bf T}^2_{t, qq} \, = \, \left[
    \begin{array}{cc}
    0 & 0 \\
    0 & N_c \\
    \end{array}
  \right] \, , \qquad
  {\bf T}^2_{s, qq} \, = \,  \frac{\sqrt{N_c^2 - 1}}{N_c} \, \left[
    \begin{array}{cc}
    \sqrt{N_c^2 - 1} & 1 \\
    1 & \frac{N_c^2 - 3}{\sqrt{N_c^2 - 1}}
    \end{array}
  \right].
\label{TsTtqqqq}  
\eeq


\subsection{Gluon-gluon scattering}
\label{gggg}

The colour structure for gluon-gluon scattering is more intricate. In this case, the basis 
is composed of six colour tensors, which again we label with their $SU(3)$ quantum 
numbers, while the expressions we give are for generic $N_c$. We choose
\beqa
\label{cgggg}
  c^{(1)}_{gg} & = & \frac{1}{N_c^2 - 1} \, {\delta^{a_4}}_{a_1} \, {\delta^{a_3}}_{a_2} \, ,
  \nonumber \\
  c^{(8_s)}_{gg} & = & \frac{N_c}{N_c^2 - 4} \, \frac{1}{\sqrt{N_c^2 - 1}} \, 
  d^{\, a_1 a_4 b} \, d^{\, a_2 a_3}_{\phantom{\, a_2 a_3} b} \, , \nonumber \\
  c^{(8_a)}_{gg} & = & \frac{1}{N_c} \, \frac{1}{\sqrt{N_c^2 - 1}} \, 
  f^{\, a_1 a_4 b} \, f^{\, a_2 a_3}_{\phantom{\, a_2 a_3} b} \, , \nonumber \\
  c^{(10 + \overline{10})}_{gg} & = & \sqrt{\frac{2}{(N_c^2 - 4)(N_c^2 - 1)}}
  \left[ \frac{1}{2} \left({\delta^{a_1}}_{a_2} \, {\delta^{a_3}}_{a_4} - 
  {\delta^{a_3}}_{a_1} \, {\delta^{a_4}}_{a_2} \right) - \frac{1}{N_c}
  f^{\, a_1 a_4 b} \, f^{\,a_2a_3}_{\phantom{\, a_2 a_3} b} \right] \, , \nonumber \\
  c^{(27)}_{gg} & = & \frac{2}{N_c \sqrt{(N_c + 3)(N_c - 1)}} \, \bigg[ - 
  \frac{N_c + 2}{2 N_c (N_c + 1)} \, {\delta^{a_4}}_{a_1} \, {\delta^{a_3}}_{a_2}
  \nonumber \\
  & & + \, \frac{N_c + 2}{4 N_c} \, \big( {\delta^{a_1}}_{a_2} \, {\delta^{a_3}}_{a_4}
  + {\delta^{a_3}}_{a_1} \, {\delta^{a_4}}_{a_2} \big) - \frac{N_c + 4}{4( N_c + 2)} \,
  d^{\, a_1 a_4 b} \, d^{\, a_2 a_3}_{\phantom{\, a_2 a_3} b} \nonumber \\
  & & + \, \frac{1}{4} \, \big( d^{\, a_1 a_2 b} \, d^{\, a_3 a_4}_{\phantom{\, a_2 a_3} b}
  + d^{\, a_1 a_3 b} \, d^{\, a_2 a_4}_{\phantom{\, a_2 a_3} b} \big) \bigg] \, , \\
  c^{(0)}_{gg} & = & \frac{2}{N_c \sqrt{(N_c - 3)(N_c + 1)}} \, \bigg[ 
  \frac{N_c - 2}{2 N_c (N_c - 1)} \, {\delta^{a_4}}_{a_1} \, {\delta^{a_3}}_{a_2} \nonumber \\
  & & + \, \frac{N_c - 2}{4 N_c} \, \big( {\delta^{a_1}}_{a_2} \, {\delta^{a_3}}_{a_4}
  + {\delta^{a_3}}_{a_1} \, {\delta^{a_4}}_{a_2} \big) + \frac{N_c - 4}{4 (N_c - 2)} \,
  d^{\, a_1 a_4 b} \, d^{\, a_2 a_3}_{\phantom{\, a_2 a_3} b} \nonumber \\
  & & - \, \frac{1}{4} \big( d^{\, a_1 a_2 b} \, d^{\, a_3 a_4}_{\phantom{\, a_2 a_3} b}  
  + d^{\, a_1 a_3 b} \, d^{\, a_2 a_4}_{\phantom{\, a_2 a_3} b} \big) \bigg] \, .
  \nonumber
\eeqa
We note that it is not necessary to treat separately the two decuplet representations
since they always contribute to the amplitude with the same coefficients. The 
tensors $c^{(8_a)}$ and $c^{(10 + \overline{10})}$ are odd under the exchanges
$a_1 \leftrightarrow a_4$ and $a_2 \leftrightarrow a_3$, while $c^{(1)}_{gg}$, 
$c^{(8_s)}_{gg}$, $c^{(27)}_{gg}$ and $c^{(0)}_{gg}$ are even. The last representation,
as suggested by its label, does not contribute for $N_c = 3$, since its dimensionality 
is given by
\beq
  {\rm dim} \left[ \, {\bf 0} \, \right] \, = \, \frac{N_c^2 (N_c - 3)(N_c + 1)}{4} \, ,
\label{dim0}
\eeq 
and it vanishes for SU(3). In the orthormal basis defined by \eq{cgggg}, the diagonal
matrix ${\bf T}^2_t$ evaluates to 
\beq
  {\bf T}^2_{t, gg} \, = \, {\rm diag} \Big( 0, \, N_c , \, N_c, \, 2 N_c, \, 2 (N_c + 1), \, 
  2 (N_c - 1) \Big) \, ,
\label{Ttgggg}
\eeq
while ${\bf T}_{s, gg}$ is symmetric and reads
\beq
  {\bf T}^2_{s, gg} \, = \, \left[
    \begin{array}{cccccc}
    2 N_c & 0 & {\cal T}_ {1, 8_a} & 0 & 0 & 0 \\
    0 & 3 N_c/2 & {\cal T}_ {8_s, 8_a} & {\cal T}_ {8_s, 10} & 0 & 0 \\
    {\cal T}_ {1, 8_a} & {\cal T}_ {8_s, 8_a} & 3 N_c/2 & 0 & {\cal T}_ {8_s, 27} &
    {\cal T}_ {8_s, 0} \\
    0 & {\cal T}_ {8_s, 10} & 0 & N_c & {\cal T}_ {10, 27} & {\cal T}_ {10, 0} \\
    0 & 0 & {\cal T}_ {8_s, 27} & {\cal T}_ {10, 27} & N_c - 1 & 0 \\
    0 & 0 & {\cal T}_ {8_s, 0} & {\cal T}_ {10, 0} & 0 & N_c + 1
    \end{array}
  \right] \, ,
\label{Tsgggg}
\eeq
where
\beqa
\label{elmatgggg}
  {\cal T}_ {1, 8_a} & = & - \frac{2 N_c}{\sqrt{N_c^2 - 1}} \, , \quad 
  {\cal T}_ {8_s, 8_a} \, = \, - \frac{N_c}{2} \, , \quad {\cal T}_ {8_s, 10} \, = \, 
  - N_c \sqrt{\frac{2}{N_c^2 - 4}} \, , \nonumber \\  
  {\cal T}_ {8_s, 27} & = & - \sqrt{\frac{N_c + 3}{N_c + 1}} \, , \quad 
  {\cal T}_ {8_s, 0} \, = \, - \sqrt{\frac{N_c - 3}{N_c - 1}} \, , \nonumber \\
  {\cal T}_ {8_s, 27} & = & - \sqrt{\frac{N_c + 3}{N_c + 1}} \, , \\
  {\cal T}_ {10, 27} & = & - \sqrt{\frac{(N_c + 3)(N_c + 1)(N_c - 2)}{2 (N_c + 2)}} \, ,
  \nonumber \\
  {\cal T}_ {10, 0} & = & - \sqrt{\frac{(N_c - 3)(N_c - 1)(N_c + 2)}{2 (N_c - 2)}} \, .
  \nonumber
\eeqa


\subsection{Quark-gluon scattering}
\label{qgqg}

We conclude by discussing quark-gluon scattering. In this case the $t$-channel exchange 
takes place between a quark line and a gluon line, so in order to enumerate the relevant 
representations one must consider the intersection of the tensor product ${\bf 3} \otimes
{\bf \overline 3} = {\bf 1} \oplus {\bf 8}$ with ${\bf 8} \otimes {\bf \overline 8} = {\bf 1} \oplus 
{\bf 8} \oplus {\bf 8} \oplus {\bf 10} \oplus {\bf \overline{10}} \oplus {\bf 27} \oplus {\bf 0}$,
including copies of equivalent representations. This leaves the singlet and the two copies of 
the adjoint representation. An orthonormal basis of color tensors in this space is given by
\beqa
\label{cqgqg}
  c^{(1)}_{qg} & = & \frac{1}{\sqrt{N_c (N_c^2 - 1)}} \, {\delta^{\a_4}}_{\a_1} \,
  {\delta^{a_3}}_{a_2} \, , \nonumber \\
  c^{(8s)}_{qg} & = & \sqrt{\frac{2 N_c}{(N_c^2 - 4)(N_c^2 - 1)}} \, 
  {( T^b )^{\a_4}}_{\a_1} \, d_b^{\, \, \, a_3 a_2} \, , \\
  c^{(8a)}_{qg} & = & {\rm i} \, \sqrt{\frac{2}{N_c (N_c^2 - 1)}} \, 
  { ( T^b )^{\a_4}}_{\a_1} \, f_b^{\, \, \, a_3 a_2} \, .
\eeqa
The operators ${\bf T}^2_t$ and ${\bf T}^2_s$ in this basis take the form
\beq
  {\bf T}^2_{t, qg} \, = \, \left[
    \begin{array}{ccc}
    0 & 0 & 0 \\
    0 & N_c & 0 \\
    0 & 0 & N_c
    \end{array}
  \right] \, , \qquad 
  {\bf T}^2_{s, qg} \, = \, \left[
    \begin{array}{ccc}
    \frac{3 N_c^2 - 1}{2 N_c} & 0 & - \sqrt{2} \\
    0 & \frac{2 N_c^2 - 1}{2 N_c} & - \frac{\sqrt{N_c^2 - 4}}{2} \\
    - \sqrt{2} & - \frac{\sqrt{N_c^2 - 4}}{2} & \frac{2 N_c^2 - 1}{2 N_c}
    \end{array}
  \right] \, .
\label{TtTsqgqg}
\eeq


\section{Anomalous dimensions}
\label{AppAnDim}

The discussion of Regge factorization in \secn{coirf} led us to the prediction of the 
divergent part of the Regge trajectory and of the impact factors, in terms of the operators 
appearing in the infrared factorization formula \eq{Zfact1}. In the text we focused mostly
on the formal relations connecting the two factorizations, as for example in \eq{alpha1}
and \eq{imp1}, however one is ultimately interested in explicit results. For completeness,
we give here the values of all the relevant anomalous dimensions, up to three loops.

In order to construct the infrared operators relevant to the Regge limit, defined in 
\eq{widetildeZ} and \eq{Z1}, we need the functions $K(\as)$, $D(\as)$ and $B_i(\as)$, 
defined in \eq{cusp} and \eq{intandim} respectively. Performing the relevant
integrals, we find
\beqa
\label{formalint}
  K (\as) & = & \frac{\alpha_s}{\pi} \, 
  \frac{\widehat{\gamma}_K^{(1)}}{4 \epsilon} \, + \left( \frac{\alpha_s}{\pi} \right)^2 \,
  \left( \frac{\widehat{\gamma}_K^{(2)}}{8 \epsilon} -
  \frac{b_0 \, \widehat{\gamma}_K^{(1)}}{32 \epsilon^2} \right) \nonumber \\
  & & + \, \left( \frac{\alpha_s}{\pi} \right)^3 \left( \frac{\widehat{\gamma}_K^{(3)}}{12 
  \epsilon} - \frac{b_0 \, \widehat{\gamma}_K^{(2)} + b_1 \, 
  \widehat{\gamma}_K^{(1)}}{48 \epsilon^2} + \frac{b_0^2 \, 
  \widehat{\gamma}_K^{(1)}}{192 \epsilon^3} \right) \, + \ord (\alpha_s^4) \, ,
  \nonumber \\
  D (\as) & = & - \frac{\as}{\pi} \, \frac{\widehat{\gamma}_K^{(1)}}{4 \eps^2}
  + \left( \frac{\as}{\pi} \right)^2 \bigg[ \frac{3 b_0 \widehat{\gamma}_K^{(1)}}{64 \epsilon^3}
  - \frac{\widehat{\gamma}_K^{(2)}}{16 \epsilon^2} \bigg] \\
  & & + \, \left( \frac{\as}{\pi} \right)^3 \bigg[ - \frac{11 b_0^2 
  \widehat{\gamma}_K^{(1)}}{1152 \epsilon^4}
  + \frac{1}{\epsilon^3} \left( \frac{b_1 \widehat{\gamma}_K^{(1)}}{36}
  + \frac{5 b_0 \widehat{\gamma}_K^{(2)}}{288} \right)
  - \frac{\widehat{\gamma}_K^{(3)}}{36 \epsilon^2} \bigg] + \ord( \as^4 ) \, ,
  \nonumber 
\eeqa
where $b_i$ are the coefficients of the perturbative expansion of the beta function which,
in our normalizations, read
\beq
  b_0 \, = \, \frac{11 C_A - 4 T_R n_f}{3} \, , \qquad 
  b_1 \, = \, \frac{17 C_A^2 - ( 10 C_A + 6 C_F) T_R n_f}{6} \, ,
\label{betacoeff}
\eeq
while $\widehat{\gamma}_K^{(i)}$ are the perturbative coefficients of the light-like
cusp anomalous dimension, divided by the quadratic Casimir eigenvalue of the
relevant representation. This is a universal (representation-independent) function
at least up to three loops, given by
\beqa
\label{hatgammaK}
  \widehat{\gamma}_K (\as) & = & 2 \, \frac{\as}{\pi}
  + \left( \frac{\as}{\pi} \right)^2 \bigg[ \left( \frac{67}{18} - \zeta(2) \right) C_A
  - \frac{10}{9} T_R n_f \bigg] \, + \nonumber \\ 
  & & \, \left( \frac{\as}{\pi} \right)^3 \bigg[ \frac{C_A^2}{96} 
  \left( 490 - \frac{1072}{3} \zeta(2) + 88 \zeta(3) + 264 \zeta(4) \right)
  \nonumber \\
  & & \hspace{1.5cm} + \, \frac{C_A T_R n_f}{96} \left(-\frac{1672}{9} + \frac{320}{3} \zeta(2)
  - 224 \zeta(3) \right) \\
  & & \hspace{1.5cm} + \, \frac{C_F T_R n_f}{32} \left( - \frac{220}{3} + 64 \zeta(3) \right)
  - \frac{2 T_R^2 n_f^2}{27} \bigg] \, + \, \ord(\as^4) \, . \nonumber 
\eeqa 
Finally we note that, up to a factor of 2, $B_i (\as)$ is defined by the same integral, and 
therefore by the same perturbative expansion, given in \eq{formalint}, as $K(\as)$, but 
with the cusp anomalous dimension replaced by the collinear anomalous dimension of 
the relevant field, $\gamma_i$ with $i = q, g$. The perturbative coefficients of collinear 
anomalous dimensions were extracted from form factors data in \cite{Becher:2009qa} 
and they are
\beqa
\label{gammacol}
  \gamma_{q} (\as) & = & - \frac{3}{4} \, C_F \, \frac{\as}{\pi} 
  + \left( \frac{\as}{\pi}\right)^2 \Bigg[ \frac{C_F^2}{16} \left( - \frac{3}{2} + 12 \zeta(2) 
  - 24 \zeta(3) \right) \nonumber \\
  & & + \, \frac{C_A C_F}{16} \left( - \frac{961}{54} - 11 \zeta(2) + 26 \zeta(3) \right)
  + \frac{C_F T_R n_f}{16} \left( \frac{130}{27} + 4 \zeta(2) \right) \Bigg] 
  \nonumber \\
  & & + \, \left( \frac{\as}{\pi} \right)^3 \frac{1}{64} \Bigg[ C_F^3 \left( - \frac{29}{2} - 
  18 \zeta(2) - 68 \zeta(3) - 144 \zeta(4) + 32 \zeta(2) \zeta(3) + 240 \zeta(5) \right) 
  \nonumber \\
  & & + \, C_F^2 C_A \left( - \frac{151}{4} + \frac{410}{3} \zeta(2) - \frac{844}{3} \zeta(3)
  + \frac{494}{3} \zeta(4) - 16 \zeta(2) \zeta(3) - 120 \zeta(5) \right) \nonumber \\
  & & + \, C_F C_A^2 \left( - \frac{139345}{2916} - \frac{7163}{81} \zeta(2)  
  + \frac{3526}{9} \zeta(3) - 83 \zeta(4) - \frac{88}{3} \zeta(2) \zeta(3) 
  - 136 \zeta(5) \right) \nonumber \\
  & & + \, C_F^2 T_R n_f \left( \frac{2953}{27} - \frac{52}{3} \zeta(2) 
  + \frac{512}{9} \zeta(3) - \frac{280}{3} \zeta(4) \right) \nonumber \\ \nn
  & & + \, C_A C_F T_R n_f \left( - \frac{17318}{729} + \frac{5188}{81} \zeta(2)
  - \frac{1928}{27} \zeta(3) + 44 \zeta(4) \right)  \nonumber \\
  & & + \, C_F T_R^2 n_f^2 \left( \frac{9668}{729} - \frac{80}{9} \zeta(2) - 
  \frac{32}{27} \zeta(3) \right) \Bigg] \, + \, \ord(\as^4) \\ 
  \gamma_g (\as) & = & - \frac{b_0}{4} \, \frac{\as}{\pi} 
  + \left( \frac{\as}{\pi} \right)^2 \Bigg[ \frac{C_A^2}{16} \left( - \frac{692}{27}
  + \frac{11}{3} \zeta(2) + 2 \zeta(3) \right) \nonumber \\
  & & + \, \frac{C_A T_R n_f}{16} \left( \frac{256}{27} - \frac{4}{3} \zeta(2) \right) 
  + \frac{C_F T_R n_f}{4} \Bigg] \nonumber \\
  & & + \, \left( \frac{\as}{\pi} \right)^3 \frac{1}{64} \Bigg[ C_A^3
  \bigg( - \frac{97186}{729} + \frac{6109}{81} \zeta(2) + \frac{122}{3} \zeta(3) 
  - \frac{319}{3} \zeta(4) \nonumber \\ 
  & & \hspace{3.2cm} - \, \frac{40}{3} \zeta(2) \zeta(3) - 16 \zeta(5) \bigg) \nonumber \\
  & & + \, C_A^2 T_R n_f \left( \frac{30715}{729} - \frac{2396}{81} \zeta(2) 
  + \frac{712}{27} \zeta(3) + \frac{164}{3} \zeta(4) \right) \nonumber \\
  & & + \, C_A C_F T_R n_f \left( \frac{2434}{27} - 4 \zeta(2) -\frac{304}{9} \zeta(3) 
  - 16 \zeta(4) \right) \, - \, 2 C_F^2 T_R n_f \nonumber \\
  & & + \, C_A T_R^2 n_f^2 \left( - \frac{538}{729} + \frac{80}{27} \zeta(2) 
  - \frac{224}{27} \zeta(3) \right) \, - \, \frac{44}{9} C_F T_R^2 n_f^2 \Bigg]
  \, + \, \ord(\as^4) \, , \nonumber
\eeqa
which completes the list of required anomalous dimensions up to three loops.


\section{Hard functions for four-parton amplitudes in the high-energy limit}
\label{AppHard}

We have now given explicitly almost all the ingredients needed to construct the 
high-energy limit of four-parton QCD amplitudes up to two loops, and up to three loops
for infrared singular contributions. Using existing calculations, this construction can be 
achieved at the one-loop level up to $\ord(\eps^2)$, using \eq{AmpCoeff1}, at the 
two-loop level up to $\ord(\eps^0)$, using \eq{AmpCoeff2}, and at the three-loop 
level up to $\ord(\eps^{-2})$, and in some cases up to $\ord(\eps^{-1})$ using 
Eqns.~(\ref{AmpCoeff30})-(\ref{AmpCoeff33}). Specifically, 
all ingredients arising from infrared factorization have been given explicitly
to the necessary accuracy in Appendices A and B. The only missing contributions
are those arising from the hard functions $H^{(n)}$, with $n = 0,1,2$, which can only
be extracted from explicit finite-order calculations. The necessary helicity amplitudes 
for the processes $qq\to qq$, $gg\to gg$ and $qg \to qg$ with massless quarks have 
been calculated up to two loops in recent years by different groups~\cite{Bern:2002tk, 
Bern:2003ck, DeFreitas:2004tk, Glover:2004si, Glover:2003cm}.  In this Appendix we 
consider the high-energy limit of these amplitudes and we provide explicit expressions
for the hard functions $H^{(n),l,[c]}$, for $n = 0,1,2$, for all logarithmic orders and color 
components, and for each process\footnote{After the completion of this work, the hard 
functions corresponding to two-loop four-parton QCD amplitudes were extracted,
without taking the high-energy limit, in ref.~\cite{Broggio:2014hoa}.}. 
Inserting these results in \eq{AmpCoeff1} at one loop, and in \eq{AmpCoeff2} at two 
loops, one recovers the high-energy limit of the results discussed in ~\cite{Bern:2002tk,
Bern:2003ck,DeFreitas:2004tk,Glover:2004si,Glover:2003cm}, while inserting them in 
Eqns. (\ref{AmpCoeff30})-(\ref{AmpCoeff33}) one gets a complete prediction for the 
poles of three-loop four-parton amplitudes in the high-energy limit, valid to leading 
power in $t/s$ and for poles down to $\eps^{-2}$, with further predictions at single-pole 
level, as discussed in \secn{even-even}. Note that, as described below, to leading power 
in $t/s$ essentially only one helicity amplitude survives for each partonic process.


\subsection{Quark-quark scattering}
\label{Hqqqq}
 
Quark-quark scattering is the simplest process we consider, as it contains only two 
colour structures, the singlet and the octet. We write the hard coefficients of the 
amplitude ${\cal M}(q^+ q^+ \to q^+ q^+)$, which has leading power in the high-energy 
limit, by using the basis of eq. (\ref{cqqqq}). At tree level only the octet contributes to 
the amplitude, and we find
\beqa
  H^{(0),[1]}_{qq} & = & 0 \, , \nonumber \\
  H^{(0),[8]}_{qq} & = & \frac{\sqrt{N_c^2 - 1}}{x}  \,  ,
\label{Hqqqqtree}
\eeqa
where here and below $x = \frac{t}{s}$. Next we consider the one-loop amplitude, expanded 
up to $O(\epsilon^2)$. Leading logarithmic terms at this perturbative order are given only by 
the Regge trajectory: indeed, we find vanishing LL colour-singlet and octet components in
$d = 4$, as explained in the text. More precisely,
\beqa
  H^{(1),1,[1]}_{qq} & = & 0 \, , \nonumber \\
  H^{(1),1,[8]}_{qq} & = & - \, \frac{N_c \sqrt{N_c^2 - 1}}{24 x} \, \epsilon\, 
  \Big( 6 \zeta(2) + 28 \epsilon \zeta(3) \Big) \, . 
\label{Hqqqq1}
\eeqa
Turning to non-logarithmic terms at one loop, we find\footnote{Throughout 
Appendix~\ref{AppHard} we have explicitly set $T_R = 1/2$.} 
\beqa
  H^{(1),0,[1]}_{qq} & = & - \, {\rm i} \pi \, \frac{N_c^2 - 1}{24 N_c \, x} \, \epsilon \,
  \Big( 6 \zeta(2 )+ 28 \epsilon \zeta(3) \Big) \, , \nonumber \\
  H^{(1),0,[8]}_{qq} & = & \frac{\sqrt{N_c^2 - 1}}{x} \, \Bigg\{ 
  \left( \frac{13}{36} + \frac{7}{4} \zeta(2) \right) N_c
  + \left( 2 - \frac{1}{4} \zeta(2) \right) \frac{1}{N_c} - \frac{5}{18} \, n_f  \\
  & & + \, \epsilon \, \bigg[ \left(\frac{20}{27} - \frac{1}{12} \zeta(2) + 
  \frac{5}{3} \zeta(3) \right) N_c + \left( 4 - \frac{3}{8} \zeta(2) - 
  \frac{7}{6} \zeta(3) \right) \frac{1}{N_c} \nonumber \\
  & & \hspace{1cm} + \, \left( - \frac{14}{27} + \frac{1}{12} \zeta(2) \right) n_f \bigg] 
  \nonumber \\
  & & + \, \epsilon^2 \bigg[ \left( \frac{121}{81} - \frac{13}{72} \zeta(2) - 
  \frac{7}{18} \zeta(3) + \frac{35}{32} \zeta(4) \right)  N_c \nonumber \\
  & & \hspace{1cm} + \left( 8 - \zeta(2) - \frac{7}{4} \zeta(3) - 
  \frac{47}{32} \zeta(4) \right) \frac{1}{N_c} + \left( - \frac{82}{81} + \frac{5}{36} \zeta(2) + 
  \frac{7}{18} \zeta(3) \right) n_f \bigg] \nonumber \\ 
  & & + \, {\rm i} \pi \, \frac{1}{12 N_c} \, \epsilon \, \Big( 6 \zeta(2)
  + 28 \epsilon \zeta(3) \Big) \Bigg\} \, . \nonumber
\label{Hqqqq10}
\eeqa
Now we consider the two-loop quark-quark scattering amplitude, where only terms up to
$\ord(\eps^0)$ are available. The leading-logarithmic hard functions vanish again in the 
high-energy limit, as discussed in the text. Indeed we find
\beqa
  H^{(2),2,[1]}_{qq} & = & 0 \, , \nonumber \\
  H^{(2),2,[8]}_{qq} & = & 0 \, .
\label{Hqqqq2}
\eeqa
At two loops, next-to-leading logarithms in the octet component are related to the gluon 
Regge trajectory, while the singlet component vanishes, so that
\beqa
  H^{(2),1,[1]}_{qq} & = & 0 \, , \nonumber \\
  H^{(2),1,[8]}_{qq} & = & - N_c \, \frac{\sqrt{N_c^2 - 1}}{x} \, 
  \frac{ \big(27 \zeta(3) - 202 \big) N_c + 28 \, n_f}{216} \, .
\label{Hqqqq2nll}
\eeqa
Finally, the non-logarithmic hard functions $H^{(2),0,[c]}$ are given by
\beqa
\label{Hqqqq2nnll}
  H^{(2),0,[1]}_{qq} & = & {\rm i} \pi \, \frac{N_c^2 - 1}{N_c \, x} \,
  \frac{ \big( 202 + 324 \zeta(2) + 135 \zeta(3) \big) N_c - 28 \, n_f}{216} \, , \nonumber \\
  H^{(2),0,[8]}_{qq} & = & \frac{\sqrt{N_c^2 - 1}}{x} \Bigg\{
  \left( \frac{23213}{20736} + \frac{437}{144} \zeta(2) + 
  \frac{41}{72} \zeta(3) + \frac{105}{64} \zeta(4) \right) N_c^2 \\
  & & + \, \frac{30659}{5184} + \frac{833}{288} \zeta(2) - 
  \frac{205}{144} \zeta(3) - \frac{41}{32} \zeta(4) \nonumber \\
  & & + \left( \frac{511}{256} + \frac{13}{32} \zeta(2) - 
  \frac{15}{16} \zeta(3) - \frac{39}{64} \zeta(4) \right) \frac{1}{N_c^2}  \nonumber \\
  & & - \left( \frac{455}{432} + \frac{107}{144} \zeta(2) + 
  \frac{23}{72} \zeta(3) \right) N_c \, n_f \nonumber \\
  & & - \left( \frac{685}{648} + \frac{13}{144} \zeta(2) + \frac{19}{72} \zeta(3) 
  \right) \frac{n_f}{N_c} + \frac{25}{324} \, n_f^2 \nonumber \\
  & & + \, {\rm i} \pi \left( - \frac{101}{54} + \frac{1}{4} \zeta(3) + \frac{7}{27} \frac{n_f}{N_c}
  \right) \Bigg\} \, . \nonumber 
\eeqa


\subsection{Gluon-gluon scattering}
\label{Hgggg}

The gluon-gluon scattering amplitude has more structures: by using the colour basis 
described in Appendix~\ref{AppColor}, we identify two odd components (the antisymmetric
octet, and the direct sum of the decuplet and its complex conjugate), and four even 
components (the singlet, the symmetric octet, and the representations we label with 
${\bf 27}$ and ${\bf 0}$). Here we consider the scattering processes ${\cal M}(g^+ g^- \to 
g^+ g^-) = {\cal M}(g^+ g^+ \to g^- g^-)$, which are leading in the high-energy 
limit. We begin with the tree-level amplitude: at this order only the antisymmetric 
octet contributes to the high energy limit, and we find
\beqa
  H^{(0),[1]}_{gg} & = & H^{(0),[8_s]}_{gg} \, = \, H^{(0),[10 + \overline{10}]}_{gg} \, 
  = \, H^{(0),[27]}_{gg} \, = \, H^{(0),[0]}_{gg} \, = \, 0 \, , \nonumber \\
  H^{(0),[8]}_{gg} & = & - \, 2 \, \frac{N_c \sqrt{N_c^2 - 1}}{x} \, .
\label{Hgggg0}
\eeqa
Leading-logarithmic one-loop hard parts, as expected, also vanish in $d=4$. More precisely
\beqa
  H^{(1),1,[1]}_{gg} & = & H^{(1),1,[8_s]}_{gg} \, = \, H^{(1),1,[10 + \overline{10}]}_{gg} \, = \, 
  H^{(1),1,[27]}_{gg} \, = \, H^{(1),1,[0]}_{gg} \, = \, 0 \, , \nonumber \\
  H^{(1),1,[8_a]}_{gg} & = & \frac{N_c^2 \sqrt{N_c^2 - 1}}{12 x} \, \epsilon \, 
  \big( 6 \zeta(2) + 28 \epsilon \zeta(3) \Big) \, ,
\label{Hgggg1ll}  
\eeqa
The one-loop amplitude is completed by the vector $H^{(1),0}$, whose components,
expanded up to $O(\epsilon^2)$, are given by
\beqa
\label{Hgggg1nll}
  H^{(1),0,[1]}_{gg} & = & - \, {\rm i} \pi \, \frac{N_c^2}{6 x} \, \epsilon \, 
  \Big( 6 \zeta(2) + 28 \epsilon \zeta(3) \Big) \, , \nonumber \\
  H^{(1),0,[8_s]}_{gg} & = & - \, {\rm i} \pi \, \frac{N_c^2 \sqrt{N_c^2 - 1}}{24 x} \, \epsilon \, 
  \Big( 6 \zeta(2) + 28 \epsilon \zeta(3) \Big) \, , \nonumber \\
  H^{(1),0,[8_a]}_{gg} & = & \frac{N_c \sqrt{N_c^2 - 1}}{x} \, \Bigg\{ \left( \frac{67}{18}
  - 4 \zeta(2) \right) N_c - \frac{5}{9} \, n_f + \epsilon \, \bigg[ \left(
  \frac{202}{27} - \frac{17}{3} \zeta(3) \right) N_c \nonumber \\
  & & - \, \frac{28}{27} \, n_f - \frac{b_0}{4} \, \zeta(2) \bigg] + 
  \epsilon^2 \bigg[ \left( \frac{1214}{81} - \frac{67}{36} \zeta(2) 
  - \frac{77}{18} \zeta(3) - \frac{41}{8} \zeta(4) \right) N_c \nonumber \\
  & & + \, \left( - \frac{164}{81} + \frac{5}{18} \zeta(2) + \frac{7}{9} \zeta(3) \right) n_f \bigg]
  - {\rm i} \pi \, \frac{N_c}{24} \, \epsilon \, \Big( 6 \zeta(2) + 28 \epsilon \zeta(3) \Big)
  \Bigg\} \, , \nonumber \\
  H^{(1),0,[10 + \overline{10}]}_{gg} & = & 0 \, , \\
  H^{(1),0,[27]}_{gg} & = & - \, {\rm i} \pi \, \frac{N_c \sqrt{(N_c + 3)(N_c - 1)}}{12 x} \, 
  \epsilon \, \Big( 6 \zeta(2) + 28 \epsilon \zeta(3) \Big) \, , \nonumber \\
  H^{(1),0,[0]}_{gg} & = & - \, {\rm i} \pi \, \frac{N_c \sqrt{(N_c - 3)(N_c + 1)}}{12 x} \, 
  \epsilon \, \Big( 6 \zeta(2) + 28 \epsilon \zeta(3) \Big) \, . \nonumber 
\eeqa
At two loops, leading-logarithmic hard functions vanish to $\ord(\eps^0)$,
\beq
  H^{(2),2,[k]}_{gg} \, = \, 0 \, , 
\label{Hgggg2ll}  
\eeq
while at NLL accuracy we find
\beqa
\label{Hgggg2nll}
  H^{(2),1,[1]}_{gg} & = & H^{(2),1,[8_s]}_{gg} \, = \, H^{(2),1,[10 + \overline{10}]}_{gg} \, 
  = \, H^{(1),2,[27]}_{gg} \, = \, H^{(2),2,[0]}_{gg} \, = \, 0 \, , \nonumber \\
  H^{(2),1,[8_a]}_{gg} & = & - \, \frac{N_c^2 \sqrt{N_c^2 - 1}}{x} \, \bigg[ \left(
  \frac{101}{54} - \frac{1}{4} \zeta(3) \right) N_c - \frac{7}{27} \, n_f \bigg] \, .
\eeqa
Finally, the components of non-logarithmic hard function $H^{(2),0}$ are
\beqa
\label{Hgggg2nnll}
  H^{(2),0,[1]}_{gg} & = & {\rm i} \pi \, \frac{1}{x} \, \Bigg\{ \bigg[ \left(
  \frac{265}{54} + \frac{5}{2} \zeta(3) \right) N_c^3 - \frac{139}{216} \, N_c^2 \, n_f + 
  \frac{7}{6} \, n_f + \frac{1}{8} \, \frac{n_f}{N_c^2} \bigg] + 2 \zeta(2) b_0 \Bigg\}, 
  \nonumber \\
  H^{(2),0,[8_s]}_{gg} & = & {\rm i} \pi \, \frac{\sqrt{N_c^2 - 1}}{x} \, 
  \Bigg[ \left( \frac{101}{108} - \frac{1}{8} \zeta(3) \right) N_c^3 - \frac{7}{54} \, N_c^2 \, n_f  
  \nonumber \\
  & & \hspace{2.5cm} + \, \left( \frac{29}{12} + \frac{4}{3} \zeta(2) \right) n_f + 
  \frac{1}{4} \, \frac{n_f}{N_c^2} \Bigg] \, , \nonumber \\
  H^{(2),0,[8_a]}_{gg} & = & \frac{\sqrt{N_c^2 - 1}}{x} \, \Bigg[ \left( \frac{11093}{1296} 
  - \frac{67}{72} \zeta(2) - \frac{22}{9} \zeta(3) - \frac{37}{8} \zeta(4) \right) N_c^3 
  \nonumber \\
  & & \hspace{2cm} + \left( - \frac{4849}{2592} + \frac{5}{36} \zeta(2) - 
  \frac{1}{18} \zeta(3) \right) N_c^2 \, n_f + \left( \frac{55}{96} - \frac{1}{2} \zeta(3)) 
  \right) n_f \nonumber \\
  & & \hspace{2cm} + \, {\rm i} \pi \, \frac{N_c^2}{216} \, \Big( \big( 202 - 27 \zeta(3) 
  \big) N_c - 28 \, n_f \Big) \Bigg] \, , \nonumber \\
  H^{(2),0,[10 + \overline{10}]}_{gg} & = & 0 \, , \\
  H^{(2),0,[27]}_{gg} & = & {\rm i} \pi \, \frac{\sqrt{(N_c + 3)(N_c - 1)} N_c}{x} \, 
  \Bigg[ \frac{5}{8} \, N_c^2 + \left(  \frac{683}{216} - \frac{11}{3} \zeta(2) - 
  \frac{7}{4} \zeta(3) \right) N_c \nonumber \\
  & & - \, \frac{11}{16} \, N_c \,n_f + \frac{1}{12} - \frac{22}{3} \zeta(2) - 3 \zeta(3) + 
  \left( - \frac{23}{54} + \frac{2}{3} \zeta(2) \right) n_f - 
  \frac{1}{16} \, \frac{n_f}{N_c} \Bigg] \, , \nonumber \\
  H^{(2),0,[0]}_{gg} & = & {\rm i} \pi \, \frac{\sqrt{(N_c - 3)(N_c + 1)} N_c}{x} \, 
  \Bigg[ - \frac{5}{8} \, N_c^2 + \left( \frac{683}{216} - \frac{11}{3} \zeta(2) -
  \frac{7}{4} \zeta(3) \right) N_c \nonumber \\
  & & + \, \frac{11}{16} \, N_c \, n_f - 
  \frac{1}{12}  + \frac{22}{3} \zeta(2) + 3 \zeta(3)
  + \left( -\frac{23}{54} + \frac{2}{3} \zeta(2) \right) n_f + 
  \frac{1}{16} \, \frac{n_f}{N_c} \Bigg] \, . \nonumber 
\eeqa


\subsection{Quark-gluon scattering}
\label{Hqgqg}

To conclude, we provide the hard functions up to two loops for quark-gluon scattering 
amplitudes. In this case there are three color components, corresponding to a singlet 
and two octets, since one has to take the intersection of the vector spaces defined by
the tensor products ${\bf 3} \otimes {\bf \overline 3}$ and  ${\bf 8} \otimes {\bf 8}$,
including separately all equivalent representations. The helicity amplitudes which 
are leading in the high-energy limit are ${\cal M}(q^+ g^- \to q^+ g^-) = - {\cal M}(q^+ 
g^+ \to q^+ g^+)$, and in the following we give the hard functions for the process 
${\cal M}(q^+g^- \to q^+g^-)$. The tree-level amplitude is given by
\beqa
\label{Hqgqg0}
  H^{(0),[1]}_{qg} & = & H^{(0),[8_s]}_{qg} \, = \, 0 \, , \nonumber \\
  H^{(0),[8_a]}_{qg} & = & \frac{\sqrt{2 N_c (N_c^2 - 1)}}{x} \, .
\eeqa
At one loop we have the leading logarithmic functions
\beqa
\label{Hqgqg1ll}
  H^{(1),1,[1]}_{qg} & = & H^{(1),1,[8_s]}_{qg} \, = \, 0 \, , \nonumber \\
  H^{(1),1,[8_a]}_{qg} & = & - \, \frac{N_c \sqrt{2 N_c (N_c^2 - 1)}}{24 x} \, 
  \epsilon \, \Big( 6 \zeta(2) + 28 \epsilon \zeta(3) \Big) \, ,
\eeqa
while the NLL result, up to $\ord{\eps^2}$, is given by
\beqa
\label{Hqgqg1nll}
  H^{(1),0,[1]}_{qg} & = & {\rm i} \pi \, \frac{\sqrt{N_c (N_c^2 - 1)}}{12 x} \, 
  \epsilon \, \Big( 6 \zeta(2) + 28 \epsilon \zeta(3) \Big) \, , \nonumber \\
  H^{(1),0,[8_s]}_{qg} & = & {\rm i} \pi \, \frac{\sqrt{2 N_c (N_c^2 - 1)(N_c^2 - 4)}}{48 x} \, 
  \epsilon \, \Big( 6 \zeta(2) + 28 \epsilon \zeta(3) \Big) \, , \\
  H^{(1),0,[8_a]}_{qg} & = & \frac{\sqrt{2 N_c (N_c^2 - 1)}}{x} \, \Bigg\{ 
  \left( - \frac{3}{4} + \frac{15}{8} \zeta(2) \right) N_c + 
  \left(1 - \frac{1}{8} \zeta(2) \right) \frac{1}{N_c} \nonumber \\
  & & + \, \epsilon \,\bigg[ \left(- \frac{3}{2} + \frac{3}{16} \zeta(2) + \frac{9}{4} \zeta(3) \right) N_c 
  + \left( 2 - \frac{3}{16} \zeta(2) - \frac{7}{12} \zeta(3) \right) \frac{1}{N_c} \bigg] \nonumber \\
  & & + \, \epsilon^2 \, \bigg[ \left( 3 + \frac{3}{8} \zeta(2) + \frac{7}{8} \zeta(3) + 
  \frac{117}{64} \zeta(4)  \right) N_c \nonumber \\
  & & \hspace{1cm} + \, \left( 4 - \frac{1}{2} \zeta(2) - \frac{7}{8} \zeta(3) - 
  \frac{47}{64} \zeta(4) \right) \frac{1}{N_c} \bigg] \nonumber \\
  & & + \, {\rm i} \pi \, \frac{N_c}{48} \, \epsilon \, \Big( 6 \zeta(2) + 28 \epsilon \zeta(3) \Big) 
  \Bigg\} \, . \nonumber 
\eeqa
At two loops, leading-logarithmic contributions to the hard functions vanish to $\ord(\eps^0)$,
\beq
\label{Hqgqg2ll}
  H^{(2),2,[k]}_{qg} \, = \, 0 \, . 
\eeq
At NLL accuracy, on the other hand, the singlet and symmetric octet components vanish, 
but the antisymmetric octet component $H^{(2),1,[8_a]}$ is related to the finite part of the 
two-loop Regge trajectory, and one finds
\beqa
\label{Hqgqg2nll}
  H^{(2),1,[1]}_{qg} & = & H^{(2),1,[8_s]}_{qg} \, = \, 0 \, , \nonumber \\
  H^{(2),1,[8_a]}_{qg} & = & - \, \frac{N_c \sqrt{2 N_c (N_c^2 -1)}}{216 x} \, 
  \bigg[ \Big( - 202 + 27 \zeta(3) \Big) N_c + 28 \, n_f \bigg] \, .
\eeqa
Finally, all colour components of $H^{(2),0}$ are non vanishing, and are given by
\beqa
\label{Hqgqg2nnll}
  H^{(2),0,[1]}_{qg} & = & {\rm i} \pi \, \frac{\sqrt{N_c (N_c^2 - 1)}}{N_c \, x} \,
  \Bigg[ \left( \frac{55}{27} + \frac{10}{3} \zeta(2) - \frac{5}{4} \zeta(3) \right) N_c^2 -
  \frac{3}{16} \, \frac{1}{N_c^2} + \frac{83}{216} \, N_c \, n_f \nonumber \\ 
  & & - \, \left( \frac{1}{6} + \frac{1}{3} \zeta(2) \right) \frac{n_f}{N_c} - \frac{1}{16} 
  \Bigg], \nonumber \\
  H^{(2),0,[8_s]}_{qg} & = & {\rm i} \pi \, \frac{\sqrt{2 N_c (N_c^2 - 1)(N_c^2 - 4)}}{N_c \, x}
  \Bigg[ \left( - \frac{101}{216} + \frac{\zeta(3)}{16} \right) N_c^2 - 
  \frac{3}{32} \, \frac{1}{N_c^2} + \frac{7}{108} \, N_c \, n_f \nonumber \\
  & & - \, \left( \frac{1}{6}  + \frac{1}{3} \zeta(2) \right) \frac{n_f}{N_c} - \frac{3}{32} 
  \Bigg] \, , \\
  H^{(2),0,[8_a]}_{qg} & = & \frac{\sqrt{2 N_c (N_c^2 - 1)}}{x} \, \Bigg\{
  \left( - \frac{30377}{13824} + \frac{17}{9} \zeta(2) + \frac{43}{48} \zeta(3) + 
  \frac{501}{256} \zeta(4) \right) N_c^2 \nonumber \\
  & & \hspace{1.5cm} + \, \left( \frac{255}{512} + \frac{21}{64} \zeta(2) - 
  \frac{15}{32} \zeta(3) - \frac{83}{256} \zeta(4) \right) \frac{1}{N_c^2} \nonumber \\
  & & \hspace{1.5cm} + \left( \frac{863}{3456} - \frac{127}{288} \zeta(2) - 
  \frac{7}{48} \zeta(3) \right) \, N_c \, n_f  \nonumber \\
  & & \hspace{1.5cm} - \left(\frac{4085}{10368} + \frac{23}{288} \zeta(2) + 
  \frac{1}{144} \zeta(3) \right) \frac{n_f}{N_c} \nonumber \\ 
  & & \hspace{1.5cm} + \, \frac{19139}{10368} + \frac{985}{576} \zeta(2) - 
  \frac{205}{288} \zeta(3) - \frac{87}{128} \zeta(4) \nonumber \\
  & & \hspace{1.5cm} + \, {\rm i} \pi \, \frac{N_c}{432} \, \bigg[ 
  \Big( - 202 + 27 \zeta(3) \Big) N_c + 28 \, n_f \bigg] \Bigg\} \, . \nonumber 
\eeqa


\section{Infrared singularities for singlet exchange}
\label{singlet}

In section \ref{even-even} we used the dipole formula to investigate the structure of infrared 
singularities for the $t$-channel exchange of colour representations other than the octet, 
providing examples at three and four loops. In this appendix, as a further example of the 
dipole formula at work, and for future reference, we give explicit expressions for the infrared 
singularities in the case of singlet exchange. We begin with the poles of on-loop amplitudes. 
Leading logarithms are given by (\ref{AmpCoeff1}),
\beq
   M^{(1),1} \, = \, K^{(1)} \, {\bf O}_t \, H^{(0)} + {\cal O} (\epsilon^0) \, ,
\label{M11sing}
\eeq
where the operators ${\bf O}_t$ and hard parts $H^{(0)}$ for the three processes are given 
respectively in eqns.~(\ref{TsTtqqqq}) and (\ref{Hqqqqtree}), (\ref{Ttgggg}) and (\ref{Hgggg0}), 
(\ref{TtTsqgqg}) and (\ref{Hqgqg0}). The coefficient $K^{(1)} = \frac{1}{2 \epsilon}$ can be 
extracted by replacing \eq{hatgammaK} in \eq{formalint}. For example, in quark-quark 
scattering one finds the simple expressions
\beq
  M^{(1),1}_{qq} \, = \, \frac{1}{2 \epsilon} \left[
  \begin{array}{cc}
    0 & 0 \\
    0 & N_c 
  \end{array}
  \right] \left(
  \begin{array}{c}
  0 \\ \frac{\sqrt{N_c^2 - 1}}{x} 
  \end{array}
  \right)=\left(
  \begin{array}{c}
  0 \\
  \frac{N_c\sqrt{N_c^2 - 1}}{2 \epsilon \, x} 
  \end{array}
  \right) + {\cal O} \left( \epsilon^0 \right) \, .
\label{M11qqsing}
\eeq
The first component of the vector corresponds to the exchange of a color singlet in the 
$t$ channel and, as expected, it vanishes at leading logarithmic accuracy. The same 
result holds for gluon-gluon and quark-gluon scattering. We then use the same procedure 
for next-to-leading logarithms, by replacing the operators ${\bf O}_{s - u}$ in the proper 
representation and the anomalous dimensions we find in appendix \ref{AppAnDim} in 
the first expression of \eq{AmpCoeff1}. Singlet components in this case are
\beqa
\label{M10}
  M^{(1),0,[1]}_{qq} & = & {\rm i} \pi \, \frac{s}{t} \, \frac{N_c^2 - 1}{2 N_c} \, 
  \frac{1}{\epsilon} + {\cal O} \left( \epsilon^0 \right) \, , \nonumber \\
  M^{(1),0,[1]}_{gg} & = & {\rm i} \pi \, \frac{s}{t} \, 2 N_c^2 \, \frac{1}{\epsilon} + 
  {\cal O} \left( \epsilon^0 \right) \, , \\
  M^{(1),0,[1]}_{qg} & = & - {\rm i} \pi \, \frac{s}{t} \, \sqrt{N_c (N_c^2 - 1)} \, 
  \frac{1}{\epsilon} + {\cal O} \left( \epsilon^0 \right) \, . \nonumber
\eeqa
We next consider two-loop amplitudes, isolating leading and subleading logarithms. 
The singularities are constructed according to \eq{AmpCoeff2}. We find again that 
leading logarithms have just the (antisymmetric) octet component, while the 
next-to-leading terms are
\beqa
\label{M21}
  M^{(2),1,[1]}_{qq} & = & {\rm i} \pi \, \frac{s}{t} \, \frac{N_c^2 - 1}{8 \epsilon^2} + 
  {\cal O} \left( \epsilon^0 \right) \, , \nonumber \\
  M^{(2),1,[1]}_{gg} & = & {\rm i} \pi \, \frac{s}{t} \, \frac{N_c^3}{2 \epsilon^2} +
  {\cal O} \left( \epsilon^0 \right) \, , \\
  M^{(2),1,[1]}_{qg} & = & - {\rm i} \pi \, \frac{s}{t} \, 
  \frac{N_c \sqrt{N_c (N_c^2 - 1)}}{4 \epsilon^2} + {\cal O} \left( \epsilon^0 \right) \, .
\nonumber
\eeqa
Finally, at next-to-next-to-leading logarithmic accuracy we find
\beqa
\label{M20}
  M^{(2),0,[1]}_{qq} & = & \frac{s}{t} \, \frac{N_c^2 - 1}{N_c} \Bigg\{ - {\rm i} \pi \, 
  \frac{N_c^2 - 1}{4 N_c \, \epsilon^3} + \frac{1}{\epsilon^2} \bigg[ 
  \frac{3}{2} \frac{1}{N_c} \zeta(2) + {\rm i} \pi \, \bigg( - \frac{29}{48} N_c + 
  \frac{1}{24} n_f + \frac{3}{8} \frac{1}{N_c} \bigg) \bigg] \nonumber \\
  & & + \, {\rm i} \pi \, \frac{1}{\epsilon} \bigg[ N_c \left( \frac{31}{48} + 
  \frac{7}{8} \zeta(2) \right) - \frac{5}{24} n_f + \frac{1}{N_c} \left(
  1 - \frac{1}{4} \zeta(2) \right) \bigg] \Bigg\} \, , \nonumber \\
  M^{(2),0,[1]}_{gg} & = & \frac{s}{t} \, N_c^2 \, \Bigg\{ - 2 {\rm i} \pi \, \frac{N_c}{\epsilon^3} +
  \frac{1}{\epsilon^2} \bigg[ \frac{3}{2} N_c \, \zeta(2) + {\rm i} \pi \, \left( - \frac{55}{12} N_c +
  \frac{5}{6} n_f \right) \bigg] \\
  & & + \, {\rm i} \pi \, \frac{1}{\epsilon} \bigg[ N_c \left( - \frac{67}{36} + \frac{9}{2} \zeta(2)
  \right) + \frac{5}{18} n_f \bigg] \Bigg\} \, , \nonumber \\
  M^{(2),0,[1]}_{qg} & = & \frac{s}{t} \sqrt{\frac{N_c^2 - 1}{N_c}} \Bigg\{ {\rm i} \pi \,
  \frac{3 N_c^2 - 1}{4 \epsilon^3} + \frac{1}{\epsilon^2} \bigg[ 
  - \frac{3}{4} N_c^2 \, \zeta(2) + {\rm i} \pi \, \left( \frac{7}{4} N_c^2 - \frac{1}{4} n_f N_c -
  \frac{3}{8} \right) \bigg] \nonumber \\
  & & - \, {\rm i} \pi \, \frac{1}{\epsilon} \bigg[ N_c^2 \left( \frac{13}{72} + 2 \zeta(2) \right) - 
  \frac{5}{36} n_f N_c + 1 - \frac{1}{4} \zeta(2) \bigg] \Bigg\} \, . \nonumber
\eeqa


\bibliographystyle{JHEP}


\end{document}